\documentclass[twocolumn,pre,aps,floatfix,superscriptaddress,longbibliography]{revtex4-2}
\pdfoutput=1 
\usepackage{amsmath,amssymb,eucal,graphicx,float,epstopdf}
\usepackage{epsfig,algorithm,algcompatible}
\usepackage[utf8]{inputenc}

\usepackage[colorlinks=true, urlcolor=blue, anchorcolor=blue, citecolor=blue,filecolor=blue,linkcolor=blue,menucolor=blue]{hyperref}

\usepackage{subfigure}

\def\beq{\begin{eqnarray}}
\def\eeq{\end{eqnarray}}
\def\be{\begin{equation}}
\def\ee{\end{equation}}
\def\bel{\begin{equation} \label}
\def\beel{\begin{eqnarray} \label}

\def\bm{\begin{math}}
\def\me{\end{math}}

\def\q{\quad}
\def\qq{\qquad}

\begin{document}

\title{Supercluster states and phase transitions in aggregation-fragmentation processes}

\author{Wendy Otieno}
\affiliation{Department of Physics, Loughborough University, Loughborough LE11 3TU, UK}
\author{Nikolai V. Brilliantov}
\affiliation{Skolkovo Institute of Science and Technology, 30 Bolshoi Boulevard, Moscow, 121205, Russia}
\affiliation{Department of Mathematics, University of Leicester, Leicester LE1 7RH, UK}
\author{P. L. Krapivsky}
\affiliation{Department of Physics, Boston University, Boston, Massachusetts 02215, USA}
\affiliation{Santa Fe Institute, Santa Fe, New Mexico 87501, USA}

\begin{abstract}
We study the evolution of aggregates triggered by collisions with monomers that either lead to the attachment of monomers or the break-up of aggregates into constituting monomers. Depending on parameters quantifying addition and break-up rates, the system falls into a jammed or a steady state. Supercluster states (SCSs) are very peculiar non-extensive jammed states that also arise in some models. Fluctuations underlie the formation of the SCSs. Conventional tools, such as the van Kampen expansion, apply to small fluctuations. We go beyond the van Kampen expansion and determine a set of critical exponents quantifying SCSs. We observe continuous and discontinuous phase transitions between the states. Our theoretical predictions are in good agreement with numerical results.
\end{abstract}

\maketitle

\section{Introduction}

The addition process is aggregation with incremental growth occurring by incorporating the monomers (aggregates of minimal mass). This process occurs at various temporal and spatial scales ranging from atomic to astrophysical. At the molecular level, addition is present in coagulation of erythrocytes (blood cells) yielding rouleaux \cite{EP1926,RWS1982,RWS1984}, aggregation of bacteria via dextran induction \cite{VR1980}, island growth where monomers (called adatoms in surface science) hop on the substrate while heavier clusters are immobile \cite{MBE,BK91,Blackman91,BlackmanMarshall_JPA94,Evans_PRB,Wolf,Zinke_Allmang1999, Krapivsky_EJB1998, Krapivsky_PRB1999, Family2001}, and many other examples \cite{Prions}. Addition processes underlie self-assembly \cite{RW04,SA08,Privman2009,Erik17}, synthesis of nanocrystals \cite{Privman2010,Privman2013}, merging of point defects in solids \cite{Koiwa,JNM2011}, etc.  Aggregation is often counterbalanced by fragmentation. Aggregation and fragmentation processes play an important role in polymer physics \cite{Blatz}, they contribute to the formation of stars and planetary rings \cite{Guettler2010,PNAS,esposito2006,BBK2009}, etc.

Addition and disintegration mimic social phenomena, e.g., users joining forums which may eventually disintegrate (partially or completely). Aggregating and disintegrating objects in social networks may be also firms, enterprises, etc. \cite{Dorogov,Socnet1,Socnet2}.

\begin{figure}
\centering
\includegraphics[width = 6cm]{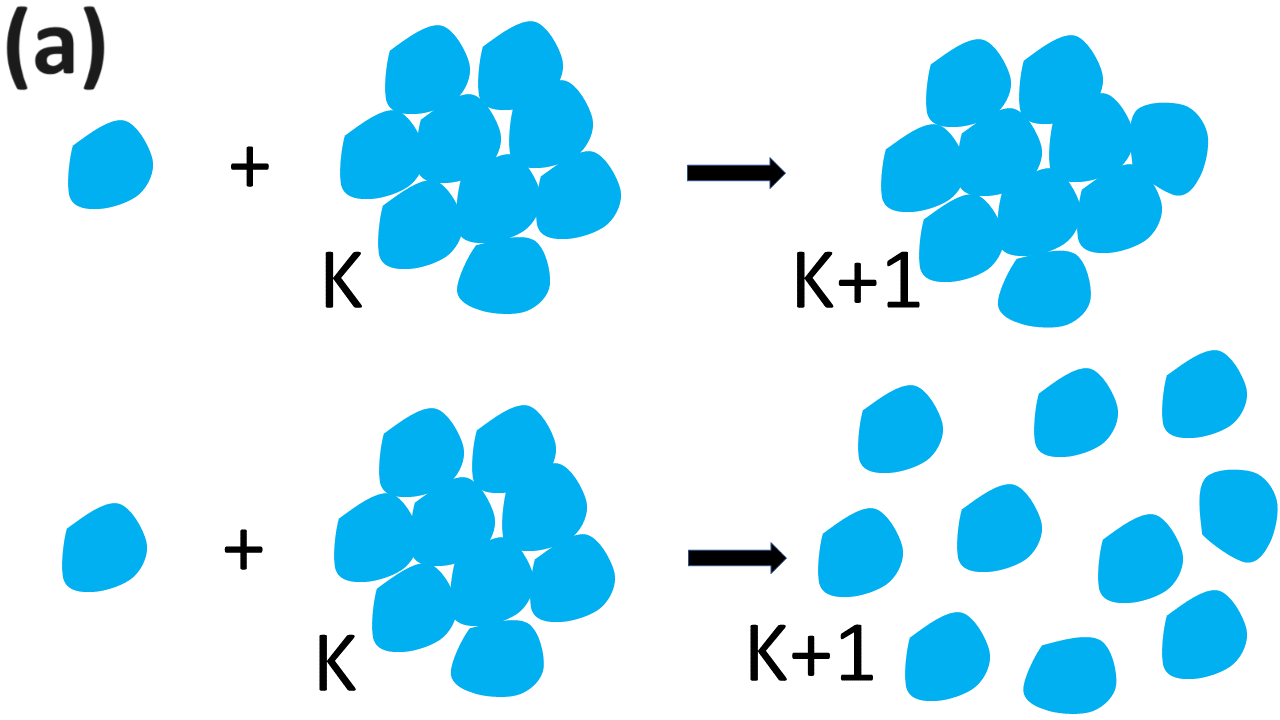}
\includegraphics[width = 6cm]{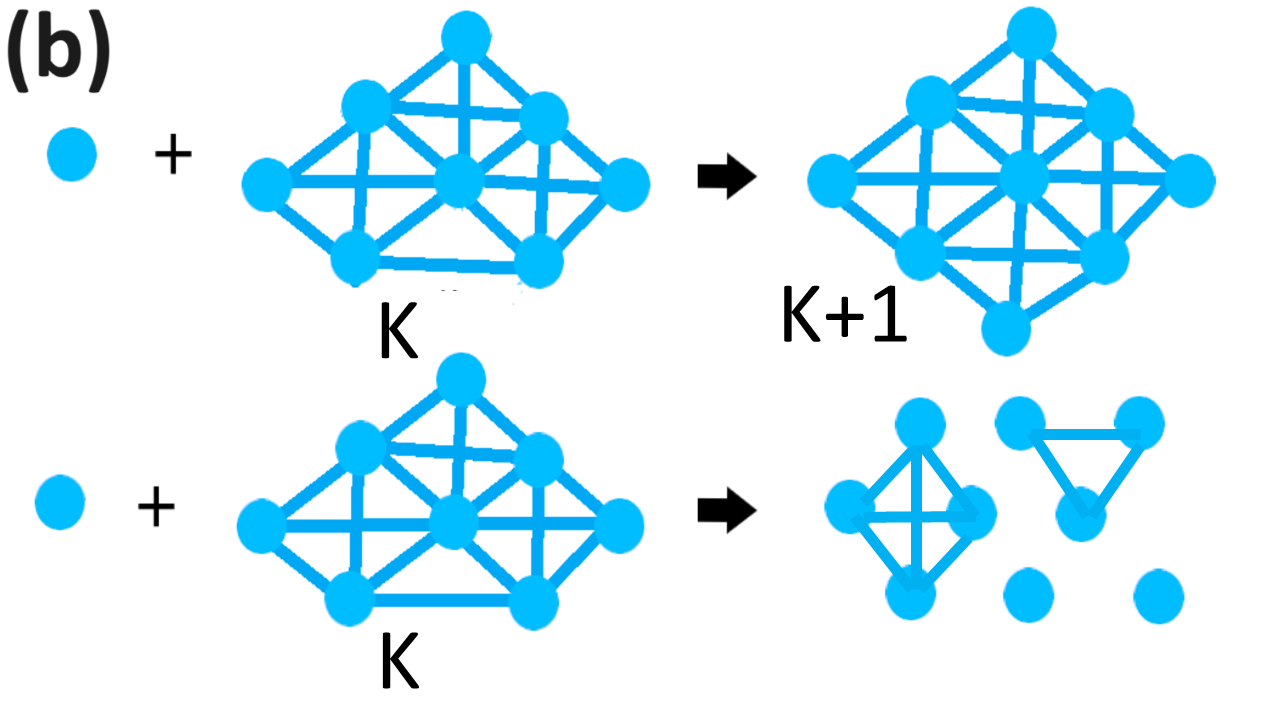}
\caption{ Addition with collision-induced disintegration: (a) illustrates material systems (complete disintegration is shown) and (b) shows networks (partial disintegration is shown).} \label{illustr}
\end{figure}

The addition process is symbolically represented by the reaction scheme (see also Fig.~\ref{illustr})
\begin{equation*}
M + I_{k} \xrightarrow{A_{k}} I_{k+1}.
\end{equation*}
Thus an elementary object (a monomer denoted by $M$) collides with another object (a monomer or a cluster) to form a cluster of larger mass.  Here $I_{k}$ denotes a cluster composed of $k$ monomers ($k$-mer), and $A_{k}$ is the merging rate. The Becker-D{\"o}ring equations \cite{Ball1986,KingWattis2002,Niethammer2003,Wattis2006} and their continuum counterpart, Lifshitz-Slyozov-Wagner model \cite{Niethammer1999,Herrmann2009}, rely on aggregation with addition mechanism.

Fragmentation works concurrently with aggregation. In fragmentation, clusters break into smaller clusters. Fragmentation can occur spontaneously or via  collisions. We consider fragmentation processes caused by collisions with monomers, i.e., dual to addition that is also caused by collisions with monomers. Schematically
\begin{equation}
\label{schemfrag}
M + I_{k} \xrightarrow{S_{k,\ell}} M + I_{j_{1}} + .... + I_{j_{\ell}}
\end{equation}
where $ \sum_{i=1}^{\ell} j _{i}= k$. We thus tacitly assume that a monomer always remains separated after collision, mimicking the situation when an energetic monomer hits a cluster and retains its identity. The collision-induced disintegration processs \eqref{schemfrag} is described in the framework of Oort-Hulst-Safranov-Dubovski models \cite{Laurencot2001,OortHulst,Laurencot2007,Dubovski1999}.  In complete disintegration, or shattering, clusters break into monomers \cite{Guettler2010,BlumErosion2011,KrapivskyBenNaim2003,PNAS,KRB}. Symbolically
\begin{equation}
M + I_{k} \xrightarrow{S_{k}} \underbrace{M + M +\ldots+ M}_{k+1}.
\end{equation}

In this work, we explore systems with addition and fragmentation and report a rich set of behaviors, including continuous and discontinuous phase transitions, the formation of steady and jammed states, and the emergence of supercluster states (SCSs) which are peculiar jammed states. In the SCS, the cluster densities vanish in the thermodynamic limit. To shed light on the SCSs, we study finite systems and  show a non-extensive nature of the SCSs manifested by sub-linear scaling of the total number of clusters with system size. To describe the SCSs, we develop a framework extending the van Kampen expansion applicable to extensive systems.

The SCSs have been detected in our Letter \cite{BOK2021}. Here we present a more detailed analysis of the SCSs. We consider a model with complete disintegration (shattering), and also with partial disintegration, and demonstrate the  emergence of SCSs in both cases. Hence we conjecture that SCSs are generic for systems with addition and fragmentation. We also present a detailed analysis and classification of phase transitions that were only briefly mentioned in  \cite{BOK2021}.

Our analytical treatment is focused on addition with shattering. Each such model is characterized by a set of rates $A_k$ and $S_k$, and some models are analytically tractable as we demonstrate in Sects.~\ref{sec:AS}--\ref{sec:nature}.

\section{Addition and shattering processes}
\label{sec:AS}

The governing equations describing addition and shattering read
\begin{subequations}
\label{ck_AS}
\begin{align}
\label{dc1awcd}
\frac{d c_{1}}{dt} &= -2A_{1}c_{1}^{2} - \sum_{j=2}^{\infty} A_{j}c_{j}c_{1} +\sum_{j=2}^{\infty} jS_{j}c_{j}c_{1}, \\
\label{dckawcd}
\frac{dc_{k}}{dt} &= A_{k-1}c_{1}c_{k-1} - A_{k}c_{k}c_{1} - S_{k}c_{1}c_{k},
\end{align}
\end{subequations}
where Eqs.~\eqref{dckawcd} apply to all $k\geq 2$. The first and second terms on the right-hand side of Eq.~\eqref{dc1awcd} describe the loss of monomers via addition due to monomer-monomer and monomer-cluster interactions. The last term in (\ref{dc1awcd}) describes the gain of monomers due to shattering. Similarly, the last two terms on the right-hand of Eqs.~(\ref{dckawcd}) represent the loss of $k$-mers due to addition and shattering, while the first term gives the gain due to addition.

In writing Eqs.~\eqref{ck_AS} we tacitly assume that the system is well-mixed, spatially homogeneous, and dilute. Even when these assumptions are satisfied, the description provided by Eqs.~\eqref{ck_AS} is mean-field in nature, so it may be erroneous in low spatial dimensions  \cite{KRB,van89}.

For many practical applications the rates of addition and shattering depend algebraically on the cluster size:
\begin{equation}
\label{ASk}
A_{k} = k^{a}, \qq S_{k} = \lambda k^{s}.
\end{equation}
Here we set the amplitude of the addition rate to unity by using the appropriate time units; the constant $\lambda$ quantifies the shattering intensity.  The dependence $A_{k} \sim k^a$ is natural since the aggregation rate is often proportional to the surface area  \cite{Ernst85a,Leyvraz2003,Wattis2006,BrilBodKrap2009}. This additionally implies that $a\leq 1$. Models with rates growing faster than mass, $a>1$, are also ill-defined in the thermodynamic limit as an infinite cluster forms at time $t=+0$, see e.g. \cite{van87,Malyshkin01,Colm11,Leyvraz12}). For networks, growth with exponent $a>0$ reminds preferential attachment \cite{Dorogov}, and the behavior also drastically changes when $a>1$, see \cite{KrapRedner2001}. The exponent $s$ is determined by the shattering mechanism and it usually satisfies the constraint $s \leq 1$. By re-scaling densities, we set the mass density to unity if not stated otherwise:
\begin{equation}
\label{mass}
\sum_{j= 1}^{\infty} jc_j=1.
\end{equation}

Using the modified time
\bel{taut}
\tau = \int_0^{t} dt' \, c_{1}(t')
\ee
we linearize Eqs.~\eqref{ck_AS}
\begin{subequations}
\label{ck_AS_tau}
\begin{align}
\frac{d c_{1}}{d \tau} &= -(1+\lambda)c_{1} - M_{a} + \lambda M_{1+s}, \label{dc1awcdlinearlized} \\
\frac{dc_{k}}{d \tau} &= (k-1)^{a} c_{k-1} - (k^{a} + \lambda k^{s} )c_{k}, \quad k\geq 2.
\label{dckawcdlinearized}
\end{align}
\end{subequations}
Hereinafter $M_\nu= \sum_{k\geq 1} k^\nu c_k$ denotes the $\nu^\text{th}$ moment.

Equation \eqref{dc1awcdlinearlized} is not closed. If the system of equations for $c_1, M_a, M_{1+s}$ is closed, one can proceed analytically. The moment $M_\nu$ evolves according to
\begin{eqnarray}
\label{Mn}
\frac{d M_\nu}{d \tau} & = & \sum_{k\geq 1} (k+1)^\nu k^a c_k \nonumber\\
& - & M_a - M_{a+\nu} +\lambda(M_{1+s}-M_{\nu+s}).
\end{eqnarray}
The term $\sum_{k\geq 1} (k+1)^\nu k^a c_k$ on the right-hand side of \eqref{Mn} can be expressed through the moments only when $\nu$ is a non-negative integer. Thus closed equations for the moments emerge when $a$ and $1+s$ are non-negative integers. In the physically acceptable range $a\leq 1$ and $s\leq 1$, there are five possibilities:  $(a,s)=(1,1)$, $(a,s)=(0,1)$, $(a,s)=(1,0)$, $(a,s)=(0,0)$  and  $(a,s)=(0,-1)$;  these models admit analytical treatment. The one-parameter class of models with exponents $(a,s)=(a,a-1)$ is also partly tractable as we shall show below.

The systems with mass-independent and linear in mass rates, $(a,s)=(0,0)$ and $(a,s)=(1,1)$, have been studied in \cite{KOB}. Some results for the models with $(a,s)=(0,1)$, $(a,s)=(0,-1)$ and $(a,s)=(a,a-1)$ appear in \cite{BOK2021}. The most interesting SCSs occur in a class of models with $(a,s)=(a,a-1)$. Therefore we begin with two more tractable models of that type: $(a,s)=(1,0)$ and $(a,s)=(0,-1)$. We then turn to the class of models with $(a,s)=(a,a-1)$ and demonstrate its peculiarity in the general class of models \eqref{ASk} with algebraic reaction rates. In this section we consider only infinite addition-shattering processes.

\subsection{The model with $(a,s) = (1,0)$}
\label{subsec:10}

For rates $A_k=k$ and $S_k=\lambda$, the governing kinetic equations read
\begin{subequations}
\label{Ncc:10}
\begin{align}
\label{dNmassdependar}
\frac{dN}{d \tau} &= - \lambda N + \lambda - 1, \\
\label{dc1massdependar}
\frac{d c_{1}}{d \tau} &= -(1+\lambda)c_{1} - 1 + \lambda,  \\
\frac{dc_{k}}{d \tau} &= (k-1) c_{k-1} - (k + \lambda)c_{k}, \quad k\geq 2.
\label{dckmassdependar}
\end{align}
\end{subequations}

If not stated otherwise, we always consider the most natural mono-disperse initial condition
\begin{equation}
\label{ckt=0}
c_k(t=0)=\delta_{k,1}.
\end{equation}
Solving \eqref{dNmassdependar}--\eqref{dc1massdependar} subject to \eqref{ckt=0} gives
\begin{subequations}
\begin{align}
\label{N:10-sol}
N(\tau) &= \frac{\lambda - 1}{\lambda} + \frac{e^{-\lambda \tau}}{\lambda},\\
\label{c1:10-sol}
c_{1}(\tau) & = \frac{2}{1+\lambda}\,e^{-(1+\lambda)\tau} - \frac{1-\lambda}{1+\lambda}.
\end{align}
\end{subequations}

Different behaviors emerge for $\lambda < 1$, $\lambda = 1$ and $\lambda > 1$. In the subcritical region, $\lambda < 1$, the monomer density vanishes at $\tau = \tau_{\rm max}(\lambda)$. Setting $c_{1}(\tau_{\rm max}) = 0$ in \eqref{c1:10-sol} and solving for $\tau_{\rm max}$ yields
\begin{equation}
\label{tau-max:10}
\tau_{\rm max}(\lambda) = \frac{1}{1+ \lambda} \ln \bigg(\frac{2}{1-\lambda}\bigg)
\end{equation}
implying that $\tau_{\rm max}(\lambda)$ is an increasing function of $\lambda$. The final modified time increases from $\tau_{\rm max}(0) = \ln(2)$ to $\tau_{\rm max}(1) = \infty$. The quantity $\tau_{\rm max}(\lambda_c)$ remained finite in models with $(a,s)=(0,0)$ and $(a,s)=(1,1)$ studied in \cite{KOB}, and also in the model with $(a,s)=(0,1)$. This is a mathematical reason for the peculiarity of the critical regime in the present model.

\begin{figure}
\centering
\includegraphics[width=8.25cm, height = 5.5cm]{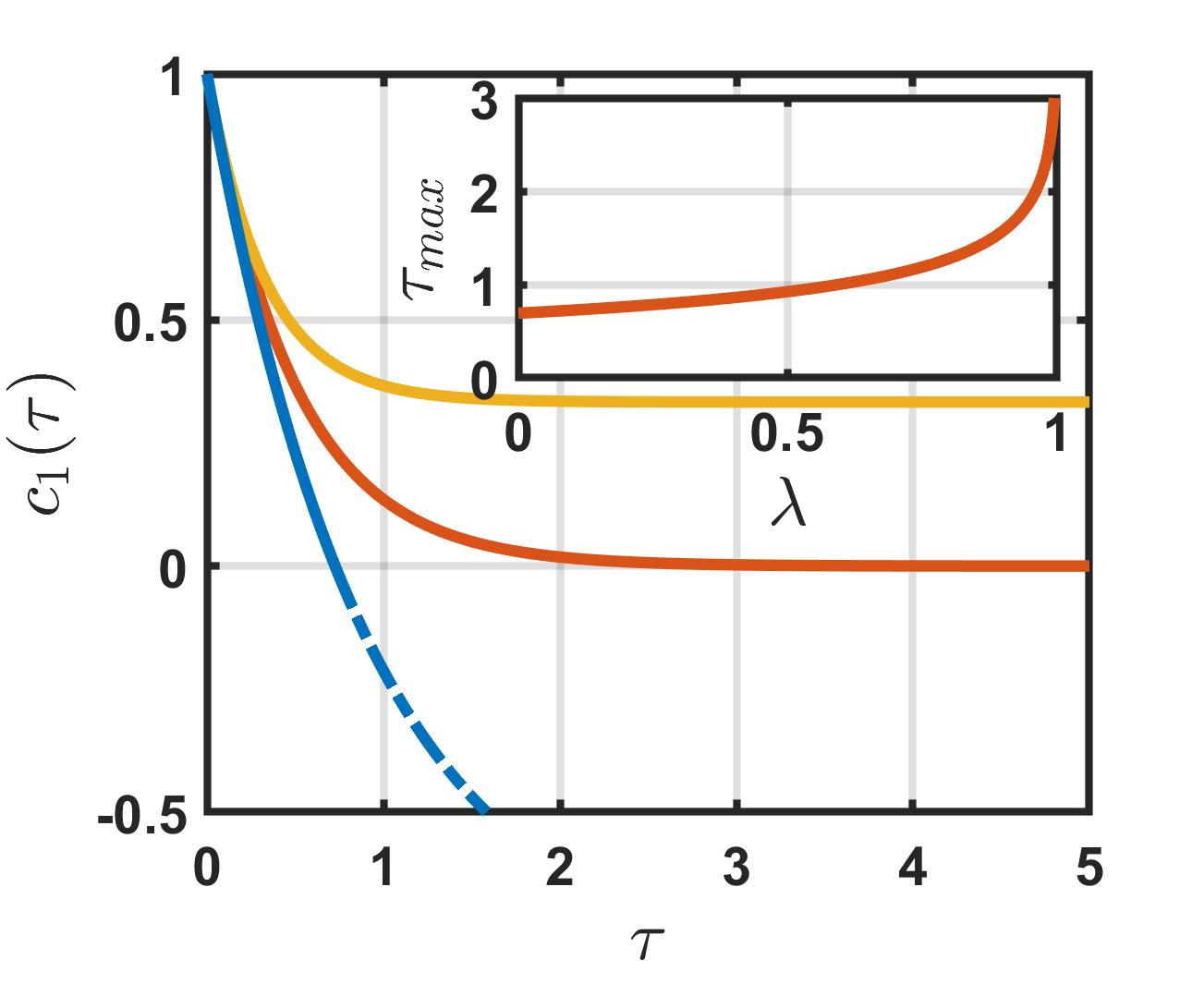}
\caption{The evolution of the monomer density for the model with $(a,s)=(1,0)$. Bottom to top: subcritical ($\lambda = 0$), critical ($\lambda =1$), and supercritical ($\lambda=1.19$) behaviors illustrating the relaxation of the monomer density in the jammed, supercluster and steady state.  Inset: $\tau_\text{max}$ is an increasing function of $\lambda$. The final modified time $\tau_{\rm max}(\lambda)$ logarithmically diverges as $\lambda\uparrow 1$, see \eqref{tau-max:10}. }
 \label{Fig:C1_tau}
\end{figure}

Combining \eqref{N:10-sol}  and  \eqref{tau-max:10} gives the total final density (valid for the mono-disperse initial condition):
\begin{equation}
\label{N:10-inf}
N_{\infty}(\lambda) = \frac{\lambda - 1}{\lambda} + \frac{1}{\lambda}\bigg[ \frac{1-\lambda}{2} \bigg]^{\frac{\lambda}{1+\lambda}}.
\end{equation}

Equations \eqref{dckmassdependar} can be solved recurrently starting with the monomer density \eqref{c1:10-sol}. One finds
\begin{eqnarray*}
c_2&=& \frac{\lambda-1}{(1+\lambda)(2+\lambda)} + \frac{2}{1+\lambda}\, e^{-(1+\lambda)\tau}  -
\frac{3}{2+\lambda}\, e^{-(2+\lambda)\tau} \\
c_3 &=& \frac{2(\lambda-1)}{(1+\lambda)(2+\lambda)(3+\lambda)} + \frac{2}{1+\lambda}\, e^{-(1+\lambda)\tau} \\
      &-& \frac{6}{2+\lambda}\, e^{-(2+\lambda)\tau} + \frac{4}{3+\lambda}\, e^{-(3+\lambda)\tau}
\end{eqnarray*}
etc. Specifying to $\tau_{\rm max}$ given by Eq.~\eqref{tau-max:10} one establishes $c_k(\infty)$. The results become more and more cumbersome, and we have not found a general compact formula valid for all $k$. We mention exact results only in the extreme case of $\lambda=0$ when shattering is absent. This pure addition process was solved in \cite{BK91}. The densities are
\begin{equation}
\label{ck-tau:10}
c_k(\tau) = e^{-\tau}(1-e^{-\tau})^{k-1} - k^{-1}(1-e^{-\tau})^k
\end{equation}
and $N = 1-\tau$. The monomer density $c_1=2 e^{-\tau}-1$ is shown in Fig.~\ref{Fig:C1_tau}.
The final densities at $\tau_{\rm max}(0) = \ln(2)$  corresponding to $t=\infty$ are
\begin{equation}
\label{ck-final:10}
c_k(\infty)=\big(1-k^{-1}\big)\,2^{-k}, \quad N_\infty =1-\ln 2.
\end{equation}

In the other extreme, namely in the proximity of the critical point, $0< 1-\lambda \ll 1$, the final densities
\begin{equation}
\label{Ck10:sub}
c_k(\infty) = \left[\frac{1}{2}-\frac{1}{k(k+1)}\right](1-\lambda) + O[(1-\lambda)^{3/2}]
\end{equation}
for $k\geq 2$. Thus the final densities vanish linearly, $c_k(\infty)\sim 1-\lambda$. The total cluster density $N_\infty$ also vanishes, but in a different manner, namely as $\sqrt{1-\lambda}$ in the $\lambda\uparrow 1$ limit. This follows from Eq.~\eqref{N:10-inf}.

The critical regime occurs at $\lambda = \lambda_{c} = 1$. Solving Eqs.~\eqref{Ncc:10} subject to \eqref{ckt=0} yields
\begin{subequations}
\label{Ncc:10-sol}
\begin{align}
\label{N:10-sol-crit}
N(\tau)      & = e^{-\tau}, \\
\label{ck:10-sol-crit}
c_k(\tau) & = e^{-2\tau}(1-e^{-\tau})^{k-1}.
\end{align}
\end{subequations}
Inverting the definition \eqref{taut} we obtain
\begin{equation}
\label{ttau}
t = \int_{0}^{\tau} \frac{d\tau'}{c_{1}(\tau')} = \frac{e^{2\tau} - 1}{2}
\end{equation}
allowing us to re-write \eqref{Ncc:10-sol} as
\begin{subequations}
\label{Ncc:10-sol-t}
\begin{align}
\label{N:10-sol-t}
N     & = \frac{1}{\sqrt{1+2t}}, \\
\label{ck:10-sol-t}
c_k & =  \frac{1}{1+2t}\bigg[1 - \frac{1}{\sqrt{1+2t}} \bigg]^{k-1}.
\end{align}
\end{subequations}
The final densities $c_{k}(\infty)$ vanish at $\lambda_{c} = 1$, yet the mass density is conserved. The same happens in pure aggregation where $c_k(t)\to 0$ as $t\to\infty$ yet the mass distribution widens, and the sum $\sum_{k\geq 1}kc_k(t)=1$ remains constant. The distinction with pure aggregation become clear if we compare the final outcomes in a finite system. All mass is engulfed by a single cluster in an aggregation process. In the addition and shattering process with rates $A_k=k$ and  $S_k=1$, the final state is very different as we show below. Monomers also disappear, but the overall number of clusters diverges with  total mass, albeit sub-linearly. We call such final outcome a supercluster state, as clusters are predominantly large.

In the supercritical regime ($\lambda>1$), the cluster densities relax to the steady state
\begin{subequations}
\label{Ncc:10-SS}
\begin{align}
\label{N:10-SS}
N     & = \frac{\lambda-1}{\lambda} \\
\label{ck:10-SS}
c_k & =  (\lambda - 1) \frac{\Gamma(k) \Gamma(1+\lambda)}{\Gamma(k+1+\lambda)}
\end{align}
\end{subequations}
following from Eqs.~\eqref{Ncc:10}. Summarizing
\begin{align*}
c_{1}(\infty) &=
\begin{cases}
0 & \lambda \leq 1 \\
\frac{\lambda - 1}{\lambda + 1} & \lambda > 1,
\end{cases} \\
\vspace{0.5cm}
c_{k}(\infty) &=
\begin{cases}
c_{k}(\tau_{\rm max}(\lambda))
 & \lambda \leq 1 \\
(\lambda - 1) \frac{\Gamma(k) \Gamma(1+\lambda)}{\Gamma(k+1+\lambda)} & \lambda > 1,
\end{cases} \\
\vspace{0.5cm}
N_{\infty}(\lambda) &=
\begin{cases}
N(\tau_{\rm max}(\lambda)) & \lambda \leq 1 \\
\frac{\lambda-1}{\lambda} & \lambda > 1
\end{cases} \\
\end{align*}
with $\tau_{\rm max}(\lambda)$ depending on the initial conditions (it is given by \eqref{tau-max:10} for the mono-disperse initial condition).

\begin{figure}
\centering
\includegraphics[width=8.25cm]{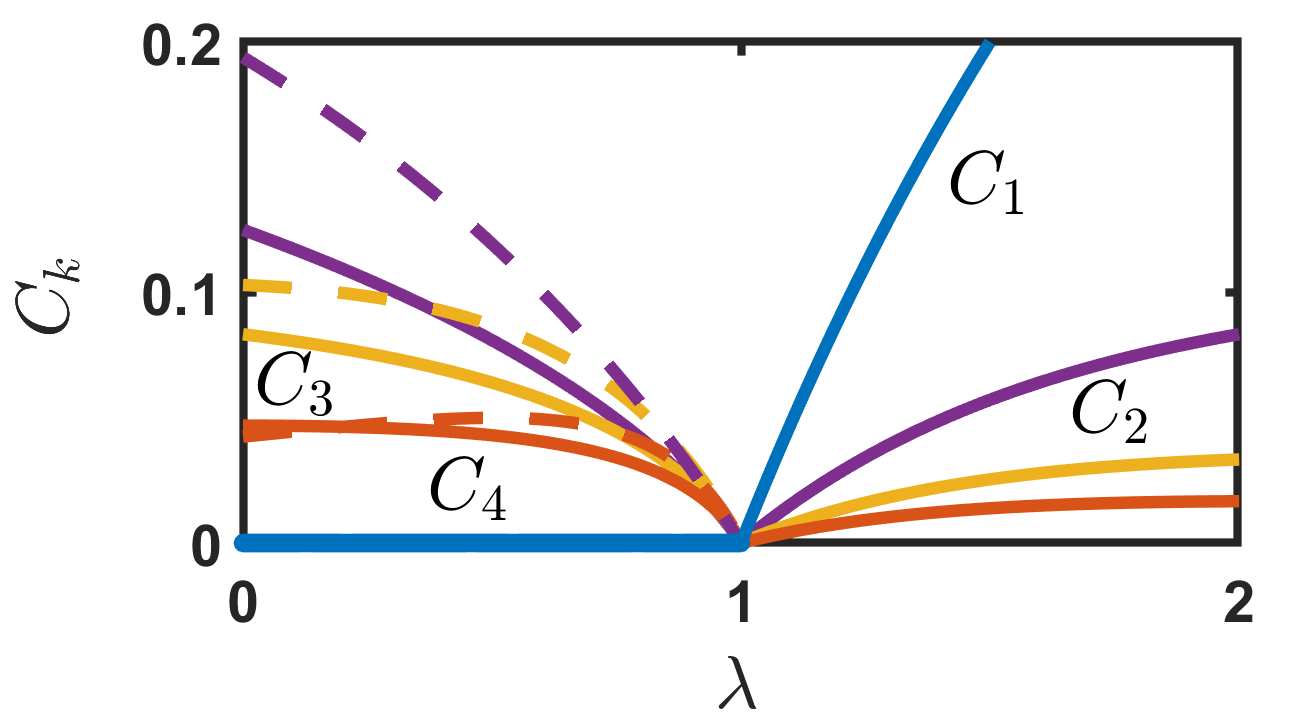}
\caption{When $(a,s)=(1,0)$, the system undergoes a continuous phase transition from a jammed state ($\lambda<\lambda_c=1$) to a steady state
($\lambda > \lambda_c=1$). The critical point, $\lambda_c=1$, corresponds to the supercluster state. The final densities in a jammed state depend on the initial conditions. Solid lines: The mono-disperse initial conditions \eqref{ckt=0}. Dashed lines: The monomer-dimer  initial conditions with $c_{1}(0) = 0.4$ and $c_{2}(0) = 0.3$.}
 \label{Fig:C1NCK}
\end{figure}

In the vicinity of the critical point
\begin{equation}
c_{k}(\infty)  \simeq |\lambda-1| \times
\begin{cases}
\frac{1}{2}-\frac{1}{k(k+1)} &\lambda\to 1-0\\
\frac{1}{k(k+1)} &\lambda\to 1+0
\end{cases}
\end{equation}
for $k\geq 2$. Thus we have a continuous phase transition from a jammed state to a steady state occurring through the critical supercluster state, see Fig.~\ref{Fig:C1NCK}.

\subsection{The model with $(a,s)=(0,-1)$}
\label{subsec:0-1}

\begin{figure}
\centering
\includegraphics[width=8.25cm]{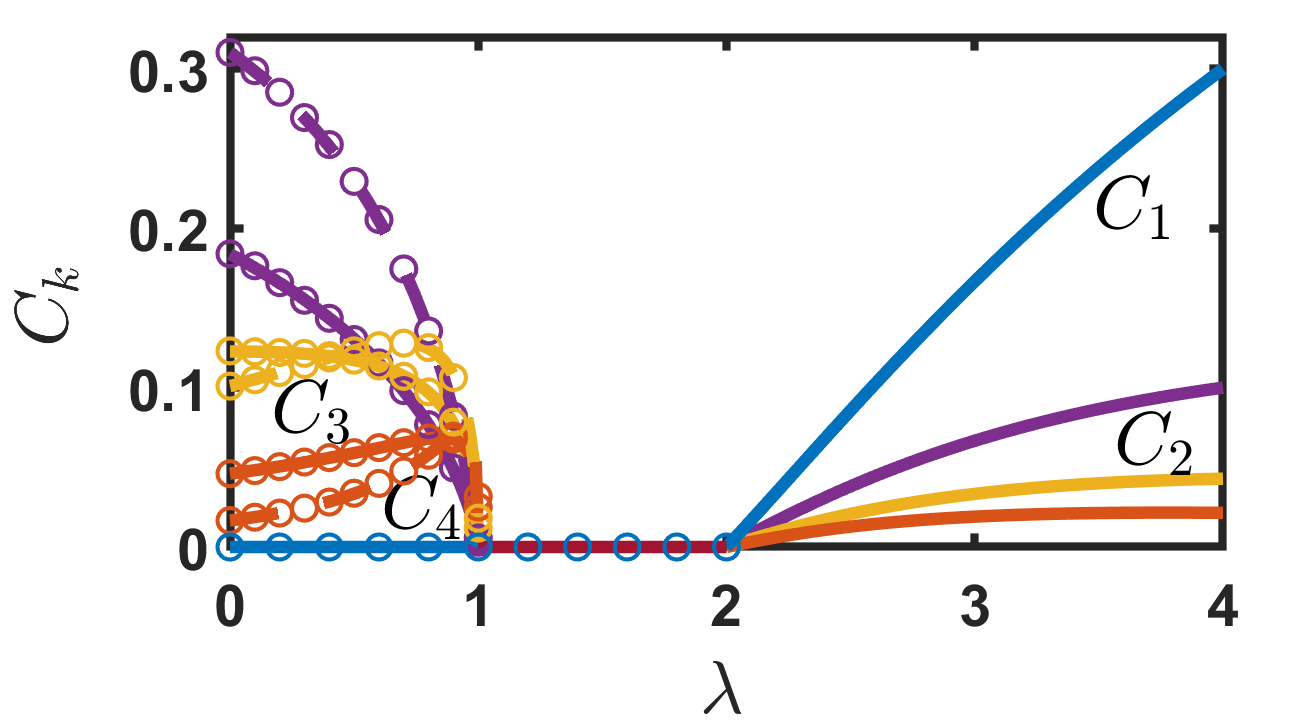}
\caption{The model with $(a,s)=(0,-1)$. A continuous phase transition from a jammed state at  $\lambda <1$ to the supercluster state for $ 1 < \lambda < 2$ is followed by a continuous phase transition from the supercluster state to a steady state at $\lambda_{\rm c,up}=2$. The final densities in a jammed state depend on the initial conditions: Shown are results for the mono-disperse initial conditions (solid lines) and the monomer-dimer initial conditions (dashed lines) with $c_{1}(0) = 0.2$ and $c_{2}(0) = 0.4$.  Curves: The analytical and numerical solution of the ODE. Dots: Monte Carlo results (in simulations, the total number of monomers was ${\cal M}=10^6$).}
 \label{Fig:C1CK0m1}
\end{figure}

In this model the evolution is governed by
\begin{subequations}
\label{c1k:0-1}
\begin{align}
\frac{d c_{1}}{d \tau} &= - (1+ \lambda)c_{1} + (\lambda - 1)N, \label{c1:0-1} \\
\frac{d c_{k}}{d \tau} &= c_{k-1} - ( 1+ \lambda/k) c_{k} .
\label{ck:0-1}
\end{align}
\end{subequations}
The critical shattering parameter $\lambda_c=1$ demarcates different regimes (Fig.~\ref{Fig:C1CK0m1}). In a subcritical regime, $\lambda <1$, the system eventually arrives at a jammed state with vanishing monomer density, $c_1( \tau_{\rm max})=0$, and cluster densities $c_k(\tau_{\rm max})$, determined by initial conditions.

When  $\lambda=\lambda_c=1$, one solves \eqref{c1k:0-1} iteratively to give
\begin{equation*}
\begin{split}
c_1  & = e^{-2 \tau}\\
c_2  & = -2e^{-2 \tau} + 2e^{-3\tau/2}\\
c_3  & = 3e^{-2 \tau}  - 12e^{-3\tau/2} + 9  e^{-4\tau/3}  \\
c_4  & = -4e^{-2 \tau}   + 48 e^{-3\tau/2} - 108  e^{-4\tau/3}  + 64 e^{-5\tau/4}
\end{split}
\end{equation*}
etc.  Equivalently
\begin{equation*}
\begin{split}
c_1  & = \frac{1}{1+2t}\\
c_2  & =  \frac{2}{(1+2t)^{3/4}} - \frac{2}{1+2t} \\
c_3  & = \frac{9}{(1+2t)^{2/3}}- \frac{12}{(1+2t)^{3/4}} + \frac{3}{1+2t}  \\
c_4  & = \frac{64}{(1+2t)^{5/8}} - \frac{108}{(1+2t)^{2/3}}  + \frac{48}{(1+2t)^{3/4}} - \frac{4}{1+2t}
\end{split}
\end{equation*}
etc. The asymptotic behavior for $t\gg  1$ and fixed $k$ reads
\begin{equation}
\label{ck:0-1-t}
c_k(t)\simeq k^{k-1} (1+2t)^{-(k+1)/2k}.
\end{equation}
All final densities vanish at the critical point (see Fig.~\ref{Fig:C1CK0m1}).

To determine the asymptotic behavior of the total density $N$ we return to \eqref{ck:0-1} with $\lambda=1$. Treating $k$ as a continuous variable we arrive at a wave equation 
\begin{equation}
\label{ck:0-1PDE}
\left(\frac{\partial }{\partial \tau} + \frac{\partial }{\partial k}\right)(kc_k) = 0
\end{equation}
which is solved to give
\begin{equation}
\label{ck-f}
k c_k(\tau) = f(\tau-k).
\end{equation}
The dominant contribution to the sum $N=\sum_{k\geq 1}c_k$ is gathered near $k_*\approx \tau$ where  $c_k(\tau)$ has a sharp maximum. This observation allows us to compute $N$ relying only on the sharpness of the maximum and mass conservation 
\begin{eqnarray}
\label{N:0-1}
N &=& \sum_{k\geq 1}c_k\simeq k_*^{-1}  \sum_{k\geq 1}  kc_k \nonumber \\
 &=& k_*^{-1}  \simeq \tau^{-1}=\frac{2}{\ln(1+2t)}.
\end{eqnarray}

In the general case of arbitrary $\lambda$ we apply the Laplace transform to  Eq.~\eqref{ck:0-1} and obtain
$$
\widehat{c}_{k}(p)=\widehat{c}_{k-1}(p)/(p+1+\lambda k^{-1})
$$
which is iterated to find
\begin{equation}
\widehat{c}_{k}(p) = \frac{\Gamma(2+\lambda \epsilon)\, k!}{\Gamma(k+1+\lambda \epsilon)}
\epsilon^{k-1} \widehat{c}_{1}(p), \quad  \epsilon  = \frac{1}{1+p}.
\label{eq:ckpc1p}
\end{equation}

Applying the Laplace transform to $\sum_{k \geq 1} k c_{k}(\tau) =1$ gives $\sum_{k \geq 1} k\widehat{c}_{k}(p) =1/p$. Using Eq.~\eqref{eq:ckpc1p} and
$$
\sum_{k=1}^{\infty} \frac{k\,k!\, \epsilon^k}{\Gamma(k+b)} = \frac{\epsilon F\left[2,2;1+b; \epsilon\right]}{\Gamma(1+b)}
$$
which is the definition of the (ordinary) hypergeometric function \cite{Knuth} we obtain
\begin{equation}
\label{eq:c1pgen}
 p\widehat{c}_{1}(p) =\frac{1}{F\left[2,2;2+\lambda\epsilon; \epsilon\right]}\,.
\end{equation}
Substituting \eqref{eq:c1pgen} into  \eqref{eq:ckpc1p}  we arrive at
\begin{equation}
\label{ckp:0-1}
p\widehat{c}_k(p)=\frac{\epsilon^{k-1}}{F\left[2,2;2+\lambda\epsilon;
\epsilon\right]}\, \frac{k!\,\Gamma(2+\lambda \epsilon)}{\Gamma(k+1+\lambda\epsilon)}\,.
\end{equation}
Using \eqref{ckp:0-1} we express the Laplace transform of the total cluster density through the ratio of hypergeometric functions:
\begin{equation}
\label{Np:0-1}
p\widehat{N}(p)=\frac{F\left[1,2;2+\lambda\epsilon; \epsilon\right]}{F\left[2,2;2+\lambda\epsilon; \epsilon\right]}.
\end{equation}

The $\tau \to \infty$ behaviors (corresponding to $t \to \infty$) are encoded in the $p \to 0$ behaviors of the Laplace transforms. To extract such behaviors we use the integral representations of the hypergeometric functions appearing in Eqs.~\eqref{ckp:0-1}--\eqref{Np:0-1}:
\begin{subequations}
\begin{align}
\label{int1}
\frac{F\left[1,2;2+\lambda\epsilon; \epsilon\right]}{\lambda\epsilon(1+\lambda\epsilon)} &= \int_0^1
dx\,\frac{x(1-x)^{\lambda\epsilon-1}}{1-x\epsilon},\\
\label{int2}
\frac{F\left[2,2;2+\lambda\epsilon; \epsilon\right]}{\lambda\epsilon(1+\lambda\epsilon)} &= \int_0^1
dx\,\frac{x(1-x)^{\lambda\epsilon-1}}{(1-x\epsilon)^2}.
\end{align}
\end{subequations}

If $\lambda >2$, these hypergeometric functions are regular at $p=0$ (that is, at $\epsilon=1$) and equal to
\begin{subequations}
\begin{align}
\label{F1}
F\left[1,2;2+\lambda; 1\right] &=\frac{\lambda+1}{\lambda-1}\\
\label{F2}
F\left[2,2;2+\lambda; 1\right] &=\frac{\lambda(\lambda+1)}{(\lambda-2)(\lambda-1)}
\end{align}
\end{subequations}
Thus all $\widehat{c}_k(p)$ have a simple pole, $\widehat{c}_k(p) \to c_k(\infty)/p$, indicating the existence of a steady state size when $\lambda>2$. Using \eqref{ckp:0-1} we find
\begin{equation}
\label{ck-inf:0-1}
c_k(\infty) = (\lambda-2)(\lambda-1)\Gamma(\lambda)\,\frac{k!}{\Gamma(k+1+\lambda)} .
\end{equation}
Summing $c_k(\infty)$, or using \eqref{Np:0-1}, we find the total cluster density in the steady state ($\lambda >2$):
\begin{equation}
\label{N-inf:0-1}
N_\infty = \frac{\lambda-2}{\lambda} .
\end{equation}

Albeit all cluster densities relax to stationary values when $\lambda >2$, the moments $M_\nu(\tau)$ with $\nu\geq \lambda-1$ grow indefinitely. Indeed, Eq.~\eqref{ck-inf:0-1} gives
\begin{equation}
\label{ck-inf-asymp:0-1}
c_k(\infty) \simeq (\lambda-2)(\lambda-1)\Gamma(\lambda)\,k^{-\lambda}
\end{equation}
when $k\gg 1$, so $M_\nu(\infty)$ exist only when $\nu<\lambda - 1$. For instance, cluster densities are stationary in the range $2<\lambda\leq 3$, but the second moment $M_2(\infty)$ diverges, so $M_2(\tau)$ grows indefinitely. We now show that
\begin{equation}
\label{M2:0-1:23}
M_2(\tau) \simeq
\begin{cases}
 \frac{(\lambda-2)(\lambda-1)\,\Gamma(\lambda)}{3-\lambda}\,\tau^{3-\lambda}  & 2<\lambda<3\\
 4 \ln\tau                                                                                                                 & \lambda=3
 \end{cases}
\end{equation}
for $\tau\gg  1$. To establish the asymptotic behaviors \eqref{M2:0-1:23} we rely on the Laplace transform
\begin{equation}
\label{M2p:0-1}
p\widehat{M}_2(p)=\frac{F\left[2,2,2;1,2+\lambda\epsilon; \epsilon\right]}{F\left[2,2;2+\lambda\epsilon; \epsilon\right]}
\end{equation}
of the second moment derived using Eqs.~\eqref{ckp:0-1}. As usual, the large $\tau$ behavior of $M_2(\tau)$ is encoded in the $p \to 0$ behavior of $\widehat{M}_2(p)$. The hypergeometric function in the denominator in the right-hand side of Eq.~\eqref{M2p:0-1} is regular at $p=0$ [its value is given by \eqref{F2}]. The hypergeometric function in the nominator is defined via \cite{Knuth}
\begin{equation*}
F\left[2,2,2;1,2+\lambda\epsilon; \epsilon\right]  = \sum_{n\geq 1}\frac{\Gamma(2+\lambda\epsilon)\,\Gamma(1+n)}{\Gamma(1+\lambda\epsilon+n)}\,n^2 \epsilon^{n-1}.
\end{equation*}
When $p \downarrow 0$, we have $\epsilon\uparrow 1$ and observe that the sum is dominated by the large $n$ behavior. Hence summation can be replaced by integration. Using the asymptotic $\frac{\Gamma(1+n)}{\Gamma(1+\lambda\epsilon+n)}\simeq n^{-\lambda\epsilon}$ and $\epsilon^{n-1}\simeq e^{-pn}$, and setting $\epsilon=1$ in regular terms, we obtain
\begin{eqnarray}
\label{F5}
F\left[2,2,2;1,2+\lambda\epsilon; \epsilon\right]  &\simeq& \Gamma(2+\lambda)\int_0^\infty dn\,n^{2-\lambda}e^{-pn}\nonumber \\
&=& \frac{\Gamma(2+\lambda)\,\Gamma(3-\lambda)}{p^{3-\lambda}}.
\end{eqnarray}
Plugging \eqref{F2} and \eqref{F5} into \eqref{M2p:0-1} we get
\begin{equation}
\label{M2p:0-1-asymp}
\widehat{M}_2(p)\simeq \frac{(\lambda-2)(\lambda-1)\,\Gamma(\lambda)\,\Gamma(3-\lambda)}{p^{4-\lambda}}
\end{equation}
from which we deduce an algebraic growth of the second moment given in \eqref{M2:0-1:23} in the $2<\lambda<3$ range. In the marginal case of $\lambda=3$ we similarly derive
\begin{equation}
\label{M2tau:0-1-marg}
M_2(\tau) \simeq 4 \ln\tau
\end{equation}
from which we deduce a logarithmic growth given in \eqref{M2:0-1:23}.

The time dependence \eqref{M2:0-1:23} suggests that when $\tau\gg 1$ the densities are stationary and given by Eq.~\eqref{ck-inf:0-1} up to a crossover size $k_*\sim \tau$, while for $k > k_*$ the densities quickly vanish. This implies that the moments diverge algebraically when  $\nu>\lambda - 1$, viz. as
\begin{equation}
\label{M-nu-tau:0-1}
M_\nu(\tau) \sim \sum_{k<\tau} k^{\nu-\lambda}  \sim \tau^{\nu+1-\lambda}
\end{equation}
and logarithmically in the marginal case: $M_{\lambda-1}\sim \ln \tau$. In terms of the physical time we thus have
\begin{equation}
\label{M-nu:0-1}
M_\nu(t) \sim
\begin{cases}
\text{finite}              &  \nu<\lambda - 1\\
\ln t                         &  \nu=\lambda - 1\\
t^{\nu+1-\lambda}  &\nu > \lambda-1
\end{cases}
\end{equation}
when $\lambda>2$ and $t\gg 1$.

We now turn to the critical region $1\leq \lambda\leq 2$. Inside this region, $F\left[2,2;2+\lambda\epsilon; \epsilon\right]$ diverges at $p \to 0$ implying that $c_k \to 0$ for $\tau \to \infty$, see Eq.~\eqref{ckp:0-1}. We conclude that the system possesses a critical \emph{ interval } with the lower $\lambda_{\rm c,l}=1$ and upper $\lambda_{\rm c,up}=2$ critical points. Again, vanishing cluster densities in conjunction with mass conservation indicate the formation of the supercluster state discussed below. Hence the final densities of clusters and monomers read:
\begin{equation*}
\label{eq:Cka0sm1}
c_k(\infty) =
\begin{cases}
c_k(\tau_{\rm max})(1-\delta_{k,1})  & \lambda < 1 \\
0                                            & 1 \leq \lambda\leq 2\\
\frac{\Gamma(k+1)}{\Gamma(k+1+\lambda)} \,(\lambda-1)(\lambda-2)\Gamma(\lambda) & \lambda > 2,
\end{cases}
\end{equation*}
with $\delta_{k,1}$ ensuring that $c_1(\infty) = 0$ in the jammed regime. The densities $c_k(\tau_{\rm max})$ depend on the initial conditions.

Thus the system undergoes a continuous phase transition from a jammed state into a supercluster state at $\lambda_{\rm c,l}=1$, and a continuous phase transition from a supercluster state to a steady state at $\lambda_{\rm c,up}=2$ (see Fig.~\ref{Fig:C1CK0m1}).

Analyzing the singularity of $F\left[2,2;2+\lambda\epsilon; \epsilon\right]$ as $p \to 0$, one can find the temporal relaxation of the cluster densities in the $1 < \lambda <2$ range. With $\epsilon=(1+p)^{-1}$ and $p \to 0$ we can replace $\epsilon$ by $1$ in the right-hand side of Eq.~\eqref{int2}, apart from the denominator $(1-x \epsilon)^{-2}$. Writing $1-x \epsilon =1-x + x(1- \epsilon)$ and analyzing the integral, we find that its dominant part gathers in the region $1-x = {\cal O} (1-\epsilon)$. Since $1-\epsilon= p+ {\cal O} (p^2)$ we write $1-x =py$ to recast Eq.~\eqref{int2} into
$$
F\left[2,2; 2 +\lambda \epsilon; \epsilon \right] \simeq \lambda (1+\lambda) p^{\lambda-2} \int_0^{\infty} dy
\frac{y^{\lambda-1}}{(1+y)^2}\,.
$$
Computing the integral we obtain
\begin{equation}
\label{eq:F222}
F\left[2,2; 2 +\lambda \epsilon; \epsilon \right] \simeq \lambda (1+\lambda) \frac{\pi (1-\lambda)}{
\sin( \pi \lambda)}p^{\lambda-2}\,.
\end{equation}
Inserting \eqref{eq:F222} into \eqref{eq:c1pgen} we find
\begin{equation}
\label{eq:c1ppto0}
\widehat{c}_{1}(p) = \frac{\sin( \pi \lambda)}{\pi (1-\lambda)}
\frac{p^{1-\lambda}}{\lambda(1+\lambda)}
\end{equation}
from which we extract the asymptotic
\begin{equation}
\label{eq:C1tau}
c_1(\tau) \simeq -  \frac{\sin( \pi \lambda)}{\pi \Gamma(2+\lambda)} \frac{1}{\tau^{2-\lambda} }\,.
\end{equation}
The total cluster density follows from Eqs.~\eqref{c1:0-1} and \eqref{eq:C1tau}:
\begin{equation}
\label{eq:Ntau}
N\simeq \frac{\lambda+1}{\lambda -1} c_1
\simeq -\frac{\sin(\pi \lambda)}{\pi (\lambda-1)\,\Gamma(\lambda+1)}\,\frac{1}{\tau^{2-\lambda} }\,.
\end{equation}
Using
\begin{equation}
\label{ttau10}
t=\int_0^{\tau} \frac{d\tau'}{c_1(\tau')} \simeq -\frac{\pi \Gamma(2+\lambda)} { (3-\lambda) \sin(\pi \lambda)}\,\tau^{3-\lambda}
\end{equation}
we can re-express previous results via physical time, e.g.
\begin{equation}
\label{c1:10-SCS}
c_1  \simeq \left[-\frac{\sin(\pi \lambda)}{\pi\,\Gamma(2+\lambda)}\right]^\frac{1}{3-\lambda} [(3-\lambda)t]^{-\frac{2-\lambda}{3-\lambda}}.
\end{equation}

When $2-\lambda\to +0$, an algebraic exponent $\frac{2-\lambda}{3-\lambda}$ in \eqref{c1:10-SCS} vanishes suggesting a logarithmic decay at the upper critical point  $\lambda=\lambda_{\rm c,up}=2$. Indeed, using the integral representation
\begin{equation}
\label{int-2}
F\left[2,2;2+2\epsilon; \epsilon\right] = 2\epsilon(1+2\epsilon) \int_0^1
dx\,\frac{x(1-x)^{2\epsilon-1}}{(1-x\epsilon)^2}
\end{equation}
of the relevant hypergeometric function we deduce the asymptotic $F\left[2,2;2+2\epsilon; \epsilon\right]\simeq 6 \ln(1/p)$ leading to $\widehat{c}_{1}(p)\simeq  [6p \ln(1/p)]^{-1}$ as $p\to 0$, from which
\begin{equation}
\label{c1:2-t}
c_1\simeq  \frac{1}{6 \ln \tau}\simeq  \frac{1}{6 \ln t}.
\end{equation}
Using  \eqref{c1:2-t} and Eqs.~\eqref{ck:0-1} with $\lambda=2$  we arrive at the asymptotic
\begin{equation}
\label{ck:2-t}
c_k\simeq    \frac{1}{(k+1)(k+2)}\,\frac{1}{\ln t}
\end{equation}
valid for any fixed $k\geq  1$ in the  $t\to\infty$ limit.

\subsection{A class of models with $(a,s)=(a,a-1)$}
\label{subsec:aa-1}

Here we consider a class of models with $A_k=k^a$ and $S_k=\lambda k^{a-1}$. The governing equations read
\begin{subequations}
\begin{align}
\frac{d c_{1}}{d \tau} &= -(1+\lambda)c_{1} - M_{a} + \lambda M_{a}, \label{dc1awcdlinearlizedsc} \\
\frac{dc_{k}}{d \tau} &= (k-1)^{a} c_{k-1} - (k^{a} + \lambda k^{a-1} )c_{k}. \label{dckawcdlinearizedsc}
\end{align}
\end{subequations}

In Sects.~\ref{subsec:10} and \ref{subsec:0-1} we discussed  two representatives of this class of models: $(a,s)=(1,0)$ and $(a,s)=(0,-1)$. We expect the formation of the supercluster states in the region $\lambda_{\rm c,l} \leq \lambda \leq \lambda_{\rm c,up} $. We now argue that
\begin{equation}
\label{lambda-mm}
\lambda_{\rm c,l}  = 1, \quad \lambda_{\rm c,up} = 2-a
\end{equation}
when $0\leq a\leq 1$. For $a\leq 0$, the lower and upper critical points merge: $\lambda_{\rm c,l}  =  \lambda_{\rm c,up} =1$.

The lower critical point is universal: $\lambda_{\rm c,l} = 1$ for any $a\leq 1$. To prove this assertion we set $c_{1}(\tau_{\rm max}(\lambda_{c})) = 0$ and $\dot{c}_{k}(\tau_{\rm max}(\lambda_{c})) = 0$ in Eq.~\eqref{dc1awcdlinearlizedsc} to give
\begin{equation}
\label{lambda-min}
(\lambda_{\rm c,l} - 1)M_{a}(\tau_{\rm max}) = 0.
\end{equation}
We have $M_{a}(\tau_{\rm max}) = \sum_{k\geq 1} k^{a}c_{k}(\tau_{\rm max}) \neq 0$ because $c_{k}(\tau_{\rm max}) \neq 0$ for $k \geq 2$ in the subcritical region. Therefore Eq.~\eqref{lambda-min} gives $\lambda_{\rm c,l} = 1$.

Since $\tau_{\rm max} = \infty$ above the critical point, we can apply  the Laplace transform for $\lambda \geq 1$. Then from Eqs.~\eqref{dckawcdlinearizedsc} we iteratively obtain
\begin{equation}
\label{Ckp_C1p}
\widehat{c}_k(p)=\widehat{c}_1(p) \prod_{j=2}^k \frac{(j-1)^a}{j^a +\lambda j^{a-1} +p}.
\end{equation}
Using again $\sum_{k \geq 1}k\widehat{c}_k(p)=1/p$ following from mass conservation, we find $p\widehat{c}_1(p) =  1/G(p,\lambda)$ with
\begin{equation}
\label{Cp-lambda}
G(p,\lambda)=\sum_{k\geq 1} k\prod_{j=2}^k \frac{(j-1)^a}{j^a+\lambda j^{a-1}+p}.
\end{equation}
The Laplace transform $\widehat{c}_1(p)$ has  a simple pole at $p=0$ giving  the steady-state density $c_1(\infty)=1/G(0,\lambda)$. Setting $p=0$ on the right-hand side of Eq.~\eqref{Cp-lambda} and massaging the sum we obtain
\begin{equation}
\label{C0-lambda1}
G(0,\lambda) = \sum_{k\geq 1} k^{1-a}\,\frac{\Gamma(k+1)\Gamma(2+\lambda)}{\Gamma(k+1+\lambda)}.
\end{equation}
The summand behaves as $\Gamma(2+\lambda)\,k^{-(\lambda+a-1)}$ in the large $k$ limit, so the sum on the right-hand side of \eqref{C0-lambda1} converges when $\lambda>2-a$. If $\lambda \leq 2-a$, the sum  in \eqref{C0-lambda1} diverges yielding vanishing final densities. Hence $\lambda_{\rm c, up}=2-a$ as stated in \eqref{lambda-mm}.

The final densities $c_k (\infty)= \lim_{p\to 0} p\widehat{c}_k(p)$ are found by combining Eqs.~\eqref{Ckp_C1p} and \eqref{C0-lambda1} with $p\widehat{c}_1(p) = 1/G(p,\lambda)$. This yields for $\lambda > \lambda_{\rm c,up} = 2-a $
$$
c_k(\infty)= \frac{k^{-a} k!/\Gamma(k+\lambda+1) }{\sum_{n \geq1} n^{1-a}n! /\Gamma(n+\lambda+1)} .
$$
The above steady-state  densities do not depend on the initial conditions. Summarizing
\begin{equation}
\label{eq:Ck_gen_a}
c_k(\infty) =
\begin{cases}
(1-\delta_{k,1}) c_k(\tau_\text{max})  & \lambda < 1 \\
0                                                        & 1 \leq \lambda\leq 2-a\\
\frac{k^{-a} k!/\Gamma(k+\lambda+1) }{\sum_{n \geq1} n^{1-a}n! /\Gamma(n+\lambda+1)} & \lambda > 2-a.
\end{cases}
\end{equation}
A continuous phase transition from the jammed state to the supercluster state occurs at the lower critical point, $\lambda_{\rm c,l}=1$. Then a continuous phase transition from the supercluster state to the steady state takes place at the upper critical point, $\lambda_{\rm c,up} = 2-a$.

Consider the critical region $1<\lambda < 2-a$. First, we re-write $G(p,\lambda)$  given by Eq.~\eqref{Cp-lambda} as
\begin{eqnarray*}
G(p,\lambda) &=& \sum_{k\geq 1} k\prod_{j=2}^k \frac{(j-1)^a}{j^a+\lambda j^{a-1}+p}\\
&=& \sum_{k\geq 1} k^{1-a}\,\frac{\Gamma(k+1)\Gamma(2+\lambda)}{\Gamma(k+1+\lambda)}
\prod_{j=2}^k\frac{1}{1+\frac{p}{j^a+\lambda j^{a-1}}}.
\end{eqnarray*}
Re-writing the product as
\begin{eqnarray*}
\prod_{j=2}^k \frac{1}{1+\frac{p}{j^a+\lambda j^{a-1}}}=\exp\!\left[-\sum_{j=2}^k\ln\left(1+\frac{p}{j^a+\lambda j^{a-1}}\right)\right]
\end{eqnarray*}
and expanding in the $p\to 0$ limit gives
\begin{subequations}
\begin{eqnarray}
\label{1-asymp}
\prod_{j=2}^k \frac{1}{1+\frac{p}{j^a+\lambda j^{a-1}}} &\simeq& \exp\!\left[-p\sum_{j=2}^k \frac{1}{j^a+\lambda j^{a-1}}\right] \nonumber  \\
 &\simeq&
 \begin{cases}
 \exp\!\left[-\frac{p\,k^{1-a}}{1-a}\right] & a<1\\
 k^{-p}                                                  & a=1
 \end{cases}
\end{eqnarray}
with the second asymptotic  valid when $k\gg 1$. We also use the asymptotic
\begin{equation}
\label{2-asymp}
k^{1-a}\,\frac{\Gamma(k+1)\Gamma(2+\lambda)}{\Gamma(k+1+\lambda)}\simeq  \frac{\Gamma(2+\lambda)}{k^{a+\lambda-1}}
\end{equation}
\end{subequations}
valid when $k\gg 1$. In the $p\to 0$ limit, the main contribution to $C(p,\lambda)$ is gathered when $k\gg 1$. This allows us to replace summation over $k$ by integration and use \eqref{1-asymp}--\eqref{2-asymp}. When $a<1$ we get
\begin{eqnarray}
\label{GpL}
G(p,\lambda)
&\simeq& \int_{1}^\infty \frac{dk}{k^{a+\lambda-1}}\,\Gamma(2+\lambda)\,\exp\!\left[-\frac{p\,k^{1-a}}{1-a}\right] \nonumber \\
&=& \Gamma(2+\lambda)\,\Gamma(1-\Lambda)\,(1-a)^{-\Lambda}\,p^{\Lambda-1}
\end{eqnarray}
where $\Lambda \equiv \frac{\lambda-1}{1-a}$. This parameter varies in the range $0<\Lambda<1$ in the critical region $1<\lambda < 2-a$.

We have $\widehat{c}_1(p)=1/pG(p,\lambda)$ and take its inverse Laplace transform to extract the large time asymptotic
\begin{equation}
\label{c1Lam}
c_1 \simeq \frac{(1-a)^\Lambda\sin(\pi \Lambda)}{\pi \Gamma(2+\lambda)}\,\,\tau^{-(1-\Lambda)}.
\end{equation}
In terms of the physical time
\begin{equation}
\label{c1Lam-t}
c_1 = \left[\frac{(1-a)^\Lambda\sin(\pi \Lambda)}{\pi \Gamma(2+\lambda)} \right]^\frac{1}{2-\Lambda}  [(2-\Lambda) t]^{-\frac{1-\Lambda}{2-\Lambda}}.
\end{equation}

The total cluster density exhibits the same temporal behavior as the density of monomers. Asymptotically,
\begin{equation}
\label{eq:ratio}
\lim_{t\to\infty}\frac{N(t)}{c_1(t)} = R(a,\lambda)=\sum_{k\geq 1} k^{-a}\,\frac{\Gamma(k+1)\Gamma(2+\lambda)}{\Gamma(k+1+\lambda)}.
\end{equation}
For $a=0$, we recover $R(0,\lambda)=(\lambda+1)/(\lambda-1)$ in the critical region $1<\lambda <2$.

The asymptotic behaviors \eqref{c1Lam}--\eqref{eq:ratio} are valid inside the critical region $1<\lambda <2-a$. More peculiar behaviors occur at the boundaries $\lambda=1$ and $\lambda=2-a$. In Eqs.~\eqref{GpL}--\eqref{c1Lam-t}, we have also assumed that $0<a<1$. The behaviors at the boundaries again require more careful treatment. We have analyzed these behaviors: the model with $a=0$ (Sec.~\ref{subsec:0-1}) and the model with $a=1$ (Sec.~\ref{subsec:10}).

At the lower critical point, the ratio
\begin{equation}
\label{Ra1}
R(a, 1)=\sum_{k\geq 1}\frac{2}{(k+1) k^a}
\end{equation}
converges when $a>0$ and diverges when $a\leq 0$. The decay of the density of monomers at $\lambda=1$ can be extracted  from \eqref{c1Lam-t} by taking the $\lambda\to 1+0$ limit. Using $c_1\sim t^{-\frac{1-\Lambda}{2-\Lambda}}$ and  $\Lambda =\frac{\lambda-1}{2-a}\to 0$ we get  $c_1\sim t^{-\frac{1}{2}}$. This asymptotic disagrees with $c_1\sim t^{-1}$ decay at $\lambda=1$ and $a=0$ due to already mentioned peculiarities at the extreme values of the parameters. For instance,  the amplitude in \eqref{c1Lam-t} is singular when $a=0$ reflecting these peculiarities.

At the upper critical point, $\Lambda \to 1$ and \eqref{c1Lam-t} suggests a logarithmic decay $c_1\sim (\ln t)^{-1}$. This logarithmic decay agrees with \eqref{c1:2-t} that was carefully derived at the upper critical point $\lambda=2$ in the model with $a=1$.

\subsection{Models with arbitrary exponents $(a,s)$}
\label{subsec:as}

Addition-and-shattering processes with algebraic rates $A_k=k^a$ and $S_k=\lambda k^s$ appear analytically intractable when the exponents $(a,s)$ are arbitrary. The only exception is the steady-state regime. Below we combine analytical results for the steady states with simulations and general expectations gained from analytically tractable addition-shattering models studied earlier.

The governing equations for the monomer density $c_{1}(\tau)$, cluster densities $c_{k}(\tau)$, and moments $M_{a}(\tau)$ and $M_{1+s}(\tau)$ are given by \eqref{dc1awcdlinearlized}, \eqref{dckawcdlinearized} and \eqref{Mn}. Numerically, we observe only  two regimes---the system reaches a jammed state or a steady state. These regimes are demarcated by the critical shattering rate $\lambda_{c}$ that depends on the exponents $(a, s)$ and initial conditions.

\begin{figure}[htp]
\centering
\includegraphics[width=7.5cm]{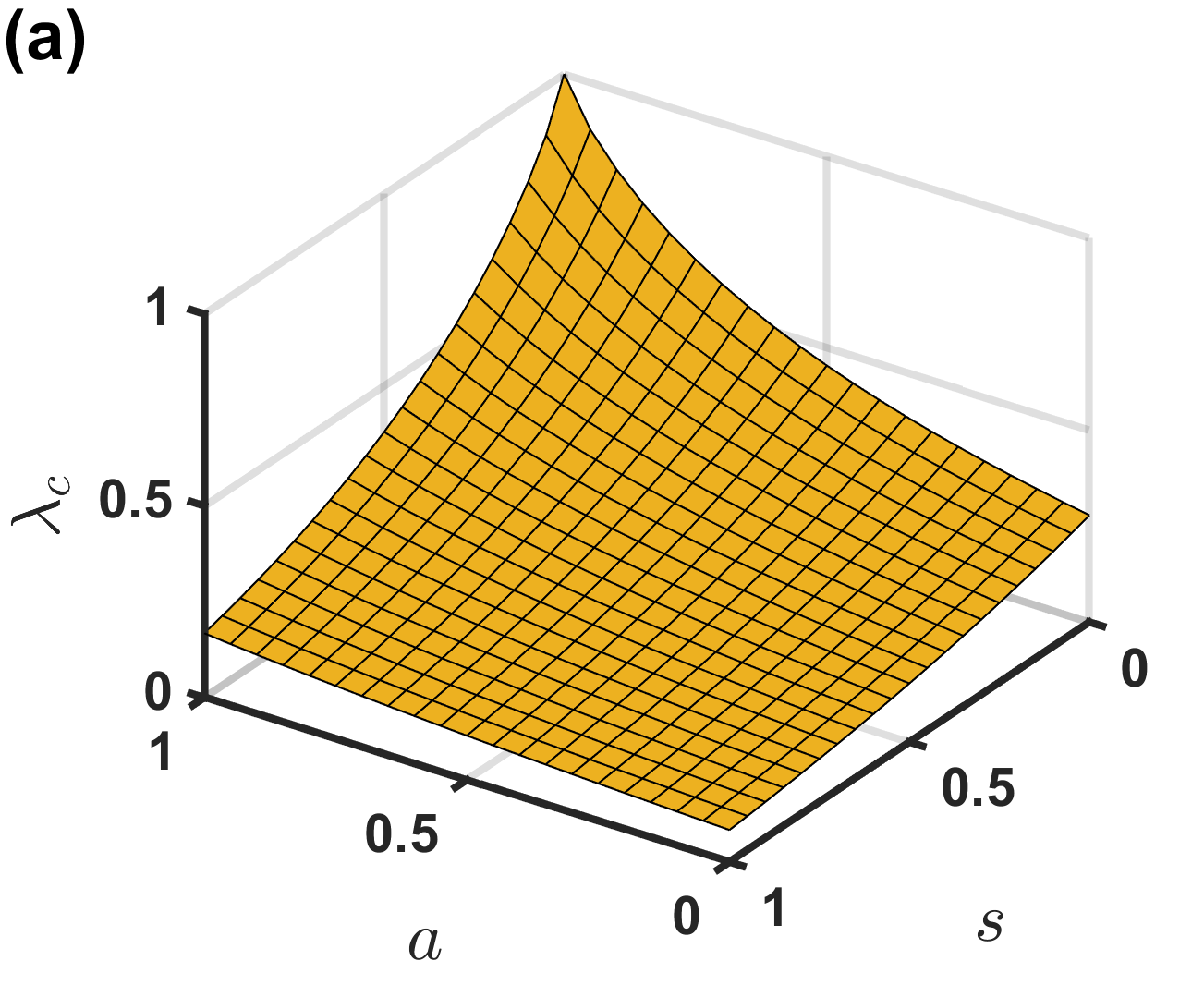} \qq
\includegraphics[width=7.5cm]{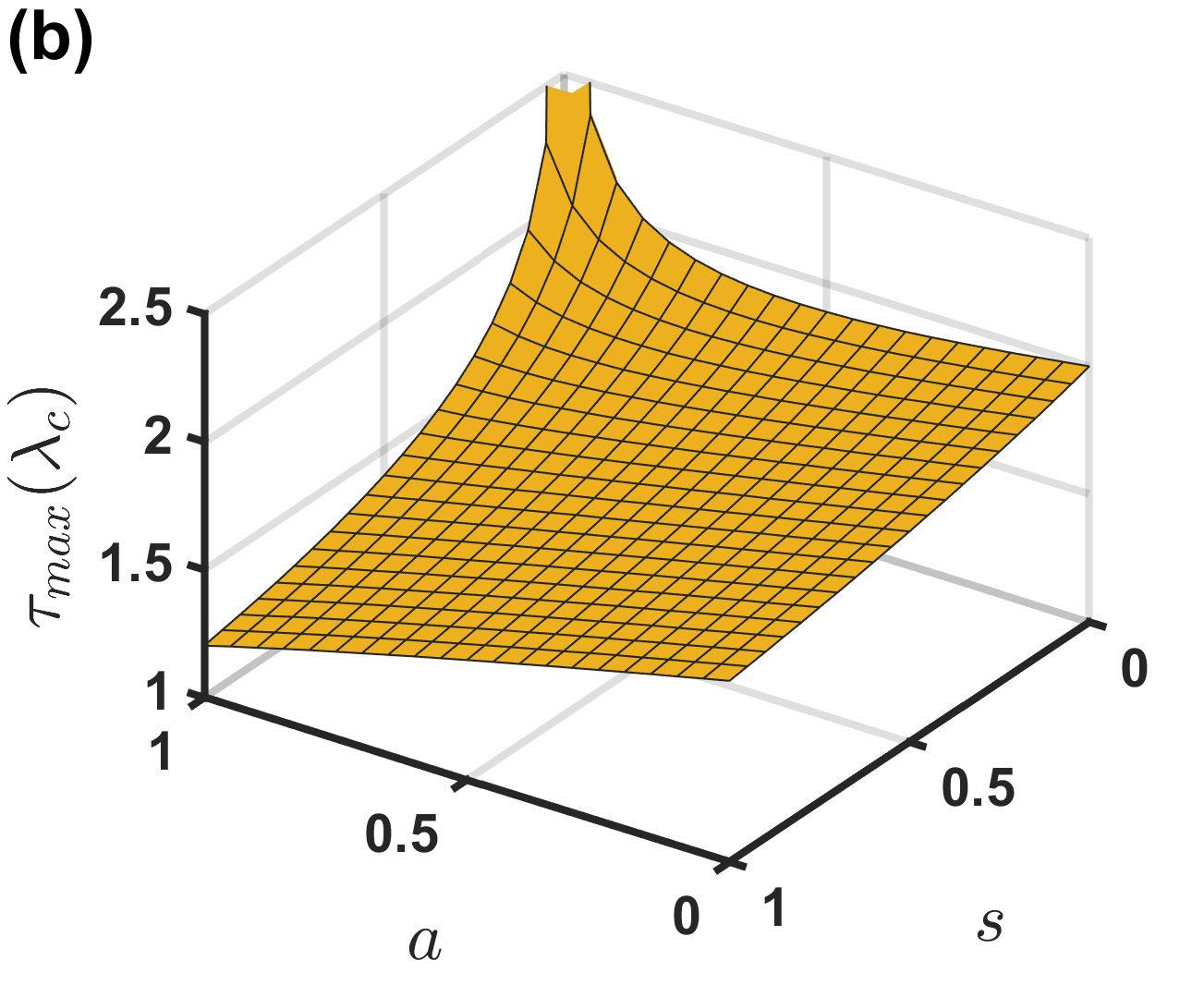}
\caption{(a) The critical shattering rate $\lambda_{c}$ as a function of $a$ and $s$. (b) The maximal modified time in the critical regime, $\tau_{\rm max}(\lambda_{c})$, as a function of $a$ and $s$. The kinetic rates $A_k=k^a$ and $S_k =\lambda k^s$; the mono-disperse initial condition is used.}
 \label{Fig:lambtau_v_s}
\end{figure}

As expected, the critical shattering rate is an increasing function of the exponent $a$ and a decreasing function of the exponent $s$, see Fig.~\ref{Fig:lambtau_v_s}(a). The maximal modified time $\tau_{\rm max}$ is a decreasing function of the exponent $s$ but has a more complicated dependence on the exponent $a$: For large $s$, the time  $\tau_{\rm max}$  is an increasing function of $a$; for small $s$,  it may be a non-monotonous function of $a$, see Fig.~\ref{Fig:lambtau_v_s}(b). Our simulations show that the critical shattering rate $\lambda_{c}$ depends on the initial conditions.

The stationary densities $c_1(\infty)$ and $c_k(\infty)$ are found from \eqref{dc1awcdlinearlized}--\eqref{dckawcdlinearized}:
\begin{equation}
\label{eq:CkC1ss}
c_k(\infty) = \frac{c_1(\infty) }{k^a\,D_k(\lambda,a-s)}.
\end{equation}
Here we shortly write
\begin{equation}
\label{Dk:def}
D_k(\lambda,\sigma)=\prod_{j=2}^k \left(1 + \frac{\lambda}{j^\sigma}\right).
\end{equation}
Combining mass conservation, $\sum_{k \geq 1} kc_{k}(\infty) = 1$, with \eqref{eq:CkC1ss}, we fix the final density of monomers:
\begin{equation}
\label{eq:exmass1}
\frac{1}{c_1(\infty)} = \sum_{k=1}^{\infty} \frac{k^{1-a}}{D_k(\lambda,a-s)}\,.
\end{equation}
The product \eqref{Dk:def} exhibits qualitatively different large $k$ behaviors depending on whether $\sigma<1$  or $\sigma>1$:
\begin{subequations}
\begin{align}
\label{Dk:min}
& \ln D_k(\lambda,\sigma) \simeq \frac{\lambda}{1-\sigma}\,k^{1-\sigma} \qquad (\sigma<1),\\
\label{Dk:crit}
&D_k(\lambda,1)              =\frac{\Gamma(k+1+\lambda)}{k! \,\Gamma(2+\lambda)}\simeq \frac{k^\lambda}{\Gamma(2+\lambda)}, \\
\label{Dk:plus}
&D_k(\lambda,\sigma)     \to D_\infty(\lambda,\sigma)  \qquad \qquad (\sigma>1).
\end{align}
\end{subequations}

If $s>a-1$, we should use \eqref{Dk:min} and then \eqref{eq:exmass1} gives $c_1(\infty)>0$. Thus when  $s>a-1$, the steady state is possible. When $s<a-1$, we should use \eqref{Dk:plus} implying that the sum in \eqref{eq:exmass1} diverges [recall that $a\leq 1$]. Thus the system falls into a jammed state if $s<a-1$.

These simple arguments explain why interesting behaviors in models with algebraic rates occur when the exponents are related via $s=a-1$. Equation \eqref{Dk:crit} implies that when $\lambda>2-a$ the sum in \eqref{eq:exmass1} converges; $c_1(\infty)>0$ and the system is in the steady state when $s=a-1$ and $\lambda>2-a$. When $\lambda<2-a$, the sum in \eqref{eq:exmass1} diverges implying that $c_1(\infty)=0$ and the system is either in the jammed state or in the SCS. The analysis for the class of models with $s=a-1$ presented before shows that the jammed state arises when $\lambda<1$ while the SCS emerges in the range $1<\lambda<2-a$.

The final densities in the steady state  for the model with $s=a-1$ simplify when $a$ is an integer. We already know the answers when $(a,s)=(1,0)$ and $(0,-1)$. In the next example $(a,s)=(-1,-2)$, the densities in the steady state ($\lambda>3$) are
\begin{equation}
c_k(\infty) = \frac{(\lambda-1)(\lambda-2)(\lambda-3)}{1+\lambda}\,
\frac{k\cdot k!\,\Gamma(\lambda)}{\Gamma(k+1+\lambda)}\,.
\end{equation}

\begin{figure}
\centering
\includegraphics[width=7cm]{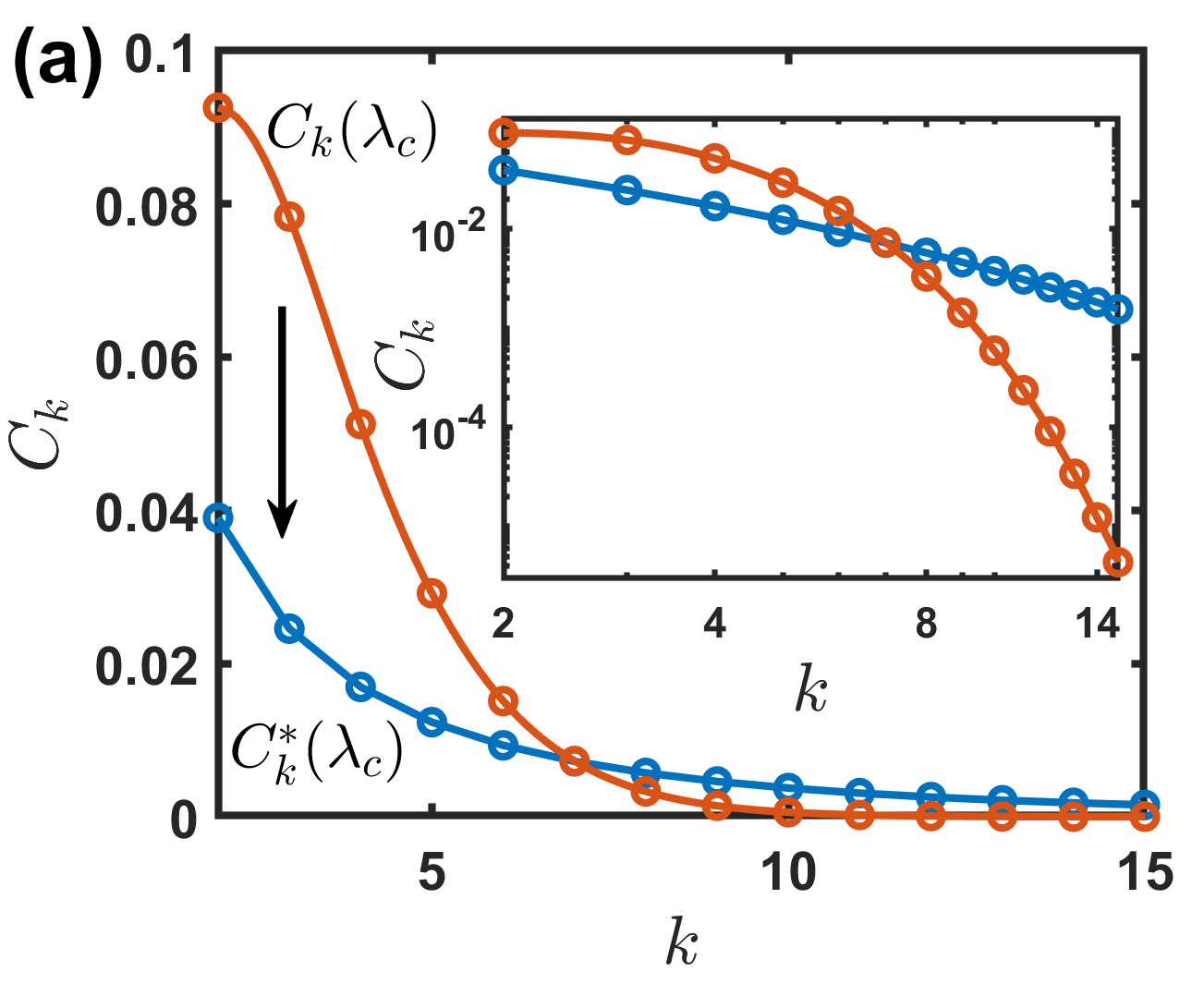}\q
\includegraphics[width=7.cm]{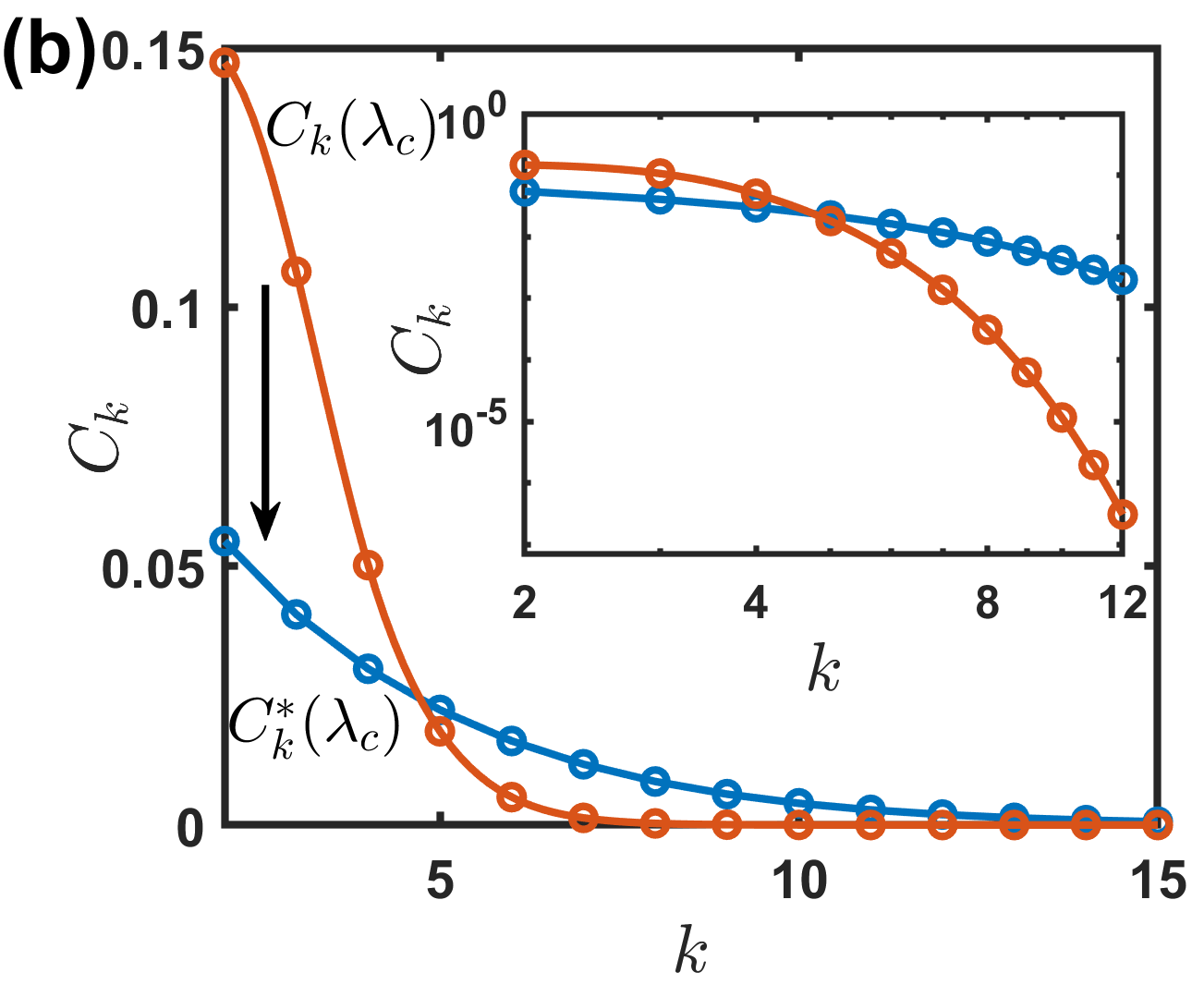}
\caption{The final cluster densities undergo a  discontinuous phase transition at $\lambda=\lambda_c(a,s)$ as illustrated for $a=0.6$ and $s=0.2$ (top panel) and for $a=0.2$, $s=0.6$ (bottom panel). The mono-disperse initial condition is used.}
 \label{Fig:ckCkPhTr}
\end{figure}

Cluster densities undergo a discontinuous (first order) phase transition at $\lambda=\lambda_c$  so that the jammed densities $c_k(\infty)$, for $\lambda=\lambda_c$, differ from the equilibrium densities $c_k(\infty)$ for $\lambda= \lambda_c+0$. This is illustrated in Fig.~\ref{Fig:ckCkPhTr}.

\section{Continuous, discontinuous and weak phase transitions}
\label{sec:CDW}

For addition-and-shattering processes with algebraic rates, $A_k=k^a$ and $S_k=\lambda k^s$, two regimes depending on the parameters $(a,s,\lambda)$ generically arise. In the jamming regime, monomers disappear, and evolution stops. In the steady-state regime, the evolution continues forever, but the densities are stationary (in the infinite size limit). For instance, if $s>a-1$, the jamming regime occurs when $\lambda<\lambda_c$ while the steady-state regime occurs in  the complimentary $\lambda>\lambda_c$ range. The transition between these regimes is discontinuous. The magnitude of the critical shattering parameter $\lambda_c$  is non-universal (it depends on the initial condition).

Models with $(a,s)=(a,a-1)$ exhibit particularly rich behaviors. There are three different regimes:
\begin{enumerate}
\item The jammed regime in the range $\lambda < \lambda_{\rm c, l}$. The jammed regime is non-universal as it depends on the initial condition. The magnitude of the lower critical shattering parameter is universal: $\lambda_{\rm c, l}=1$.
\item The supercluster states (SCSs) occur in the range $\lambda_{\rm c, l}\leq \lambda  \leq \lambda_{\rm c, up}$. The SCSs will be discussed in detail below.
\item The steady state regime in the range $\lambda > \lambda_{\rm c, up}$. The steady state densities are universal (independent on the initial condition).
\end{enumerate}

We now briefly discuss weak phase transitions in models with $(a,s)=(a,a-1)$ occurring on the boundary between jammed and supercluster states at the low critical point $\lambda = \lambda_{\rm c, l} =1$. There are infinitely many critical values $a_p$ where transitions occur. The critical values are defined by
\begin{equation}
\label{ap}
p^{a_p}+p^{a_p-1} = (p-1)^{a_p}+(p-1)^{a_p-1}.
\end{equation}
It is convenient to set $a_1=1$, so $a_2$ is defined by
\begin{equation}
\label{ap-2}
2^{a_2}+2^{a_2-1} =2.
\end{equation}
The critical values $a_p$ decrease as $p$ increases. Numerical values are
\begin{equation*}
\begin{split}
&a_1=1; \qquad ~~~~~~~~a_2\approx 0.41503; \qquad
a_3\approx 0.29048; \\
&a_4\approx 0.22433; \qquad a_5\approx 0.18294 ; \qquad a_6\approx 0.15451;
\end{split}
\end{equation*}
etc. Using \eqref{ap} one finds that when $p\gg 1$, the critical values approach to zero according to
\begin{equation*}
a_p=\frac{1}{p} - \frac{1}{2p^2}+\frac{5}{12p^3} - \frac{7}{24 p^4} + \ldots
\end{equation*}

\begin{figure}
\centering
\includegraphics[width=7.5cm]{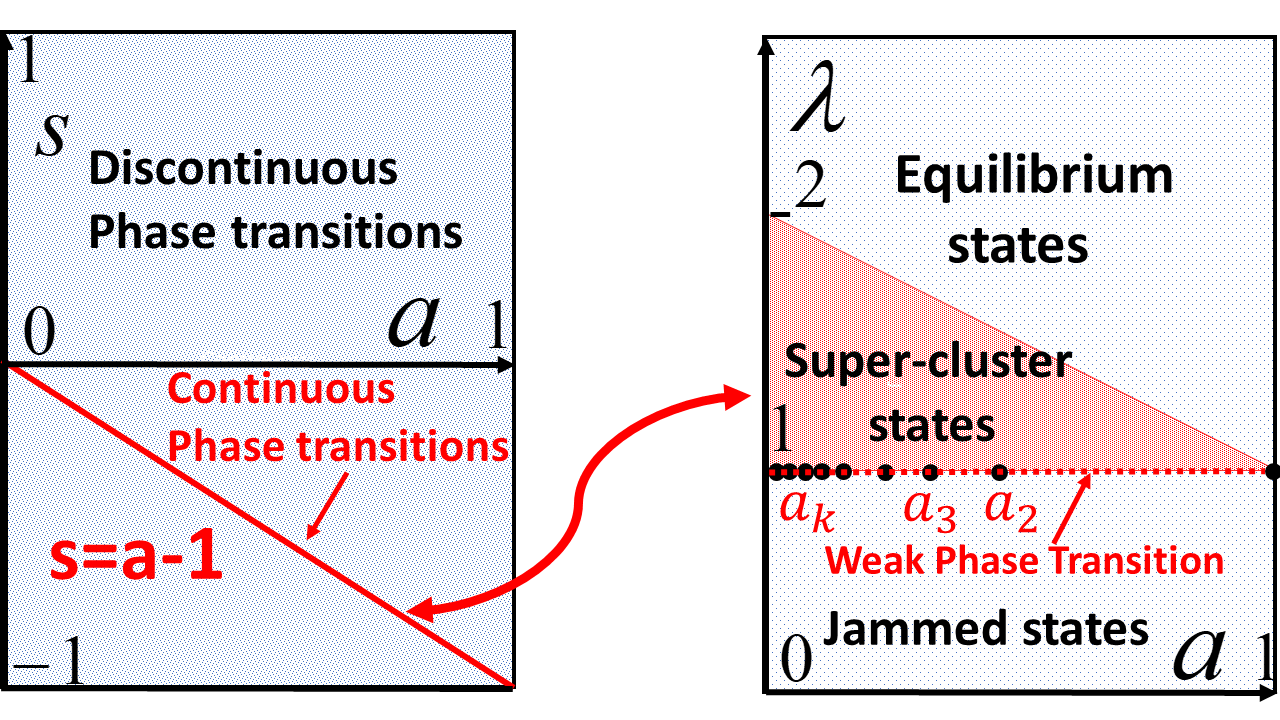}
\caption{Illustration of the phase diagram of the infinite system with aggregation and shattering. Left panel: For general addition and shattering exponents $(a,s)$, the system undergoes a discontinuous phase transition from a jammed to a steady state at $\lambda=\lambda_c$. On the $s=a-1$ line, the system undergoes a continuous phase transition. The right panel details the behaviors on the line $s=a-1$ of the left panel. The continuous phase transition from a jammed to a supercluster state occurs at $\lambda=1$. The continuous phase transition from a supercluster state to the steady state occurs at $\lambda= 2-a$.  On the $\lambda_{\rm c,l}=1$ line, there is an infinite set of weak phase transitions at $a=a_1,a_2, \ldots a_p, \ldots$.}
 \label{Fig:phasediag}
\end{figure}

To appreciate the emergence of phase transitions at these values we start with the governing equations for the model with exponents $(a, a-1)$ and $\lambda=1$:
\begin{subequations}
\begin{align}
\label{c1:aa}
\frac{d c_{1}}{d \tau} &= -2c_{1}, \\
\label{ck:aa}
\frac{dc_{k}}{d \tau} &= (k-1)^{a} c_{k-1} - \big(k^{a} + k^{a-1}\big)c_{k}, \quad k\geq 2.
\end{align}
\end{subequations}
Equation \eqref{c1:aa} yields $c_1=e^{-2\tau}$, and then from \eqref{ck:aa} with  $k=2$ one gets
\begin{equation}
\label{c2:a}
c_2 = \frac{e^{-2\tau}-e^{-(2^a+2^{a-1})\tau}}{2^a+2^{a-1}-2}.
\end{equation}
If $2^a+2^{a-1}>2$, equivalently $a>a_2$, the asymptotic behavior of the density of dimers is $c_2\simeq {\cal B}_2 e^{-2\tau}$ with ${\cal B}_2=1/[2^a+2^{a-1}-2]$. All densities decay similarly and only amplitudes vary:
\begin{equation}
\label{Bk:def}
c_k\simeq {\cal B}_k e^{-2\tau}
\end{equation}
when $k\geq 1$ and $a>a_2$. To determine the amplitudes we specialize the Laplace transform \eqref{Ckp_C1p} to $\lambda=1$ and use the Laplace transform $\widehat{c}_1(p)=\frac{1}{p+2}$ of $c_1=e^{-2\tau}$. We get
\begin{equation}
\label{Bk-p}
\widehat{c}_k(p) = \frac{1}{p+2} \prod_{j=2}^k \frac{(j-1)^a}{j^a +j^{a-1} +p}.
\end{equation}
This Laplace transform has a simple pole $p=-2$ with residue
\begin{equation}
\label{Bk}
\mathcal{B}_k = \prod_{j=2}^k \frac{(j-1)^a}{j^a +j^{a-1} -2}
\end{equation}
which is the amplitude in \eqref{Bk:def}.

When $a=a_2$, the density of dimers is $c_2 = \tau e^{-2\tau}$, and generally
\begin{equation}
\label{ck:a2}
c_k\simeq \mathcal{B}_k \tau e^{-2\tau}\quad\text{for}\quad k\geq 2.
\end{equation}
Similarly to \eqref{Ckp_C1p} we derive
\begin{equation}
\label{Bk-p-2}
\widehat{c}_k(p) = \frac{1}{(p+2)^2} \prod_{j=3}^k \frac{(j-1)^{a_2}}{j^{a_2} +j^{a_2-1} +p}
\end{equation}
where we have used the Laplace transform $\widehat{c}_2(p)=\frac{1}{(p+2)^2}$ of $c_2=\tau e^{-2\tau}$. Thus $\widehat{c}_k(p)$ has a pole of order 2 at  $p=-2$ with amplitude
\begin{equation}
\mathcal{B}_k = \prod_{j=3}^k \frac{(j-1)^{a_2}}{j^{a_2} + j^{a_2-1}-2}
\end{equation}
which yields the amplitude appearing in \eqref{ck:a2}.

Similarly in the $a_3 < a < a_2$ range
\begin{equation}
\label{cka23} c_k \simeq \mathcal{B}_k e^{-\left(2^a +2^{a-1}\right) \tau}
\end{equation}
for $k\geq 2$ with amplitudes
\begin{equation}
\label{Bka23}
\mathcal{B}_k = \frac{1}{2-2^a - 2^{a-1}} \prod_{j=3}^k \frac{(j-1)^{a}}{j^{a} + j^{a-1}-2^a - 2^{a-1}}.
\end{equation}

Continuing these calculations we find the decay laws for the densities.  In terms of the physical time
\begin{equation}
\label{cktotal}
c_k \sim
\begin{cases}
t^{-1}                                     &a_2 <a \leq a_1 =1\quad (k\geq 1)\\
t^{-1} \ln t                             & a=a_2 \qquad  ~\qquad\quad (k\geq 2)\\
t^{-\alpha_p }                       &a_{p+1} <a < a_p \qquad (k>p>2) \\
t^{-\alpha_p} \ln  t                &a = a_p ~\qquad \qquad\quad (k>p>2)
\end{cases}
\end{equation}
with $\alpha_p=(p^{a_p}+p^{a_p-1})/2$. In contrast to the phase transition between jamming and steady-state regimes where the final densities are finite and undergo a jump across the transition point, the cluster densities vanish in the present case, and only the decay exponent jumps from $\alpha_{p-1}$ to $\alpha_p$ when the transition point $a_p$ is crossed. Therefore we call these phase transitions {\em weak}.

\begin{figure}
\centering
\includegraphics[width=7cm]{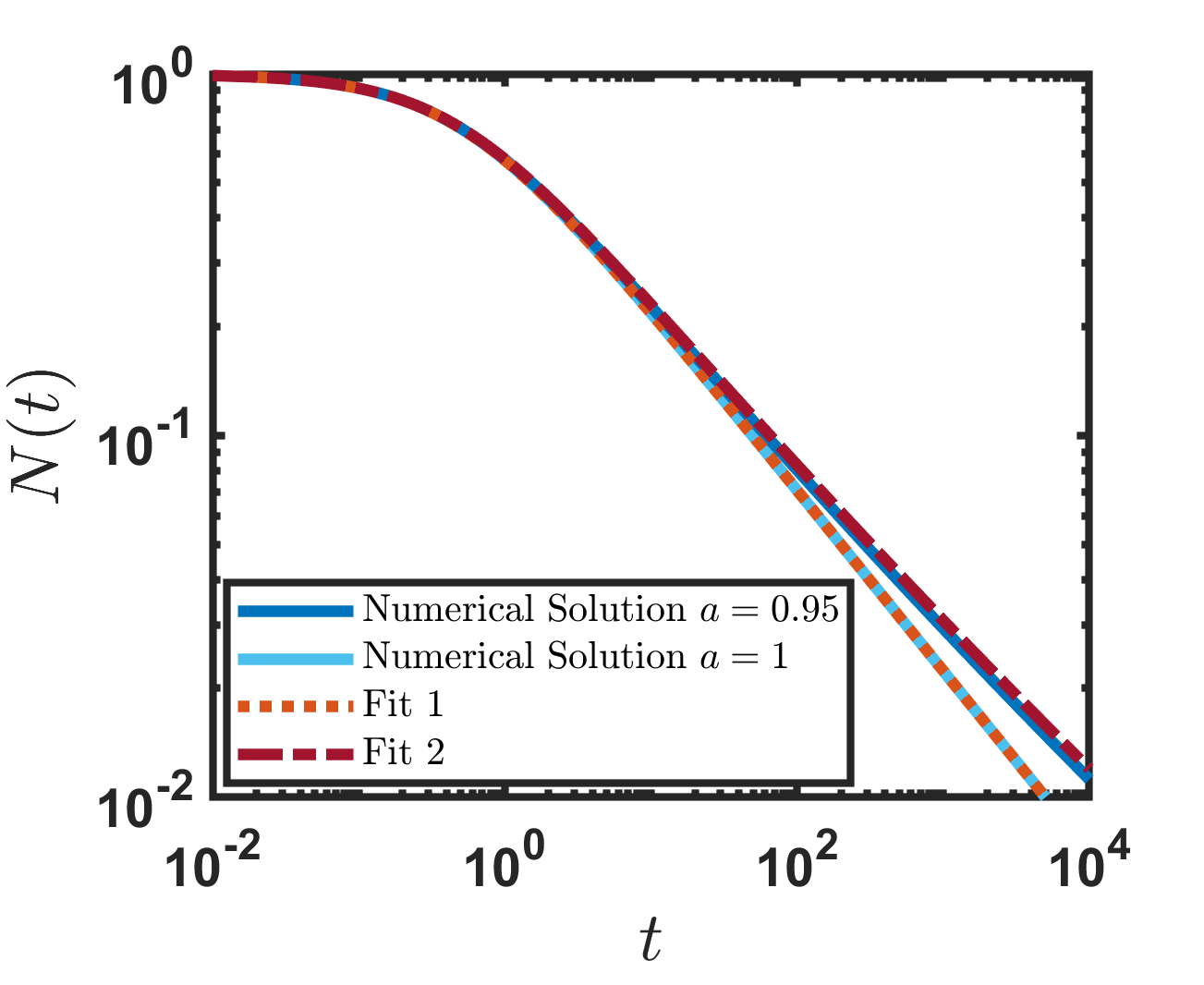}\q
\caption{Evolution of the total cluster density $N(t)$ in the SCS regime at the lower critical point $\lambda=1$ for the model with exponents $(a,s)=(0.95,-0.05)$ -- upper curve. We show numerical solutions and theoretical predictions. For comparison, $N(t)$ for $\lambda=1$ and $(a,s)=(1,0)$ is also shown -- bottom curve.  Fit 1: The theoretical prediction \eqref{N:10-sol-t}, i.e., $N(t) =(1+2t)^{-1/2}$ for $\lambda=1$ and $a=1$ (dotted curve); Fit 2:  $N(t)= [1+0.025 \ln (1+2t)]^{-1/0.05}$ agreeing with the initial condition $N(0)=1$ and the asymptotic \eqref{Na1} when $a=0.95$ (dash-dotted curve).  }
 \label{Fig:Supll_7}
\end{figure}

Finally, we consider the total cluster density at the lower critical point ($\lambda=1$). When $a=1$, the individual cluster densities decay algebraically according to \eqref{ck:10-sol-t}, and the total cluster density also decays algebraically, Eq.~\eqref{N:10-sol-t}. An algebraic decay \eqref{cktotal} of cluster densities when $a<1$ suggests a similar behavior of $N(t)$, yet much slower logarithmic decay
\begin{equation}
\label{Na1}
N \simeq \big[1+\tfrac{1-a}{2}\,\ln(1+2t)\big]^{-\frac{1}{1-a}}
\end{equation}
occurs for all $a<1$. In the $a\uparrow 1$ limit, the decay law \eqref{Na1} reduces to the exact solution for the cluster density, $N =(1+2t)^{-1/2}$, of the model with $a=1$ and $\lambda=1$, see Fig.~\ref{Fig:Supll_7}. When $a=0$, the asymptotic \eqref{Na1} reduces to \eqref{N:0-1}. In the $a<1$ range, the logarithmic decay law \eqref{Na1} is derived below, Eq.~\eqref{Na1-tt}, using essentially the same arguments as in the derivation of \eqref{N:0-1}.

 \section{The nature of the supercluster states}
 \label{sec:nature}

We have shown that addition-and-shattering processes with rates $A_k=k^a$  and $S_k=\lambda k^{a-1}$ exhibit intriguing  behaviors in the range $1\leq \lambda \leq 2-a$. All cluster densities decay to zero, so it is neither a steady state where final densities remain positive nor a jammed state where only monomers disappear in the final state. The above results refer to an infinite system. To shed light on the nature of supercluster states we analyze a finite system initially composed of ${\cal M} \gg 1$ monomers. The evolution stops when the last monomer disappears. A naive criterion
\begin{equation}
\label{naive:stop}
{\cal M} c_1(t^*)=1
\end{equation}
gives an estimate of  the time $t^*$ when the last monomer disappears. Equation \eqref{eq:ratio} tells us that $N(t)/c_1(t)$ remains finite when $1<\lambda<2-a$. This apparently implies that the total number of clusters $ \mathcal{N}_{\infty}$ remains finite:
\begin{equation}
\label{naive:N}
 \mathcal{N}_{\infty}  = \mathcal{M} N(t^*) = {\cal O}(1).
\end{equation}

\begin{figure}
\centering
\includegraphics[width=8.25cm]{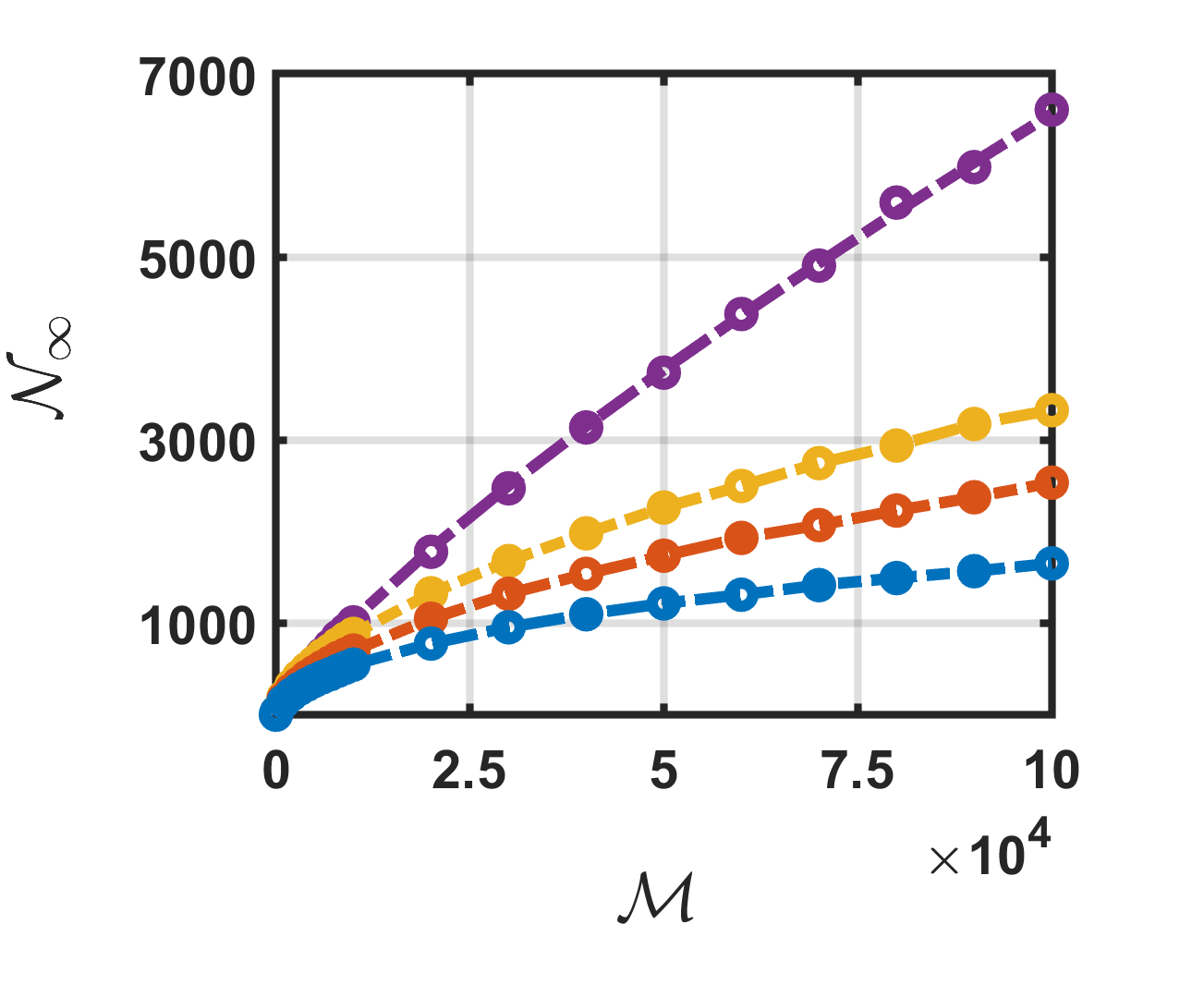}
\caption{The total number of clusters in the final SCS as a function of the system size ${\cal M}$ for different system parameters: ${\cal N}_{\infty} \sim {\cal M}^{\delta}$ with $\delta <1$. The curves from top to bottom correspond to $(a,s,\lambda)$ equal $(1,0,1)$ where $\delta=4/5$ [in agreement with the prediction \eqref{Ntot}]; $(0,-1,1.25)$ where $\delta=0.571$;  $(0,-1,1.35)$ where $\delta=0.599$; and $(0,-1,1.5)$ where $\delta=0.5$.  }
 \label{totclus}
\end{figure}

Simulations disagree with \eqref{naive:N} and indicate that $\mathcal{N}_{\infty}$ diverges with system size (see Fig.~\ref{totclus}) :
\begin{equation}
\label{delta:def}
 \mathcal{N}_{\infty}  \sim \mathcal{M}^\delta
\end{equation}
In the jamming and steady-state regimes ${\cal N}_{\infty}  \sim \mathcal{M}$, while in the SCS the growth is sub-linear: $\delta<1$.  The final mass distribution in the CSC has a scaling form
\begin{equation}
\label{SCS:scaling}
 \langle C_k\rangle  \sim \mathcal{M}^\gamma\,\Phi(\kappa), \quad \kappa = \frac{k}{\mathcal{M}^\alpha}
\end{equation}
describing the average $\langle C_k\rangle$ of the total number of clusters of mass $k$. The total numbers $C_k(\infty)= C_k(t_*)$ are non-self-averaging random quantities (they significantly vary from realization to realization) in the SCS. Combining the scaling form \eqref{SCS:scaling} with  \eqref{delta:def} and mass conservation leads to relations
\begin{equation}
\label{alpha-gamma}
\alpha + \gamma=\delta, \quad  \alpha + 2\gamma = 1
\end{equation}
from which we express through $\delta$ the exponents characterizing the scaled final mass distribution: $\gamma=1-\delta$ and $\alpha=2\delta-1$. Therefore one anticipates more tie bounds on the exponent  $\delta$, viz. $\frac{1}{2}\leq \delta\leq 1$, which were indeed observed in simulations (see Fig.~\ref{totclus}).

The time $t^*$ when the last monomer disappears also scales algebraically with system size
\begin{equation}
\label{beta:def}
t_*  \sim \mathcal{M}^\beta.
\end{equation}

\begin{figure}
\centering
\includegraphics[width=4.27cm]{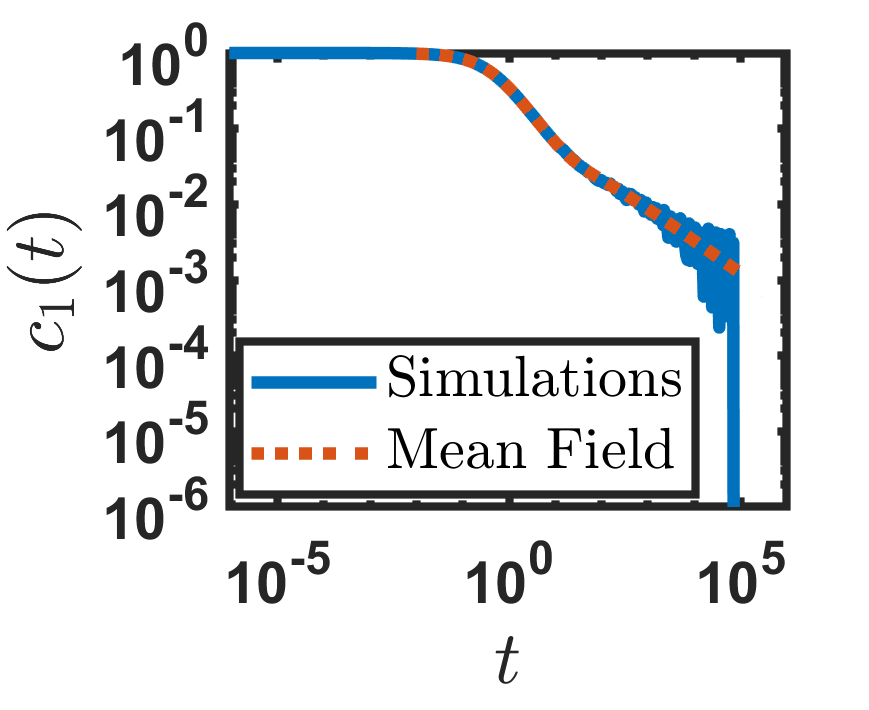}
\includegraphics[width=4.27cm]{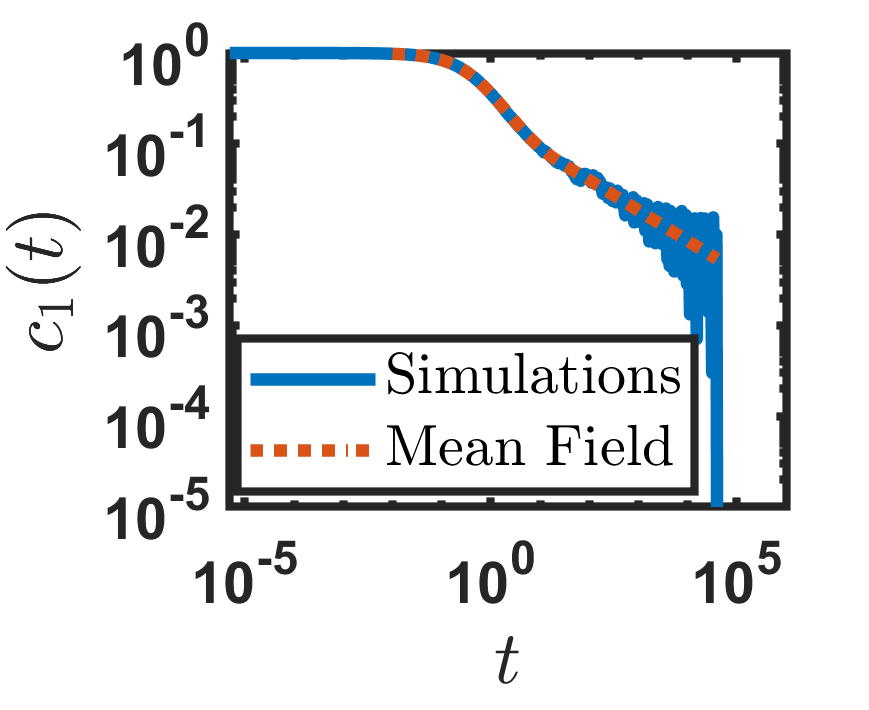}
\caption{Evolution of the monomer density for the model with $(a, s)=(0.1,-0.9)$. The SCSs arise when $1\leq \lambda\leq 1.9$. Left panel: $ \lambda = 1.2$ and $\mathcal{M}= 10^6$. Right panel: $ \lambda = 1.45$ and $\mathcal{M}= 10^5$. Approaching the SCS, the monomer density sharply drops to zero. We show simulation results (Gillespie algorithm \cite{Gillespie},  full line) together with the mean-field prediction corresponding to the numerical solution of the rate equations (dotted line). When the system approaches the SCS, density fluctuations become significant.}
 \label{FigSCS_mono}
\end{figure}

To deduce $t_*$ one should not use the naive criterion \eqref{naive:stop}. The evolution of monomers just before the system reaches the SCS (Fig.~\ref{FigSCS_mono}) hints at the flaws in reasoning based on \eqref{naive:stop}. Significant fluctuations in the number of monomers close to $t_*$ indicate that relying on the deterministic average number of monomers ${\cal M} c_1(t)$ is questionable. Secondly, just before the system drops into the SCS, the number of monomers is still very large instead of being of the order of one as posited by Eq.~\eqref{naive:stop}.

To account for fluctuations in finite systems it is customary to employ the van Kampen expansion \cite{vanKampen}. Van Kampen expansions have been used in several areas (see, e.g., \cite{vanKampen,KRB,McKane04,Mauro07,McKane08}) including aggregation and annihilation processes \cite{Lushnikov78,Spouge,Dani86,Ernst87a,Ernst87b,BK-van}. The idea is to decompose quantities of interest into extensive deterministic components and sub-extensive stochastic components. In the present case
\begin{equation}
\label{eq:confluc}
 C_{k}(t) = \mathcal{M} c_k(t) + \sqrt{\mathcal{M}}\, \eta_k (t).
\end{equation}
The terms linear in $\mathcal{M}$ are deterministic: the densities $c_k(t)$ obey the rate equations describing infinite systems. The terms proportional to $\sqrt{\mathcal{M}}$ are are stochastic: $\eta_k (t)$ are (evolving) random variables. The magnitude  $\sqrt{\mathcal{M}}$ of stochastic terms agrees with the basic tenets of statistical physics and probability theory.

Fluctuations are negligible in the thermodynamic limit (see Fig.~\ref{Fig:C1CK0m1} comparing simulation results for a large system with the mean-field solution corresponding to the infinite system). In our system, the van Kampen expansion is consistent in the jamming regime. The formation of the SCS, however, is dominated by fluctuations. The van Kampen expansion is quantitatively incorrect at the late stage, but it predicts qualitative features such as the exponents $\delta$ and $\beta$. The steady-state regime is quasi-steady when $\mathcal{M}$ is finite. The van Kampen expansion is applicable for times that scale exponentially with $\mathcal{M}$, but it does not describe a rare fluctuation that eventually leads to the extinction of monomers.

Analysis of finite systems is, in principle, straightforward but technically involved. Instead of rate equations describing the infinite system, we must rely on stochastic rules describing addition or shattering events. The state of the system is quantified by $\textbf{C} = ({C_{1}, C_{2}, C_{3}, ...})$, where $C_k(t)$ is the number of clusters of size $k$. The quantities $C_k(t)$ are non-negative integers satisfying
\begin{equation}
\label{mass-full}
\sum_{k\geq 1} k C_k(t) =  \mathcal{M}.
\end{equation}
In an elementary reaction, $\textbf{C}$ may transform as follows:
\begin{equation}
\begin{cases}
C_{1}-2,C_{2}+1                     & C_{1}(C_{1}-1)/\mathcal{M} \\
C_{1}-1,C_{k}-1,C_{k+1}+1   &  k^{a}C_{1}C_{k}/\mathcal{M} \\
C_{1}+k,C_{k}-1,C_{k+1}       & \lambda k^{a-1} C_{1}C_{k}/\mathcal{M}.
\end{cases}
\label{Cconfig}
\end{equation}
We have shown only the components of $\textbf{C}$  that changed and presented reaction rates. The reaction channel shown at the top describes the formation of dimers. The factor $\mathcal{M}^{-1}$ is necessary to recover Eqs.~\eqref{dc1awcd}--\eqref{dckawcd}  in the $\mathcal{M} \rightarrow \infty$ limit. The following reaction channels in \eqref{Cconfig} describe addition and shattering involving clusters with $k\geq 2$. To numerically investigate the supercluster state in finite systems, we use the Monte Carlo (MC) technique, namely, the Gillespie algorithm \cite{Gillespie}.

Figure \ref{FigSCS_mono} shows the evolution of the monomer density obtained by MC together with the mean-field (MF) behavior. At the beginning of the evolution, the MC and MF dependencies almost coincide. At latter times, $c_1(t)= {C_{1}(t)}/{\mathcal{M}}$ experiences large fluctuations and abruptly drops to zero. The intensity of fluctuations depends on the exponent $a$ and the shattering coefficient $\lambda$. Large fluctuations reflect that close to the emergence of the SCS, clusters tend to be large. The shattering of such clusters produces a large number of monomers. This ``duel" between addition and shattering eventually leads to the disappearance of monomers.

Using stochastic rules \eqref{Cconfig} we deduce
\begin{subequations}
\label{C12k:av}
\begin{eqnarray}
\label{C1:av}
\mathcal{M} \frac{d\langle C_{1} \rangle}{dt} &=& -2 \langle C_{1} (C_{1}-1) \rangle \nonumber\\
&+& (\lambda-1) \sum_{k=2}^{\infty} k^a \langle C_{1}C_{k} \rangle,
\end{eqnarray}
\begin{eqnarray}
\label{C2:av}
\mathcal{M} \frac{d\langle C_{2} \rangle}{dt} &=& \langle C_{1}(C_{1}-1) \rangle  \nonumber\\
&-& (2^a + \lambda 2^{a-1}) \langle C_{1}C_{2} \rangle,
\end{eqnarray}
\begin{eqnarray}
\label{Ck:av}
\mathcal{M} \frac{d\langle C_{k} \rangle}{dt} &=& (k-1)^a \langle C_{1}C_{k-1} \rangle  \nonumber\\
 & -&(k^a + \lambda k^{a-1}) \langle C_{1}C_{k} \rangle.
\end{eqnarray}
\end{subequations}
Equations \eqref{Ck:av} are valid for $k\geq 3$. The evolution equations \eqref{C12k:av} depend on the second order moments. One  can write exact equations for these moments but they depend on the third order moments. This hierarchical patterns continues making the system analytically intractable.

Some progress is possible, however, upon relying on the van Kampen expansion. Consider the simplest moment $\langle C_{1}^{2} \rangle$. Using \eqref{Cconfig} we deduce the exact evolution equation
\begin{align}
\label{C1-2:av}
\mathcal{M} \frac{d\langle C_{1}^2 \rangle}{dt} &=
-4 \langle C_{1} (C_{1}-1)^2 \rangle -\sum_{k=2}^{\infty} k^a \langle C_{1} C_{k} (2C_{1}-1) \rangle \nonumber \\
& + \lambda \sum_{k=2}^{\infty} k^a \langle C_{1} C_{k} (2C_{1}+k) \rangle.
\end{align}
Combining \eqref{C1:av}  and \eqref{C1-2:av} we find the exact equation for the variance 
\begin{align}
\mathcal{M} \frac{d ( \langle C_{1}^2 \rangle - \langle C_{1} \rangle^2)}{d\tau} &= 4 \langle C_{1} (C_{1}-1)  (\langle C_{1} \rangle - C_{1}+ 1) \rangle  \nonumber \\
& + 2(\lambda-1) \sum_{k=2}^{\infty} k^{a} \langle C_{1}C_{k}(2C_{1}-1)  \rangle \nonumber  \\
&+ \lambda \sum_{k=2}^{\infty} k^{a} \langle C_{1}C_{k}(2C_{1}+k)  \rangle.
\label{V1}
\end{align}

Using the van Kampen expansion and the shorthand notations
\begin{equation*}
\langle \eta_{1}^2 \rangle =V_{1}, \quad \langle \eta_{1} \eta_{k} \rangle = W_{1k}
\end{equation*}
we re-write the terms appearing in \eqref{V1} as
\begin{equation*}
\begin{split}
\langle C_{1}^2 \rangle - \langle C_{1} \rangle^2  &=  \mathcal{M} V_{1}  \\
\langle C_{1} C_{k} \rangle & = \mathcal{M}^{2} c_{1}c_{k} + \mathcal{M} W_{1k} \\
\langle C_{1} C_{k} (C_{1} - \langle C_{1} \rangle ) \rangle  &  = \mathcal{M}^{2} [c_{1} W_{1k}  + c_{k} V_1] \\
&+ \mathcal{M}^{3/2} \langle \eta_{1}^{2} \eta_{k} \rangle\\
\langle C_{1} (C_{1}-1)  (\langle C_{1} \rangle - C_{1}+ 1) \rangle & =  \mathcal{M}^{2} [c_{1}^{2}  -2 c_{1}V_{1}]\\
& - \mathcal{M}^{3/2} \langle \eta_{1}^{3} \rangle +\mathcal{M} [2V_{1} - c_{1}]
\end{split}
\end{equation*}
where we have taken into account that $\langle \eta_{k}\rangle = 0$. Plugging these expansions into  \eqref{V1} and equating the leading terms of the order $\mathcal{O}(\mathcal{M}^{2})$  we arrive at
\begin{align}
&\frac{dV_{1}}{d\tau} + 8 \bigg(1 - \frac{\lambda-1}{4c_{1}}  \sum_{k=2}^{\infty} k^a c_{k} \bigg) V_{1} = (3-\lambda)c_{1} \nonumber \\
&+ \lambda M_{a+1} + M_{a}+ 2(\lambda-1) \sum_{k=2}^{\infty} k^a W_{1k}.
\label{rateV1}
\end{align}

To close this equation one needs equations for covariances $W_{1k}$ with all  $k\geq 2$. The only exception is the case of $\lambda=1$ when the term with covariances vanishes.

For the model with $(a,s)=(1,0)$  [Sec.~\ref{subsec:10}], the SCS occurs at $\lambda=1$, so this is the only interesting value of the shattering parameter. Equation \eqref{rateV1} becomes
\begin{equation}
\label{V1tau}
\frac{dV_{1}}{d\tau} + 8V_{1} = 2c_{1}+M_2+1.
\end{equation}
Using $c_1=e^{-2\tau}$  and $M_2=2e^{\tau}-1$ following from the exact solution \eqref{ck:10-sol-crit} we reduce \eqref{V1tau} to
\begin{equation*}
\frac{dV_{1}}{d\tau} + 8V_{1} = 2e^{\tau} + 2e^{-2\tau}
\end{equation*}
from which $V_{1} = \tfrac{2}{9} e^{\tau} + \tfrac{1}{3} e^{-2\tau} -\tfrac{5}{9} e^{-8\tau}$. In terms of the physical time
\begin{equation}
\label{V1fin}
V_{1}(t) = \tfrac{2}{9} \sqrt{1+2t} +\tfrac{1}{3} (1+2t)^{-1} - \tfrac{5}{9}(1+2t)^{-4}.
\end{equation}
The variance diverges with time while the density vanishes: $V_1 \sim t^{1/2}$ and $c_1 \sim t^{-1}$. Thus fluctuations eventually dominate.

For the model with $(a,s,\lambda)=(1,0,1)$  the total number of monomers is
\begin{equation}
\label{C1V}
C_1(t) = \mathcal{M}(1+2t)^{-1}+ \sqrt{\mathcal{M}} \eta_1(t).
\end{equation}
The deterministic part decreases with time while the stochastic part increases as $V_1= \langle \eta_1^2 \rangle$ is a growing function of time. Both  contributions become comparable at time $t_*$ estimated from
\begin{equation}
\label{fluct:stop}
\mathcal{M} c_1(t^*)=  \sqrt{\mathcal{M}} \sqrt{V_1(t_*)}
\end{equation}
rather than the naive  criterion \eqref{naive:stop}. Plugging $V_1 \sim t^{1/2}$ and $c_1 \sim t^{-1}$ into \eqref{fluct:stop} leads to \eqref{beta:def} with $\beta=2/5$.

Recalling that $N(t)=1/\sqrt{1+2t}$, see \eqref{N:10-sol-t}, we get
\begin{equation}
\label{Ntot}
\mathcal{N}_\infty \sim \mathcal{M} N(t_*)\sim \mathcal{M}^{4/5}
\end{equation}
i.e., $\delta=\frac{4}{5}$. The exponents $\alpha$ and $\gamma$ describing the scaled final mass distribution are therefore $\alpha=\frac{3}{5}$ and  $\gamma=\frac{1}{5}$.

Simulations suggest that when the system approaches the SCS, fluctuations rapidly drive it towards the final jammed state without  monomers (see Fig.~\ref{FigSCS_mono}). Thus we estimate the final mass distribution in the SCS and final jammed state as $\langle C_k (\infty) \rangle  \simeq \langle C_k (t_*) \rangle$ for $k\geq 2$. Using \eqref{ck:10-sol-crit} we obtain
$$
\langle C_k (\infty) \rangle \simeq  \mathcal{M} c_k(t_*) + \sqrt{\mathcal{M}}\,\langle \eta_k \rangle = {\cal M} (2t_*)^{-1} e^{-k/\sqrt{t_*}}.
$$
Setting $t_{*} \sim \mathcal{M}^{\frac{2}{5}} $ we confirm the values of exponents
\begin{equation}
\langle C_{k} \rangle \sim \mathcal{M}^\frac{3}{5}\Phi(\kappa), \quad  \kappa =\frac{k}{{\cal M}^\frac{1}{5}}
\end{equation}
and even get $\Phi(\kappa) = e^{-b \kappa}$  with some unknown amplitude $b$. The prediction for the scaled average mass distribution is uncontrolled as it relies on the mean-field solution in the region  where fluctuations become important, but it is in a fair agreement with simulations.

As another example, consider the model with exponents $(a,s)=(0,-1)$. The SCS occurs [Sec.~\ref{subsec:0-1}] when $1\leq \lambda\leq 2$. When $\lambda=1$, Eq.~\eqref{rateV1} becomes
\begin{equation}
\label{V1:0}
\frac{dV_{1}}{d\tau} + 8V_{1} = 2c_1+ 1 + N.
\end{equation}
Using $c_1=e^{-2\tau}$ and $N\simeq \tau^{-1}$, see \eqref{N:0-1}, we find
\begin{equation}
\label{V1:0-sol}
V_{1} = \tfrac{1}{8}+ \tfrac{1}{8}\tau^{-1}+\ldots
\end{equation}
Plugging this into \eqref{fluct:stop} yields \eqref{beta:def} with $\beta=1/2$. Another exponent has the maximal value $\delta=1$, albeit the scaling law \eqref{delta:def} acquires a logarithmic correction
\begin{equation}
\label{N-inf:0-1-finite}
 \mathcal{N}_{\infty} \sim \mathcal{M}N(t_*) \sim \frac{\mathcal{M}}{\ln \mathcal{M}}
\end{equation}

Generally if $\lambda=1$ and $a<1$, Eq.~\eqref{rateV1} simplifies to
\begin{equation}
\label{V1:a}
\frac{dV_{1}}{d\tau} + 8V_{1} = 2c_1+ M_{a+1} + M_{a}.
\end{equation}
The moment $M_{a+1}$ dominates the right-hand side when $\tau\gg 1$. To establish its asymptotic we return to the governing equation \eqref{ck:aa} which we re-write in the form
\begin{align}
\label{ck:aa-PDE}
\left(\frac{\partial }{\partial \tau} + k^a\,\frac{\partial }{\partial k} \right) (k^a c_k)= -\frac{k^{a} c_{k}}{k^{1-a}}
\end{align}
obtained by treating $k$ as a continuous variable as we have done in deriving Eq.~\eqref{ck:0-1PDE} which follows from Eq.~\eqref{ck:aa-PDE} when $a=0$. Introducing an auxiliary variable
\begin{equation}
\label{K:def}
K=\frac{k^{1-a}}{1-a}
\end{equation}
we re-write Eq.~\eqref{ck:aa-PDE} as
\begin{align}
\label{cK:PDE}
\left(\frac{\partial }{\partial \tau} + \frac{\partial }{\partial K} \right) (k^a c_k)= -\frac{1-a}{K}\,(k^a c_k)
\end{align}
from which
\begin{align}
\label{ckK}
k^a c_k= \frac{1}{K^{1-a}}\,f(\tau-K).
\end{align}
To determine the asymptotic behavior of $M_{a+1}$ we again rely on a key feature of $c_k(\tau)$, namely that it has a sharp maximum at $K_*\simeq \tau$, i.e., $k_*\simeq [(1-a)\tau]^\frac{1}{1-a}$. Thus
\begin{eqnarray}
\label{Ma1}
M_{a+1} & = &  \sum_{k\geq 1}k^{a+1}  c_k \simeq (k_*)^a  \sum_{k\geq 1}k  c_k  \nonumber \\
&=& (k_*)^a   \simeq [(1-a)\tau]^\frac{a}{1-a}
\end{eqnarray}
and hence Eq.~\eqref{V1:a} gives
\begin{equation}
V_1\simeq \tfrac{1}{8}\,[(1-a)\,\tau]^\frac{a}{1-a} \simeq \tfrac{1}{8}\,\big[\tfrac{1-a}{2}\,\ln t\big]^\frac{a}{1-a}.
\end{equation}
Plugging this into \eqref{fluct:stop} leads to the scaling law \eqref{beta:def} with $\beta=1/2$ and a logarithmic correction:
\begin{equation}
\label{t*:a}
t_* \sim \mathcal{M}^\frac{1}{2} [\ln \mathcal{M}]^{-\frac{a}{2(1-a)}}.
\end{equation}
The same steps as in computing \eqref{Ma1} give [cf. Eq.~\eqref{Na1}]
\begin{equation}
\label{Na1-tt}
N \simeq k_*^{-1}  \simeq [(1-a)\tau]^{-\frac{1}{1-a}} \simeq \big[\tfrac{1-a}{2}\,\ln t\big]^{-\frac{1}{1-a}}.
\end{equation}
Hence
\begin{equation}
\label{N-inf:a}
 \mathcal{N}_{\infty} \sim \mathcal{M}N(t_*) \sim \mathcal{M}\,[\ln \mathcal{M}]^{-\frac{1}{1-a}}.
\end{equation}

\section{Partial disintegration}
\label{sec:partial}

Here we demonstrate the emergence of the SCS in systems with partial fragmentation resulting in abundant production of monomers. One such model posits that a significant part of an aggregate (say a half) is shattered into monomers while the other part is kept whole. We analyze a more symmetric model of abundant incomplete disintegration. As previously, we postulate that a cluster-monomer leads to absorption of monomer with the rate $A_{k}$. Another possibility is fragmentation: A $k-$mer may break into $k$ monomers, or a dimer and $k-2$ monomers, or a trimer and $k-3$ monomers, etc. All these breaking events are assumed equiprobable, so a cluster of size $k$ may disintegrate into $k-1$ equiprobable ways. This type of disintegration is schematically described by
\begin{equation}
\label{fig:partialdis}
I_{k} + M \rightarrow I_{j} + \underbrace{M + M +\ldots+ M}_{k-j+1}.
\end{equation}
For the complete disintegration, we assume the homogeneous kernels, $A_k =k^a$ and $R_k=\lambda k^r$. Thus the probability that a cluster of size $k$ breaks into a chunk of size $j$ and $k-j$ monomers is  $\lambda k^r /(k-1)$ for all $j$.

The density of monomers obeys
\begin{subequations}
\label{erosion}
\begin{eqnarray}
\label{erosionc1}
\frac{dc_{1}}{dt} &=& -2A_{1}c_{1}^2 - \sum_{j=2}^{\infty} A_{j}c_{j}c_{1} + \sum_{j=2}^{\infty} \frac{j}{j-1} R_{j}c_{j}c_{1} \nonumber \\
&+& \frac{1}{2}\sum_{j=3}^{\infty} (j-2) R_{j}c_{j}c_{1}
\end{eqnarray}
while for $k\geq 2$
\begin{eqnarray}
\frac{dc_{k}}{dt} &=& A_{k-1}c_{1}c_{k-1} - A_{k}c_{1}c_{k} + \sum_{j=k+1}^{\infty} \frac{1}{j-1} R_{j}c_{j}c_{1} \nonumber \\
&-& R_{k}c_{1}c_{k}. \label{erosionck}
\end{eqnarray}
\end{subequations}
For instance, the first two terms in the right-hand side of (\ref{erosionc1}) describe the loss of monomers in addition process, the third terms represents the gain of monomers from incomplete disintegration and the last term gives the gain due to complete disintegration.

In terms of the modified time, Eqs.~\eqref{erosion} become
\begin{align*}
\frac{dc_{1}}{d \tau} &= -\bigg(1+ \frac{\lambda}{2}\bigg)c_{1} - M_{a} + \frac{\lambda}{2} M_{r+1} + \sum_{j=2}^{\infty} \frac{\lambda j^r}{j-1} c_{j}, \\
\frac{dc_{k}}{d \tau} &=  (k-1)^ac_{k-1} - k^ac_{k} + \sum_{j=k+1}^{\infty} \frac{\lambda j^r}{j-1} c_{j} - \lambda k^r c_{k}.
\end{align*}
Summing these equations we arrive at the evolution equation for the cluster density:
\begin{equation}
\label{N:ar}
\frac{dN}{d \tau}  = -M_{a} + \frac{\lambda}{2} (M_{r+1}-c_{1}).
\end{equation}

\begin{figure}
\centering
\includegraphics[width=4.27cm]{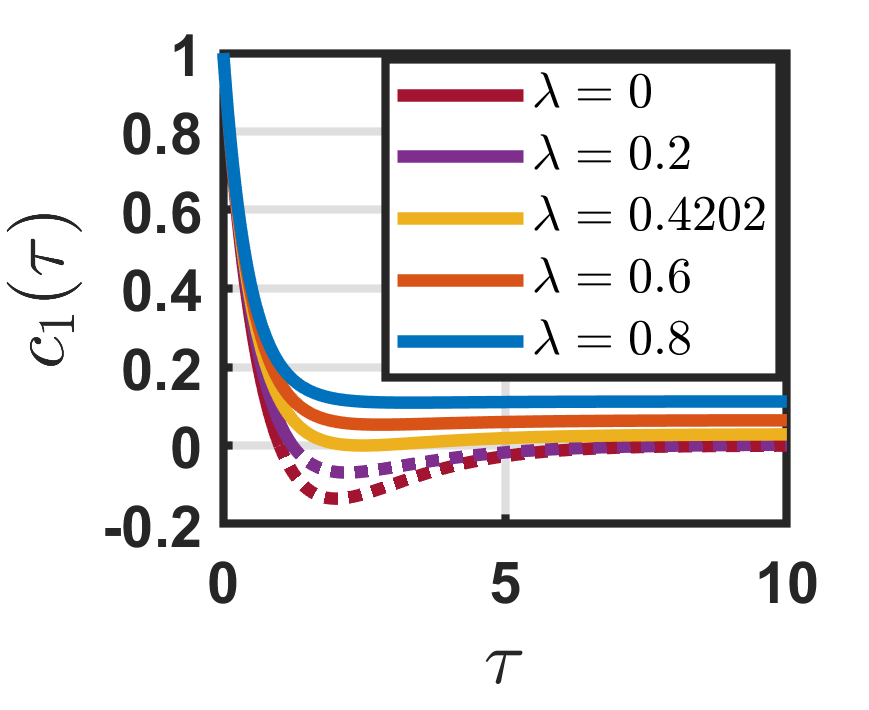}
\includegraphics[width=4.27cm]{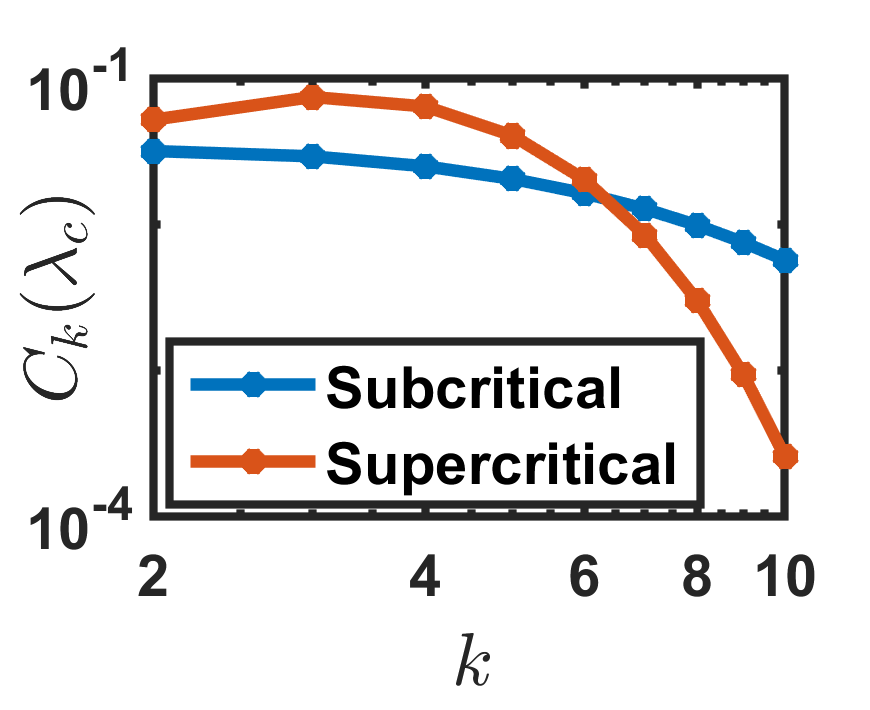}
\caption{Models with addition and abundant incomplete disintegration and exponents $(a,r)=(0,0)$. Left panel: Evolution of the monomer density for different values of $\lambda$, from bottom to top: $\lambda = 0$, $0.2$, $0.4202$, $0.6$ and $0.8$. The monomer density vanishes at $\tau=\tau_{\rm max}$. The derivative of the monomer density also vanishes in the critical regime:  $c_1(\tau_{\rm max})=0$, and $\tfrac{d}{d\tau} c_1(\tau_{\rm max})=0$. Right panel: At the critical point, a discontinuous phase transition of the final densities takes place where $c_k(\infty)|_{\lambda=\lambda_c+0}$ (upper curve) are not equal to $c_k(\infty)|_{\lambda=\lambda_c-0}$ (lower curve). }
 \label{c1ck_incom}
\end{figure}

As in the case of complete disintegration, there exists a critical value $\lambda_c$ such that for $\lambda \leq \lambda_c$, the system evolves to a jammed final state, while for $\lambda > \lambda_c$ it reaches an equilibrium state. The evolution of the monomer density is also the same: When $\lambda < \lambda_c$, it decays to zero at $\tau=\tau_{\rm max}$, where the system falls into the jammed state. When $\lambda > \lambda_c$, the monomer density is always positive, and the system reaches a steady state. When $\lambda = \lambda_c$, the system arrives at the jammed state at  $\tau=\tau_{\rm max}$, where $c_1(\tau_{\rm max})=0$, and $\tfrac{d}{d\tau} c_1(\tau_{\rm max})=0$ (see Fig.~\ref{c1ck_incom} (left)). The final cluster densities $c_k(\infty)$ with $k\geq 2$ undergo a discontinuous phase transition at  $\lambda = \lambda_c$. That is, $c_k(\infty)$ experience a final density jump at this point, so that $c_k(\infty)|_{\lambda=\lambda_c+0} \neq c_k(\infty)|_{\lambda=\lambda_c}$, see Fig.~\ref{c1ck_incom} (right).

\subsection{Supercluster states}
\label{subsec:CPT}

\begin{figure}
\centering
\includegraphics[width=4.25cm]{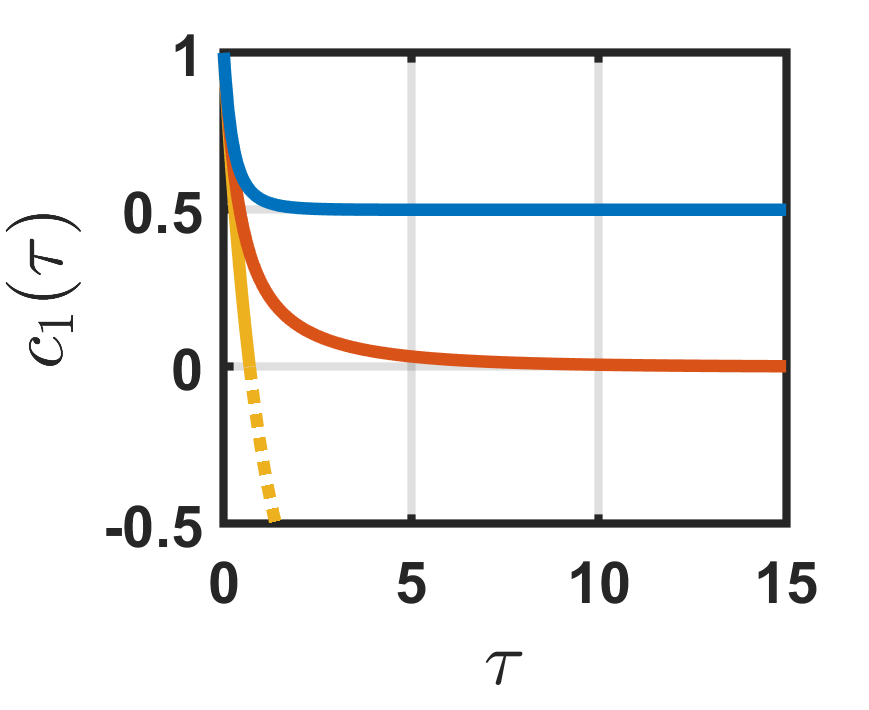}
\includegraphics[width=4.25cm]{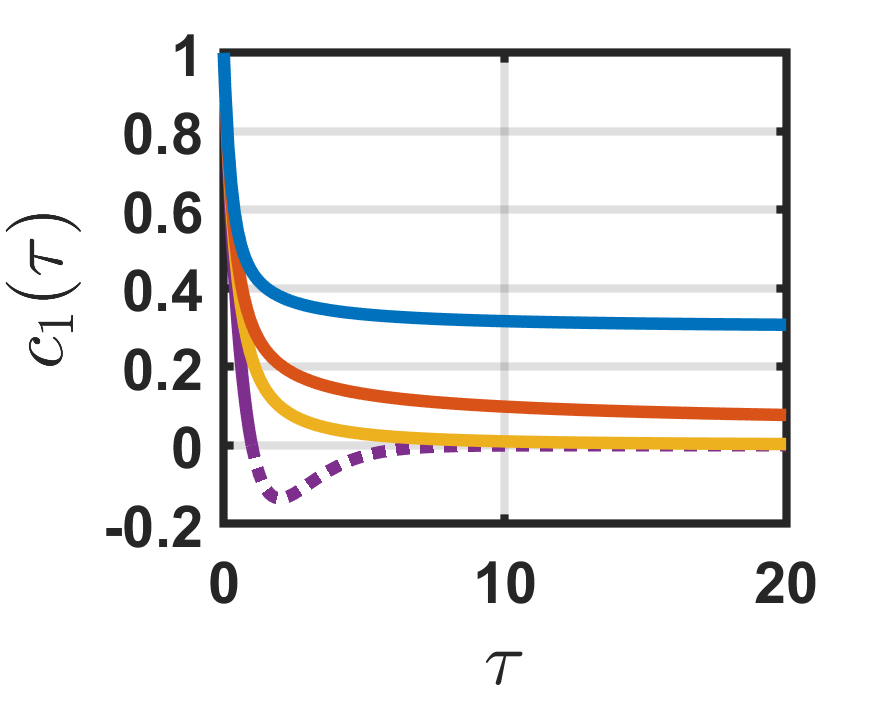}
\caption{Evolution of the monomer density for different values of $\lambda$ for $(a,r)=(1,0)$ (left panel - from bottom to top: $\lambda = 0$, $2$ and $4$) and $(a,r)=(0,-1)$ (right panel  - from bottom to top: $\lambda = 0$, $2$, $3$ and $5$). The mono-disperse initial condition is used. Shown: subcritical, critical, and supercritical curves. The critical curves in these models correspond to $\tau_{\rm max} \to \infty$. }
 \label{c1_lam_inc}
\end{figure}

Similar to systems with shattering, the SCSs exist if the exponents shifts by one: $r = a-1$. In this case $\tau_{\rm max} \to \infty$ for $\lambda_{\rm c, l} \leq \lambda \leq \lambda_{\rm c, up}$ and $c_k(\infty)=0$ for this range of $\lambda $.  At the lower critical point, the transition from the jammed state with $c_1(\infty)=0$ and $c_k(\infty) \neq 0$ for $k\geq 2$ to the SCS with $c_k(\infty)=0$ for all $k$ takes place. The transition from the SCS to the steady state occurs at the upper critical point. The evolution of the monomer density for $\lambda_{\rm c, l} \leq \lambda \leq \lambda_{\rm c, up}$ is similar to that for the case of complete disintegration (see Fig.~\ref{c1_lam_inc}).

When $r=a-1$, the governing equations  read
\begin{equation}
\label{eq:partial}
\begin{split}
\frac{dc_{1}}{d \tau} &= - \big(1+ \tfrac{\lambda}{2} \big)c_{1} + \big(\tfrac{\lambda}{2} - 1 \big)M_{a}
+ \sum_{j\geq 2} \frac{\lambda j^{a-1}}{j-1} c_{j}, \\
\frac{dc_{k}}{d \tau} &= (k-1)^{a}c_{k-1} -(k^{a} + \lambda k^{a-1})c_{k}\\
& + \sum_{j\geq k+1} \frac{\lambda j^{a-1}}{j-1} c_{j}, \quad k\geq 2.
\end{split}
\end{equation}
Equation \eqref{N:ar} becomes
\begin{equation}
\frac{dN}{d \tau} = \big(\tfrac{\lambda}{2} - 1 \big)M_{a} - \tfrac{\lambda}{2}c_{1}
\end{equation}
and hints on a special role of $\lambda=2$. Indeed, $\lambda_{c,1} = 2$ is the lower critical point.  To establish this result we set $c_{1}(\tau_{\rm max}(\lambda_{c,1})) = 0$ and $ \dot{c}_{1}(\tau_{\rm max}(\lambda_{c,1})) = 0$ in the above equation for $c_1$ and obtain
\begin{equation}
\bigg(1 - \frac{\lambda}{2} \bigg)M_{a}(\tau_{\rm max})  = \lambda\sum_{j=2}^{\infty} \frac{j^{a-1}}{j-1} c_{j}(\tau_{\rm max}).
\label{lambda-2}
\end{equation}
In the jamming regime $M_{a} > 0$ and $c_j > 0$, so Eq.~\eqref{lambda-2} can hold only for $\lambda < 2$. If $\lambda \geq 2$,  Eq.~\eqref{lambda-2} may hold only for vanishing densities corresponding to the SCS. Hence we conclude that $\lambda_{c,1}=2$.

\begin{figure}
\centering
\includegraphics[width=4cm]{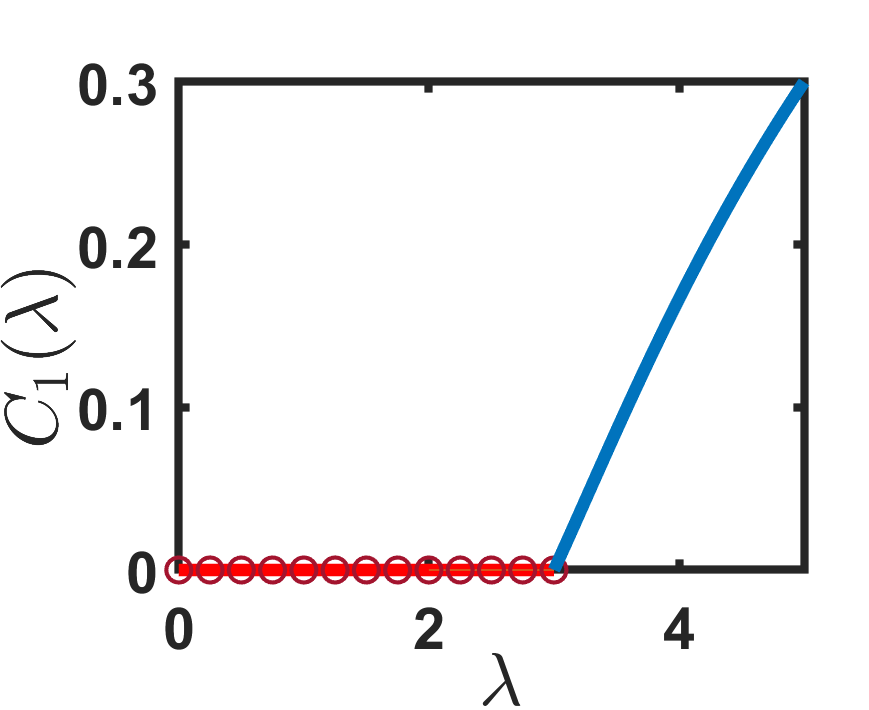}
\includegraphics[width=4cm]{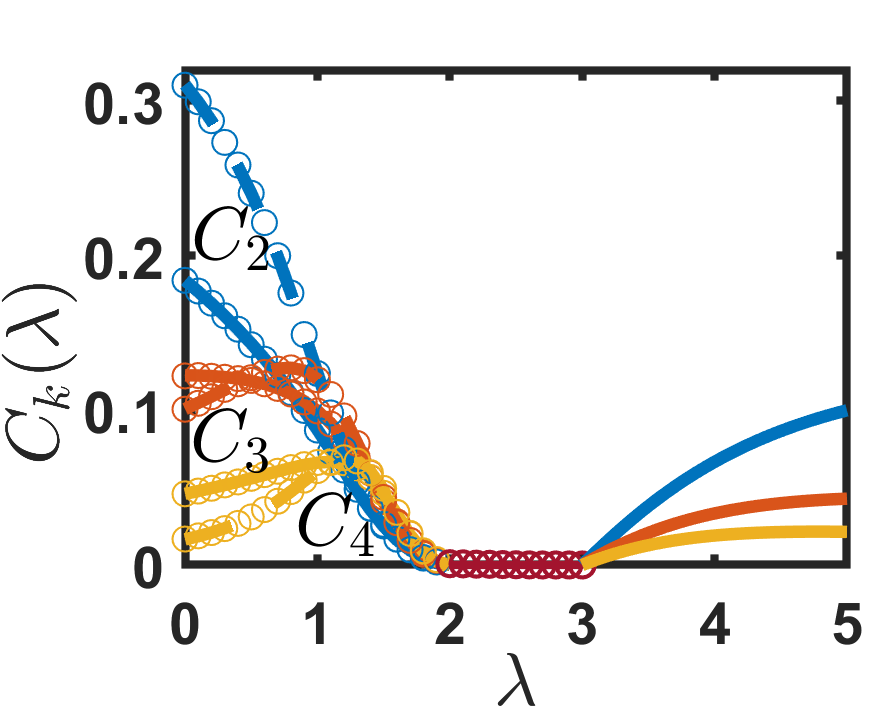}
\caption{The final densities of monomers (left panel) and clusters (right panel) for the model described by Eqs.~\eqref{eq:partial} with exponents $(a,r)=(0,-1)$. Curves from top to bottom correspond to the densities, $c_{2}$, $c_{3}$ and $c_{4}$. At $\lambda_{\rm c, l}=2$, the system undergoes a continuous phase transition from a jammed state to the SCS; at $\lambda_{\rm c, up}=3$, it undergoes a continuous phase transition from the SCS to the steady state. The final densities in the jammed state depend on the initial conditions. Solid lines: $c_{1}(0) =1$; dashed lines: $c_{1}(0) =0.2$, $c_{2}(0) = 0.4$.  Curves are solutions of rate equations; Monte Carlo results are shown by dots. }
 \label{Fig:Erosion}
\end{figure}

To find the upper critical point, we recall that the phase transition is continuous so that $c_k(\infty)$ vanish when $\lambda \to \lambda_{\rm c, up}$. Consider clusters heavier than monomers, $I=N-c_1$. Subtracting the first equation in \eqref{eq:partial} from the last we find that  $I$ satisfies
$$
\frac{dI}{d \tau} = c_1-  \sum_{j=2}^{\infty} \frac{\lambda j^{a-1}}{j-1} c_{j}.
$$
In the steady state
$$
c_1(\infty)=  \sum_{j=2}^{\infty} \frac{\lambda j^{a-1}}{j-1} c_{j}(\infty)
$$
which we use to recast the second equation in \eqref{eq:partial} into
\begin{eqnarray}
(k^{a} + \lambda k^{a-1})c_{k}(\infty) &=&(k-1)^{a}c_{k-1}(\infty) + c_1(\infty)\nonumber\\
&-&\sum_{j=2}^{k} \frac{\lambda j^{a-1}}{j-1} c_{j}(\infty)
\end{eqnarray}
for $k\geq 2$. From this recurrence we deduce
\begin{equation}
\label{ckc1_inc}
c_k(\infty)= \frac{k\,[\Gamma(k)]^a c_1(\infty)}{\prod_{n=2}^{k} \left(n^{a} + \lambda n^{a-1} +\lambda \frac{n^{a-1}}{n-1}\right)}
\end{equation}
which together with mass conservation $\sum_{k\geq 1}kc_k=1$ yields $c_{1}(\infty) = 1/G(a,\lambda)$ where $G(a,\lambda)=\sum_{k\geq 1}G_k$ with
\begin{equation}
\label{Gk}
G_k =\frac{\lambda k^2 [\Gamma(k)]^{a+1}}{\prod_{n=1}^{k}  n^{a}(n + \lambda - 1)}
= \frac{\Gamma(\lambda+1) \Gamma(k)}{\Gamma(\lambda+k)\,k^{a-2}}
\end{equation}
Since $G_k\sim k^{2-a-\lambda}$ in the large $k$ limit, the sum $G(a,\lambda)=\sum_{k\geq 1}G_k$ converges when $\lambda > 3-a$ leading to $c_k(\infty)>0$ for all  $k\geq 1$. For $\lambda \leq  3-a$, the  sum $\sum_{k\geq 1}G_k$ diverges, so $c_k(\infty)=0$ for all $k$. Hence we obtain the critical interval associated with the SCS:
\begin{equation}
\label{lam12}
\lambda_{\rm c,l} =2, \qquad \qquad \lambda_{\rm c,up} =3-a.
\end{equation}
Simulations agree with these predictions (see Fig.~\ref{Fig:Erosion}).

Closed-form expressions can be derived for integer $a$. If $(a,r)=(0,-1)$, the densities $c_k(\infty)$ and $N_\infty$ are
\begin{align*}
c_{1}(\infty) &=
\begin{cases}
0 & \lambda \leq 3\\
\frac{(\lambda-2)(\lambda-3)}{\lambda (\lambda-1)}  & \lambda > 3
\end{cases} \\
c_{k}(\infty) &=
\begin{cases}
c_{k}(\tau_{\rm max}) & 0 \leq \lambda <  2\\
0 & 2 \leq \lambda \leq 3 \\
\frac{\Gamma(k+1)\Gamma(\lambda+1)}{\Gamma(k+\lambda)} \frac{(\lambda-2)(\lambda-3)}{\lambda (\lambda-1)} &\lambda > 3
\end{cases} \\
N_\infty &=
\begin{cases}
N(\tau_{\rm max}) &  0 \leq \lambda <  2 \\
0 & 2 \leq \lambda \leq 3 \\
\frac{\lambda-3}{\lambda-1} & \lambda > 3.
\end{cases}
\end{align*}

If $(a,r)=(1, 0)$, the densities $c_k(\infty)$ and $N_\infty$ are
\begin{align*}
c_{1}(\infty) &=
\begin{cases}
0 & \lambda \leq 2\\
\frac{\lambda-2}{\lambda}  & \lambda > 2
\end{cases} \\
c_{k}(\infty) &=
\begin{cases}
c_{k}(\tau_{\rm max}) & 0 \leq \lambda <  2\\
\frac{\Gamma(k)}{\Gamma(k+\lambda)}\,(\lambda-2)\,\Gamma(\lambda)  &\lambda > 2
\end{cases} \\
N_\infty &=
\begin{cases}
N(\tau_{\rm max}) &  0 \leq \lambda <  2 \\
\frac{\lambda-2}{\lambda-1} & \lambda > 2.
\end{cases}
\end{align*}
In this model, the SCS occurs at a single point $\lambda_c=2$. The model with parameters $(a,  r, \lambda)=(1, 0, 2)$ appears most tractable for analytical exploration of the SCS in systems with abundant incomplete disintegration. The governing equations
\begin{subequations}
\label{eq:partial-2}
\begin{align}
\frac{dc_{1}}{d \tau} &= - 2c_{1} + \sum_{j\geq 2} \frac{2c_{j} }{j-1}\\
\frac{dc_{k}}{d \tau} &= (k-1)c_{k-1} -(k+2)c_k+ \sum_{j\geq k+1} \frac{2c_{j} }{j-1}
\end{align}
\end{subequations}
cannot be solved analytically, but appear amenable to asymptotic analysis leading to
\begin{equation}
\label{c1N:partial-2}
c_1\sim t^{-1}, \qquad N \sim t^{-1}\ln t
\end{equation}
and suggesting the scaling form $c_k(t) = k^{-1} t^{-1}\Phi(k/t)$.

\subsection{Finite Systems}

In a single reaction event in a finite system, the configuration $\textbf{C} = \lbrace C_{1},C_{2},\ldots \rbrace$ may transform into one of the following configurations:
\begin{equation*}
\begin{cases}
C_{1}-2,C_{2}+1 & \textnormal{rate} \; C_{1}(C_{1}-1)/\mathcal{M}, \\
C_{1}-1,C_{k}-1,C_{k+1}+1 & \textnormal{rate} \; k^{a}C_{1}C_{k}/\mathcal{M}, \\
\begin{cases}
C_{1}+1,C_{k-1}+1,C_{k}-1 \\
C_{1}+2,C_{k-2}+1,C_{k}-1 \\
\vdots \\
C_{1}+k,C_{k}-1 \\
\end{cases}  & \textnormal{rate}  \; \frac{\lambda k^{a-1}}{k-1} C_{1}C_{k}/\mathcal{M}.
\end{cases}
\label{Cdisintconfig}
\end{equation*}
We show only the components that have been altered. The top channel represents merging of two monomers, the next describes addition of monomer to clusters of mass $k\geq 2$, and the last channels describe disintegration of clusters of mass $k\geq 2$.

\begin{figure}
\centering
\includegraphics[width=2.82cm]{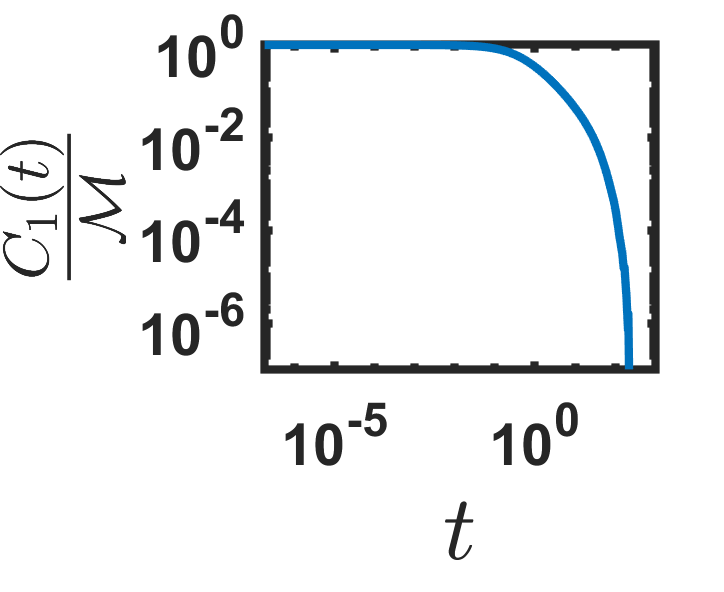}
\includegraphics[width=2.82cm]{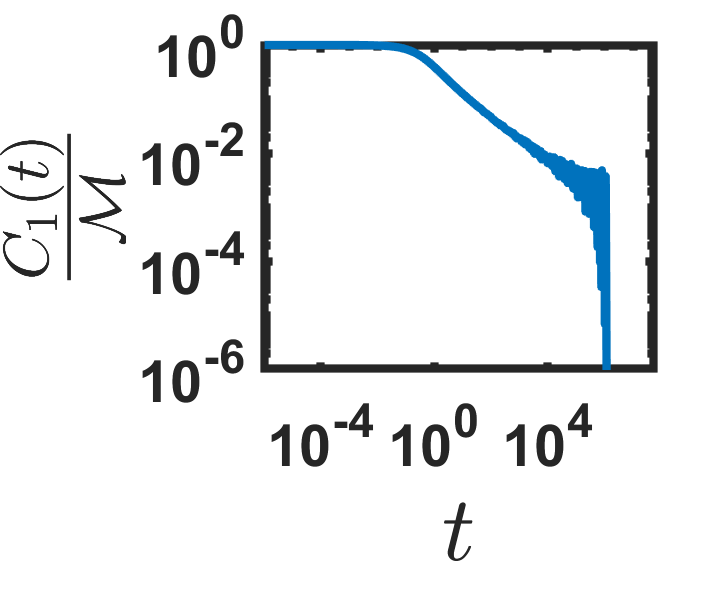}
\includegraphics[width=2.8cm]{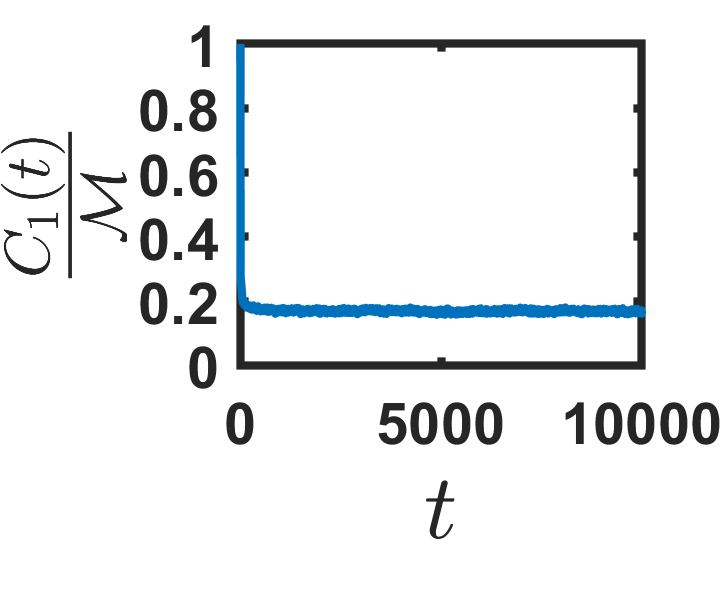}
\caption{ Evolution of the monomer density for finite systems at $\lambda=1.2 < \lambda_{\rm c,l}$ (left - $\mathcal{M} =10^7$);  $\lambda =2.4 \in (\lambda_{\rm c,l}, \lambda_{\rm c,up})$ (middle - $\mathcal{M} =10^6$); and $\lambda=4 > \lambda_{\rm c,up}$ (right - $\mathcal{M} =10^5$). In the subcritical regime, the monomer density vanishes; in the supercritical regime, it quickly reaches a positive value. These results agree with the MF predictions. In the critical regime, fluctuations dominate for large time.}
 \label{c1_lam_inc1}
\end{figure}

Figure \ref{c1_lam_inc1} shows the behavior of the monomer density $C_1(t) /{\cal M}$ in systems with $\mathcal{M} =10^5-10^7$  monomers. The dominance of fluctuations for $\lambda $ inside the critical interval associated with the SCS is visible. Outside the critical interval,  the cluster densities obtained from the MF rate equations agree with MC data. In the critical interval, fluctuations dominate over the MF predictions.

\begin{figure}
\centering
\includegraphics[width=4.27cm]{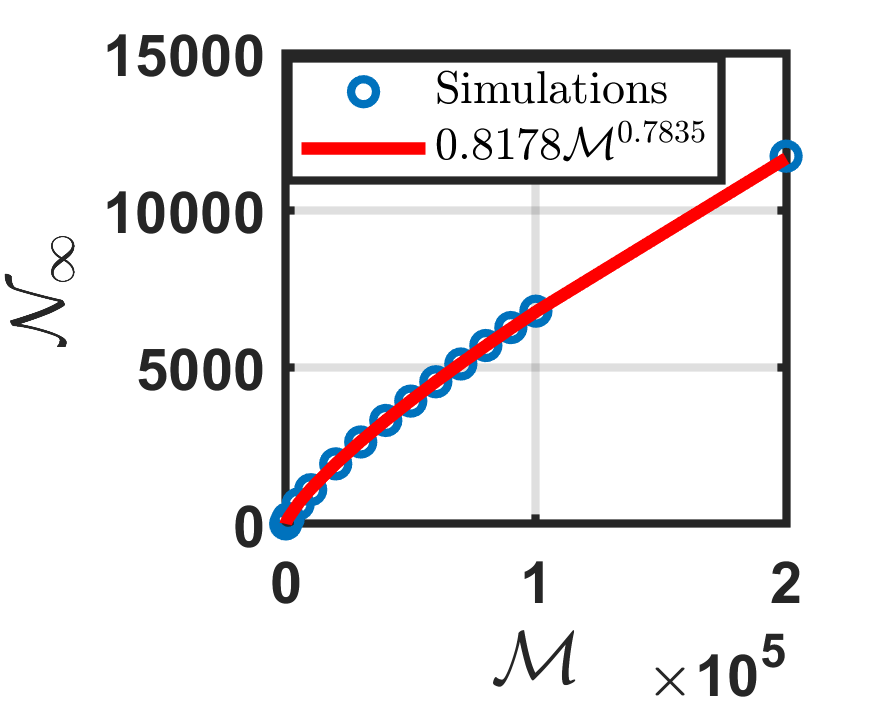}
\includegraphics[width=4.27cm]{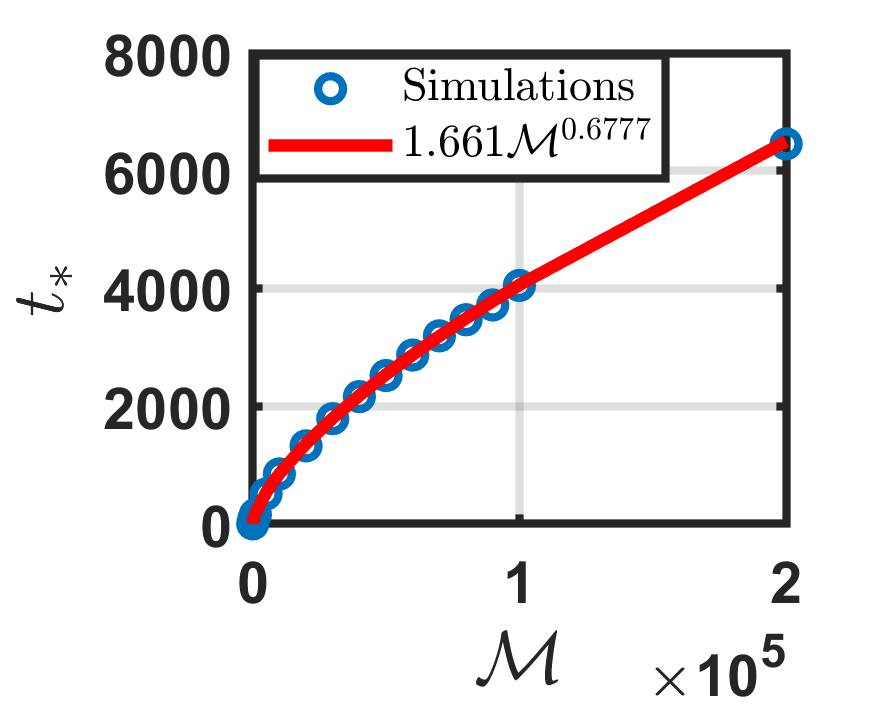}
\caption{Left panel: Final total cluster density,  ${\cal N}_{\infty}$, as a function of system  size ${\cal M}$. Dots show MC results; line is a fit to ${\cal N}_{\infty} \sim {\cal M}^{\delta}$ with $\delta = 0.6777$. Right panel: The transition time $t_*$ as a function of system  size. Dots show MC results; line is a fit to  $t_* \sim {\cal M}^{\beta }$ with $\beta = 0.7835$. The parameters are $(a,r,\lambda)=(0,-1, 2)$.  }
 \label{Ntstar_incom}
\end{figure}

We have shown that systems with abundant incomplete disintegration demonstrate qualitatively similar behaviors to systems undergoing shattering. In particular, the SCSs are also non-extensive, with the final number of clusters exhibiting a sub-linear scaling in $\mathcal{M}$,  see Fig.~\ref{Ntstar_incom}. The transition time to the SCS also has a power-law dependence on the system size (Fig.~\ref{Ntstar_incom}).

\section{Conclusions}

We have investigated the evolution of aggregates triggered by collisions with monomers. A collision may result in the attachment of a monomer or the break-up of an aggregate into constituting monomers. We have assumed that addition and shattering rates vary algebraically with size, $A_k=k^a$ and $S_k=\lambda k^s$. Without loss of generality, we set the amplitude in the addition rate to unity; the amplitude in the shattering rate is denoted by $\lambda$, so the reaction rates are parameterized by $(a,s,\lambda)$. Depending on these parameters, we observed three types of behavior:
\begin{enumerate}
\item The system falls into a jammed state where monomers disappear and the evolution stops; other cluster densities depend on the initial condition.
\item The system reaches a steady state with cluster densities independent on the initial condition.
\item The system reaches a supercluster state, a peculiar jammed state where all densities vanish in the thermodynamic  limit.
\end{enumerate}

The SCSs occur when $s=a-1$ and the shattering amplitude lies in the interval $1\leq  \lambda\leq 2-a$. The phase transitions at $\lambda_{\rm c,l}=1$ and $\lambda_{\rm c, up}=2-a$ are continuous. When  $\lambda=1$, there is also an infinite set of weak phase transitions at $a_n$ ($n=1,2, \ldots$) manifested by an abrupt change of the exponents characterizing the power-law decay of the cluster densities.

The nature of the SCSs is best revealed in finite systems initially composed of $\mathcal{M}\gg 1$ monomers. The SCSs are peculiar jammed states. The time to reach the SCS scales as $\mathcal{M}^\beta$, sometimes with a logarithmic correction [cf. Eq.~\eqref{t*:a}]. The final number of clusters in the SCS also scales algebraically, $\mathcal{N}_\infty  \sim \mathcal{M}^\delta$, sometimes with a logarithmic correction. The final number of clusters is sub-linear in system size, $0<\delta<1$. We have argued for stronger bounds $\frac{1}{2}\leq \delta\leq 1$.  The  extreme value, $\delta=1$, supplemented by a logarithmic correction, is feasible, see Eq.~\eqref{N-inf:a}, so $\mathcal{N}_\infty$ continues to scale sub-linearly with $\mathcal{M}$.

The SCSs are born in a fluctuation-dominated process. This results in non-self-averaging and non-extensive characteristics of the SCSs. The van Kampen expansion becomes dubious close to the birth of the SCS, albeit we showed how to use it to probe the basic features of the SCS, e.g., the exponents  $\beta$ and $\delta$.

We have argued that systems with abundant incomplete disintegration exhibit qualitatively similar behaviors to systems undergoing shattering. The SCSs again emerge with exponents shifted by one: $r=a-1$. We have little theoretical understanding of the SCSs in systems with abundant incomplete disintegration since we do not have analytical solutions for the time-dependent densities $c_k(t)$. The model with $(a,r)=(1,0)$ in which the SCS occurs at a single point $\lambda=2$ looks most interesting and tractable, and establishing the asymptotic behavior of $c_k(t)$ looks achievable [cf. \eqref{c1N:partial-2}]. However, even in that model, we do not see how to obtain a closed equation for the variance $\langle C_1^2 \rangle - \langle C_1 \rangle^2 $, like Eq.~\eqref{V1tau} for complete disintegration. We need such an equation for finding the exponents  $\beta$ and $\delta$.

Another possible generalization is to postulate that monomers and a few other light cluster species are mobile and consider the processes triggered by collisions with them. The pure addition processes of this kind are already analytically intractable when there are two mobile species, e.g., monomers and dimers. However, the phenomenology is the same, viz., the system quickly reaches a jammed state. Exploration of SCSs in addition-shattering processes of this type is a challenge for future work.

We emphasize that in a finite system, a steady state is {\em  quasi}-steady as monomers eventually disappear in a rare fluctuation, and the system gets jammed. The (average) lifetime $T$ of quasi-steady states is astronomically large, namely, it scales exponentially with system size
\begin{equation}
\label{T:SS}
\ln T  \simeq C\mathcal{M}.
\end{equation}
Giant adsorption times like \eqref{T:SS} arise e.g. in population dynamics where they are known as extinction times. Wentzel-Kramers-Brillouin (WKB) technique is a powerful toolbox for finding the controlling exponential behavior \eqref{T:SS}. Single-population models admit analytical treatment, see, e.g., \cite{WKB:Dykman,WKB:Kamenev,Meerson06,WKB:KS,Meerson10,WKB:KR} and a review \cite{WKB:rev}. An intriguing challenge is to develop the WKB for addition-shattering processes possessing many populations (cluster species). Systems of two interacting populations are generally intractable analytically \cite{WKB:rev}. Therefore computing the amplitude $C$ in \eqref{T:SS} could be impossible but probing it numerically is perhaps feasible in the WKB framework.

Supercluster states may arise in collision-controlled finite systems where jamming is inevitable, and the overall rates of competing merging and disintegration mechanisms are comparable. In addition-shattering processes, when the merging and disintegration rates are comparable, $A_k \sim kS_k$, the exponents must obey $s=a-1$ when rates are algebraic. Indeed, we found the supercluster states in this situation with an extra constraint on the shattering intensity: $1\leq  \lambda\leq 2-a$. In addition-chipping processes, the rates are comparable when $A_k\sim C_k$. Therefore when the exponents are equal, $s=a$, the supercluster states could emerge in addition-chipping processes, and they do emerge \cite{Sergei23} when the chipping intensity obeys $\lambda=1$. 

So far the supercluster states have been detected only in the mean-field framework. Finding them in finite dimensions is an intriguing challenge. 

\bigskip\noindent
NVB acknowledges the support of the RSF grant No.~21-11-00363.

\bibliography{references-AF}

\begin{thebibliography}{75}%
\makeatletter
\providecommand \@ifxundefined [1]{%
 \@ifx{#1\undefined}
}%
\providecommand \@ifnum [1]{%
 \ifnum #1\expandafter \@firstoftwo
 \else \expandafter \@secondoftwo
 \fi
}%
\providecommand \@ifx [1]{%
 \ifx #1\expandafter \@firstoftwo
 \else \expandafter \@secondoftwo
 \fi
}%
\providecommand \natexlab [1]{#1}%
\providecommand \enquote  [1]{``#1''}%
\providecommand \bibnamefont  [1]{#1}%
\providecommand \bibfnamefont [1]{#1}%
\providecommand \citenamefont [1]{#1}%
\providecommand \href@noop [0]{\@secondoftwo}%
\providecommand \href [0]{\begingroup \@sanitize@url \@href}%
\providecommand \@href[1]{\@@startlink{#1}\@@href}%
\providecommand \@@href[1]{\endgroup#1\@@endlink}%
\providecommand \@sanitize@url [0]{\catcode `\\12\catcode `\$12\catcode
  `\&12\catcode `\#12\catcode `\^12\catcode `\_12\catcode `\%12\relax}%
\providecommand \@@startlink[1]{}%
\providecommand \@@endlink[0]{}%
\providecommand \url  [0]{\begingroup\@sanitize@url \@url }%
\providecommand \@url [1]{\endgroup\@href {#1}{\urlprefix }}%
\providecommand \urlprefix  [0]{URL }%
\providecommand \Eprint [0]{\href }%
\providecommand \doibase [0]{https://doi.org/}%
\providecommand \selectlanguage [0]{\@gobble}%
\providecommand \bibinfo  [0]{\@secondoftwo}%
\providecommand \bibfield  [0]{\@secondoftwo}%
\providecommand \translation [1]{[#1]}%
\providecommand \BibitemOpen [0]{}%
\providecommand \bibitemStop [0]{}%
\providecommand \bibitemNoStop [0]{.\EOS\space}%
\providecommand \EOS [0]{\spacefactor3000\relax}%
\providecommand \BibitemShut  [1]{\csname bibitem#1\endcsname}%
\let\auto@bib@innerbib\@empty
\bibitem [{\citenamefont {Ponder}(1926)}]{EP1926}%
  \BibitemOpen
  \bibfield  {author} {\bibinfo {author} {\bibfnamefont {E.}~\bibnamefont
  {Ponder}},\ }\bibfield  {title} {\bibinfo {title} {On sedimentation and
  rouleaux formation--{II}},\ }\href@noop {} {\bibfield  {journal} {\bibinfo
  {journal} {Quarterly J. Exp. Physiology: Translation and Integration}\
  }\textbf {\bibinfo {volume} {16}},\ \bibinfo {pages} {173} (\bibinfo {year}
  {1926})}\BibitemShut {NoStop}%
\bibitem [{\citenamefont {Samsel}\ and\ \citenamefont
  {Perelson}(1982)}]{RWS1982}%
  \BibitemOpen
  \bibfield  {author} {\bibinfo {author} {\bibfnamefont {R.~W.}\ \bibnamefont
  {Samsel}}\ and\ \bibinfo {author} {\bibfnamefont {A.~S.}\ \bibnamefont
  {Perelson}},\ }\bibfield  {title} {\bibinfo {title} {Kinetics of rouleau
  formation. {I}. {A} mass action approach with geometric features},\ }\href
  {https://doi.org/10.1016/S0006-3495(82)84696-1} {\bibfield  {journal}
  {\bibinfo  {journal} {Biophys. J.}\ }\textbf {\bibinfo {volume} {37}},\
  \bibinfo {pages} {493} (\bibinfo {year} {1982})}\BibitemShut {NoStop}%
\bibitem [{\citenamefont {Samsel}\ and\ \citenamefont
  {Perelson}(1984)}]{RWS1984}%
  \BibitemOpen
  \bibfield  {author} {\bibinfo {author} {\bibfnamefont {R.~W.}\ \bibnamefont
  {Samsel}}\ and\ \bibinfo {author} {\bibfnamefont {A.~S.}\ \bibnamefont
  {Perelson}},\ }\bibfield  {title} {\bibinfo {title} {Kinetics of rouleau
  formation. {II}. {R}eversible reactions},\ }\href
  {https://doi.org/10.1016/S0006-3495(84)84225-3} {\bibfield  {journal}
  {\bibinfo  {journal} {Biophys. J.}\ }\textbf {\bibinfo {volume} {45}},\
  \bibinfo {pages} {805} (\bibinfo {year} {1984})}\BibitemShut {NoStop}%
\bibitem [{\citenamefont {Ryan}\ \emph {et~al.}(1980)\citenamefont {Ryan},
  \citenamefont {Hart},\ and\ \citenamefont {Schiller}}]{VR1980}%
  \BibitemOpen
  \bibfield  {author} {\bibinfo {author} {\bibfnamefont {V.}~\bibnamefont
  {Ryan}}, \bibinfo {author} {\bibfnamefont {T.~R.}\ \bibnamefont {Hart}},\
  and\ \bibinfo {author} {\bibfnamefont {R.}~\bibnamefont {Schiller}},\
  }\bibfield  {title} {\bibinfo {title} {Laser light scattering measurement of
  dextran-induced {S}treptococcus mutans aggregation},\ }\href
  {10.1016/S0006-3495(80)85043-0} {\bibfield  {journal} {\bibinfo  {journal}
  {Biophys. J.}\ }\textbf {\bibinfo {volume} {31}},\ \bibinfo {pages} {113}
  (\bibinfo {year} {1980})}\BibitemShut {NoStop}%
\bibitem [{\citenamefont {Pimpinelli}\ and\ \citenamefont
  {Villain}(1998)}]{MBE}%
  \BibitemOpen
  \bibfield  {author} {\bibinfo {author} {\bibfnamefont {A.}~\bibnamefont
  {Pimpinelli}}\ and\ \bibinfo {author} {\bibfnamefont {J.}~\bibnamefont
  {Villain}},\ }\href@noop {} {\emph {\bibinfo {title} {Physics of Crystal
  Growth}}}\ (\bibinfo  {publisher} {Cambridge University Press},\ \bibinfo
  {address} {Cambridge},\ \bibinfo {year} {1998})\BibitemShut {NoStop}%
\bibitem [{\citenamefont {Brilliantov}\ and\ \citenamefont
  {Krapivsky}(1991)}]{BK91}%
  \BibitemOpen
  \bibfield  {author} {\bibinfo {author} {\bibfnamefont {N.~V.}\ \bibnamefont
  {Brilliantov}}\ and\ \bibinfo {author} {\bibfnamefont {P.~L.}\ \bibnamefont
  {Krapivsky}},\ }\bibfield  {title} {\bibinfo {title} {Nonscaling and
  source-induced scaling behaviour in aggregation model of movable monomers and
  immovable clusters},\ }\href {https://doi.org/10.1088/0305-4470/24/20/014}
  {\bibfield  {journal} {\bibinfo  {journal} {J. Phys. A}\ }\textbf {\bibinfo
  {volume} {24}},\ \bibinfo {pages} {4787} (\bibinfo {year}
  {1991})}\BibitemShut {NoStop}%
\bibitem [{\citenamefont {Blackman}\ and\ \citenamefont
  {Wielding}(1991)}]{Blackman91}%
  \BibitemOpen
  \bibfield  {author} {\bibinfo {author} {\bibfnamefont {J.~A.}\ \bibnamefont
  {Blackman}}\ and\ \bibinfo {author} {\bibfnamefont {A.}~\bibnamefont
  {Wielding}},\ }\bibfield  {title} {\bibinfo {title} {Scaling theory of island
  growth in thin films},\ }\href {https://doi.org/10.1209/0295-5075/16/1/020}
  {\bibfield  {journal} {\bibinfo  {journal} {EPL}\ }\textbf {\bibinfo {volume}
  {16}},\ \bibinfo {pages} {115} (\bibinfo {year} {1991})}\BibitemShut
  {NoStop}%
\bibitem [{\citenamefont {Blackman}\ and\ \citenamefont
  {Marshall}(1994)}]{BlackmanMarshall_JPA94}%
  \BibitemOpen
  \bibfield  {author} {\bibinfo {author} {\bibfnamefont {J.~A.}\ \bibnamefont
  {Blackman}}\ and\ \bibinfo {author} {\bibfnamefont {A.}~\bibnamefont
  {Marshall}},\ }\bibfield  {title} {\bibinfo {title} {Coagulation and
  fragmentation in cluster-monomer reaction models},\ }\href
  {https://doi.org/10.1088/0305-4470/27/3/017} {\bibfield  {journal} {\bibinfo
  {journal} {J. Phys. A}\ }\textbf {\bibinfo {volume} {27}},\ \bibinfo {pages}
  {725} (\bibinfo {year} {1994})}\BibitemShut {NoStop}%
\bibitem [{\citenamefont {Bartelt}\ and\ \citenamefont
  {Evans}(1992)}]{Evans_PRB}%
  \BibitemOpen
  \bibfield  {author} {\bibinfo {author} {\bibfnamefont {M.~C.}\ \bibnamefont
  {Bartelt}}\ and\ \bibinfo {author} {\bibfnamefont {J.~W.}\ \bibnamefont
  {Evans}},\ }\bibfield  {title} {\bibinfo {title} {Scaling analysis of
  diffusion-mediated island growth in surface adsorption processes},\ }\href
  {https://doi.org/10.1103/PhysRevB.46.12675} {\bibfield  {journal} {\bibinfo
  {journal} {Phys. Rev. B}\ }\textbf {\bibinfo {volume} {46}},\ \bibinfo
  {pages} {12675} (\bibinfo {year} {1992})}\BibitemShut {NoStop}%
\bibitem [{\citenamefont {Kallabis}\ \emph {et~al.}(1998)\citenamefont
  {Kallabis}, \citenamefont {Krapivsky},\ and\ \citenamefont {Wolf}}]{Wolf}%
  \BibitemOpen
  \bibfield  {author} {\bibinfo {author} {\bibfnamefont {H.}~\bibnamefont
  {Kallabis}}, \bibinfo {author} {\bibfnamefont {P.~L.}\ \bibnamefont
  {Krapivsky}},\ and\ \bibinfo {author} {\bibfnamefont {D.~E.}\ \bibnamefont
  {Wolf}},\ }\bibfield  {title} {\bibinfo {title} {Island distance in
  one-dimensional epitaxial growth},\ }\href
  {https://doi.org/10.1007/s100510050505} {\bibfield  {journal} {\bibinfo
  {journal} {Eur. Phys. J. B}\ }\textbf {\bibinfo {volume} {5}},\ \bibinfo
  {pages} {801} (\bibinfo {year} {1998})}\BibitemShut {NoStop}%
\bibitem [{\citenamefont {Zinke-Allmang}(1999)}]{Zinke_Allmang1999}%
  \BibitemOpen
  \bibfield  {author} {\bibinfo {author} {\bibfnamefont {M.}~\bibnamefont
  {Zinke-Allmang}},\ }\bibfield  {title} {\bibinfo {title} {Phase separation on
  solid surfaces: nucleation, coarsening and coalescence kinetics},\ }\href
  {https://doi.org/10.1016/S0040-6090(98)01479-5} {\bibfield  {journal}
  {\bibinfo  {journal} {Thin Solid Films}\ }\textbf {\bibinfo {volume} {346}},\
  \bibinfo {pages} {1} (\bibinfo {year} {1999})}\BibitemShut {NoStop}%
\bibitem [{\citenamefont {Krapivsky}\ \emph {et~al.}(1998)\citenamefont
  {Krapivsky}, \citenamefont {Mendes},\ and\ \citenamefont
  {Redner}}]{Krapivsky_EJB1998}%
  \BibitemOpen
  \bibfield  {author} {\bibinfo {author} {\bibfnamefont {P.~L.}\ \bibnamefont
  {Krapivsky}}, \bibinfo {author} {\bibfnamefont {J.~F.~F.}\ \bibnamefont
  {Mendes}},\ and\ \bibinfo {author} {\bibfnamefont {S.}~\bibnamefont
  {Redner}},\ }\bibfield  {title} {\bibinfo {title} {Logarithmic clustering in
  sub-monolayer epitaxial growth},\ }\href
  {https://doi.org/10.1007/s100510050395} {\bibfield  {journal} {\bibinfo
  {journal} {Eur. Phys. J. B}\ }\textbf {\bibinfo {volume} {4}},\ \bibinfo
  {pages} {401} (\bibinfo {year} {1998})}\BibitemShut {NoStop}%
\bibitem [{\citenamefont {Krapivsky}\ \emph {et~al.}(1999)\citenamefont
  {Krapivsky}, \citenamefont {Mendes},\ and\ \citenamefont
  {Redner}}]{Krapivsky_PRB1999}%
  \BibitemOpen
  \bibfield  {author} {\bibinfo {author} {\bibfnamefont {P.~L.}\ \bibnamefont
  {Krapivsky}}, \bibinfo {author} {\bibfnamefont {J.~F.~F.}\ \bibnamefont
  {Mendes}},\ and\ \bibinfo {author} {\bibfnamefont {S.}~\bibnamefont
  {Redner}},\ }\bibfield  {title} {\bibinfo {title} {Influence of island
  diffusion on submonolayer epitaxial growth},\ }\href
  {https://doi.org/10.1103/PhysRevB.59.15950} {\bibfield  {journal} {\bibinfo
  {journal} {Phys. Rev. B}\ }\textbf {\bibinfo {volume} {59}},\ \bibinfo
  {pages} {15950} (\bibinfo {year} {1999})}\BibitemShut {NoStop}%
\bibitem [{\citenamefont {Amar}\ \emph {et~al.}(2001)\citenamefont {Amar},
  \citenamefont {Popescu},\ and\ \citenamefont {Family}}]{Family2001}%
  \BibitemOpen
  \bibfield  {author} {\bibinfo {author} {\bibfnamefont {J.~G.}\ \bibnamefont
  {Amar}}, \bibinfo {author} {\bibfnamefont {M.~N.}\ \bibnamefont {Popescu}},\
  and\ \bibinfo {author} {\bibfnamefont {F.}~\bibnamefont {Family}},\
  }\bibfield  {title} {\bibinfo {title} {Rate-equation approach to island
  capture zones and size distributions in epitaxial growth},\ }\href
  {https://doi.org/10.1103/PhysRevLett.86.3092} {\bibfield  {journal} {\bibinfo
   {journal} {Phys. Rev. Lett.}\ }\textbf {\bibinfo {volume} {86}},\ \bibinfo
  {pages} {3092} (\bibinfo {year} {2001})}\BibitemShut {NoStop}%
\bibitem [{\citenamefont {Poeschel}\ \emph {et~al.}(2003)\citenamefont
  {Poeschel}, \citenamefont {Brilliantov},\ and\ \citenamefont
  {Frommel}}]{Prions}%
  \BibitemOpen
  \bibfield  {author} {\bibinfo {author} {\bibfnamefont {T.}~\bibnamefont
  {Poeschel}}, \bibinfo {author} {\bibfnamefont {N.}~\bibnamefont
  {Brilliantov}},\ and\ \bibinfo {author} {\bibfnamefont {C.}~\bibnamefont
  {Frommel}},\ }\bibfield  {title} {\bibinfo {title} {Kinetics of prion
  growth},\ }\href {https://doi.org/10.1016/S0006-3495(03)74767-5} {\bibfield
  {journal} {\bibinfo  {journal} {Biophys. J.}\ }\textbf {\bibinfo {volume}
  {85}},\ \bibinfo {pages} {3460} (\bibinfo {year} {2003})}\BibitemShut
  {NoStop}%
\bibitem [{\citenamefont {Rothemund}\ \emph {et~al.}(2004)\citenamefont
  {Rothemund}, \citenamefont {Papadakis},\ and\ \citenamefont
  {Winfree}}]{RW04}%
  \BibitemOpen
  \bibfield  {author} {\bibinfo {author} {\bibfnamefont {P.~W.~K.}\
  \bibnamefont {Rothemund}}, \bibinfo {author} {\bibfnamefont {N.}~\bibnamefont
  {Papadakis}},\ and\ \bibinfo {author} {\bibfnamefont {E.}~\bibnamefont
  {Winfree}},\ }\bibfield  {title} {\bibinfo {title} {Algorithmic self-assembly
  of {DNA} {S}ierpinski triangles},\ }\href
  {https://doi.org/10.1371/journal.pbio.0020424} {\bibfield  {journal}
  {\bibinfo  {journal} {PLoS Biology}\ }\textbf {\bibinfo {volume} {2}},\
  \bibinfo {pages} {e424} (\bibinfo {year} {2004})}\BibitemShut {NoStop}%
\bibitem [{\citenamefont {Ariga}\ \emph {et~al.}(2008)\citenamefont {Ariga},
  \citenamefont {Hill}, \citenamefont {Lee}, \citenamefont {Vinu},
  \citenamefont {Charvet},\ and\ \citenamefont {Acharya}}]{SA08}%
  \BibitemOpen
  \bibfield  {author} {\bibinfo {author} {\bibfnamefont {K.}~\bibnamefont
  {Ariga}}, \bibinfo {author} {\bibfnamefont {J.~P.}\ \bibnamefont {Hill}},
  \bibinfo {author} {\bibfnamefont {M.~V.}\ \bibnamefont {Lee}}, \bibinfo
  {author} {\bibfnamefont {A.}~\bibnamefont {Vinu}}, \bibinfo {author}
  {\bibfnamefont {R.}~\bibnamefont {Charvet}},\ and\ \bibinfo {author}
  {\bibfnamefont {S.}~\bibnamefont {Acharya}},\ }\bibfield  {title} {\bibinfo
  {title} {Challenges and breakthroughs in recent research on self-assembly},\
  }\href {https://doi.org/10.1088/1468-6996/9/1/014109} {\bibfield  {journal}
  {\bibinfo  {journal} {Sci. Technol. Adv. Mater.}\ }\textbf {\bibinfo {volume}
  {9}},\ \bibinfo {pages} {014109} (\bibinfo {year} {2008})}\BibitemShut
  {NoStop}%
\bibitem [{\citenamefont {Privman}(2009)}]{Privman2009}%
  \BibitemOpen
  \bibfield  {author} {\bibinfo {author} {\bibfnamefont {V.}~\bibnamefont
  {Privman}},\ }\bibfield  {title} {\bibinfo {title} {Mechanisms of diffusional
  nucleation of nanocrystals and their self-assembly into uniform colloids},\
  }\href {https://doi.org/10.1111/j.1749-6632.2008.04323.x} {\bibfield
  {journal} {\bibinfo  {journal} {Ann. New York Acad. Sci.}\ }\textbf {\bibinfo
  {volume} {1161}},\ \bibinfo {pages} {508} (\bibinfo {year}
  {2009})}\BibitemShut {NoStop}%
\bibitem [{\citenamefont {Evans}\ and\ \citenamefont {Winfree}(2017)}]{Erik17}%
  \BibitemOpen
  \bibfield  {author} {\bibinfo {author} {\bibfnamefont {C.~G.}\ \bibnamefont
  {Evans}}\ and\ \bibinfo {author} {\bibfnamefont {E.}~\bibnamefont
  {Winfree}},\ }\bibfield  {title} {\bibinfo {title} {Physical principles for
  {DNA} tile self-assembly},\ }\href {https://doi.org/10.1039/c6cs00745g}
  {\bibfield  {journal} {\bibinfo  {journal} {Chem. Soc. Rev.}\ }\textbf
  {\bibinfo {volume} {46}},\ \bibinfo {pages} {3808} (\bibinfo {year}
  {2017})}\BibitemShut {NoStop}%
\bibitem [{\citenamefont {Gorshkov}\ and\ \citenamefont
  {Privman}(2010)}]{Privman2010}%
  \BibitemOpen
  \bibfield  {author} {\bibinfo {author} {\bibfnamefont {V.}~\bibnamefont
  {Gorshkov}}\ and\ \bibinfo {author} {\bibfnamefont {V.}~\bibnamefont
  {Privman}},\ }\bibfield  {title} {\bibinfo {title} {Models of synthesis of
  uniform colloids and nanocrystals},\ }\href
  {https://doi.org/10.1016/j.physe.2010.07.006} {\bibfield  {journal} {\bibinfo
   {journal} {Physica E}\ }\textbf {\bibinfo {volume} {43}},\ \bibinfo {pages}
  {1} (\bibinfo {year} {2010})}\BibitemShut {NoStop}%
\bibitem [{\citenamefont {Sevonkaev}\ \emph {et~al.}(2013)\citenamefont
  {Sevonkaev}, \citenamefont {Privman},\ and\ \citenamefont
  {Goia}}]{Privman2013}%
  \BibitemOpen
  \bibfield  {author} {\bibinfo {author} {\bibfnamefont {I.}~\bibnamefont
  {Sevonkaev}}, \bibinfo {author} {\bibfnamefont {V.}~\bibnamefont {Privman}},\
  and\ \bibinfo {author} {\bibfnamefont {D.}~\bibnamefont {Goia}},\ }\bibfield
  {title} {\bibinfo {title} {Growth of highly crystalline nickel particles by
  diffusional capture of atoms},\ }\href {https://doi.org/10.1063/1.4772743}
  {\bibfield  {journal} {\bibinfo  {journal} {J. Chem. Phys.}\ }\textbf
  {\bibinfo {volume} {138}},\ \bibinfo {pages} {014703} (\bibinfo {year}
  {2013})}\BibitemShut {NoStop}%
\bibitem [{\citenamefont {Koiwa}(1974)}]{Koiwa}%
  \BibitemOpen
  \bibfield  {author} {\bibinfo {author} {\bibfnamefont {M.}~\bibnamefont
  {Koiwa}},\ }\bibfield  {title} {\bibinfo {title} {On the validity of the
  grouping method -- comments on ``{A}nalysis of the clustering process of
  supersaturated lattice vacancies"},\ }\href
  {https://doi.org/10.1143/JPSJ.37.1532} {\bibfield  {journal} {\bibinfo
  {journal} {J. Phys. Soc. Jap.}\ }\textbf {\bibinfo {volume} {37}},\ \bibinfo
  {pages} {1532} (\bibinfo {year} {1974})}\BibitemShut {NoStop}%
\bibitem [{\citenamefont {Marian}\ and\ \citenamefont
  {Bulatov}(2011)}]{JNM2011}%
  \BibitemOpen
  \bibfield  {author} {\bibinfo {author} {\bibfnamefont {J.}~\bibnamefont
  {Marian}}\ and\ \bibinfo {author} {\bibfnamefont {V.~V.}\ \bibnamefont
  {Bulatov}},\ }\bibfield  {title} {\bibinfo {title} {Stochastic cluster
  dynamics method for simulations of multispecies irradiation damage
  accumulation},\ }\href {https://doi.org/10.1016/j.jnucmat.2011.05.045}
  {\bibfield  {journal} {\bibinfo  {journal} {J. Nucl. Mater.}\ }\textbf
  {\bibinfo {volume} {415}},\ \bibinfo {pages} {84} (\bibinfo {year}
  {2011})}\BibitemShut {NoStop}%
\bibitem [{\citenamefont {Blatz}\ and\ \citenamefont {Tobolsky}(1945)}]{Blatz}%
  \BibitemOpen
  \bibfield  {author} {\bibinfo {author} {\bibfnamefont {P.~J.}\ \bibnamefont
  {Blatz}}\ and\ \bibinfo {author} {\bibfnamefont {A.~V.}\ \bibnamefont
  {Tobolsky}},\ }\bibfield  {title} {\bibinfo {title} {Note on the kinetics of
  systems manifesting simultaneous polymerization-depolymerization phenomena},\
  }\href {https://doi.org/10.1021/j150440a004} {\bibfield  {journal} {\bibinfo
  {journal} {J. Phys. Chem.}\ }\textbf {\bibinfo {volume} {49}},\ \bibinfo
  {pages} {77} (\bibinfo {year} {1945})}\BibitemShut {NoStop}%
\bibitem [{\citenamefont {G\"{u}ttler}\ \emph {et~al.}(2010)\citenamefont
  {G\"{u}ttler}, \citenamefont {Blum}, \citenamefont {Zsom}, \citenamefont
  {Ormel},\ and\ \citenamefont {Dullemond}}]{Guettler2010}%
  \BibitemOpen
  \bibfield  {author} {\bibinfo {author} {\bibfnamefont {C.}~\bibnamefont
  {G\"{u}ttler}}, \bibinfo {author} {\bibfnamefont {J.}~\bibnamefont {Blum}},
  \bibinfo {author} {\bibfnamefont {A.}~\bibnamefont {Zsom}}, \bibinfo {author}
  {\bibfnamefont {C.}~\bibnamefont {Ormel}},\ and\ \bibinfo {author}
  {\bibfnamefont {C.~P.}\ \bibnamefont {Dullemond}},\ }\bibfield  {title}
  {\bibinfo {title} {The outcome of protoplanetary dust growth: {P}ebbles,
  boulders, or planetesimals?},\ }\href
  {https://doi.org/10.1051/0004-6361/200912852} {\bibfield  {journal} {\bibinfo
   {journal} {A \& A}\ }\textbf {\bibinfo {volume} {513}},\ \bibinfo {pages}
  {A56} (\bibinfo {year} {2010})}\BibitemShut {NoStop}%
\bibitem [{\citenamefont {Brilliantov}\ \emph {et~al.}(2015)\citenamefont
  {Brilliantov}, \citenamefont {Krapivsky}, \citenamefont {Bodrova},
  \citenamefont {Spahn}, \citenamefont {Hayakawa}, \citenamefont {Stadnichuk},\
  and\ \citenamefont {Schmidt}}]{PNAS}%
  \BibitemOpen
  \bibfield  {author} {\bibinfo {author} {\bibfnamefont {N.~V.}\ \bibnamefont
  {Brilliantov}}, \bibinfo {author} {\bibfnamefont {P.~L.}\ \bibnamefont
  {Krapivsky}}, \bibinfo {author} {\bibfnamefont {A.}~\bibnamefont {Bodrova}},
  \bibinfo {author} {\bibfnamefont {F.}~\bibnamefont {Spahn}}, \bibinfo
  {author} {\bibfnamefont {H.}~\bibnamefont {Hayakawa}}, \bibinfo {author}
  {\bibfnamefont {V.}~\bibnamefont {Stadnichuk}},\ and\ \bibinfo {author}
  {\bibfnamefont {J.}~\bibnamefont {Schmidt}},\ }\bibfield  {title} {\bibinfo
  {title} {Size distribution of particles in {S}aturn's rings from aggregation
  and fragmentation},\ }\href {https://doi.org/10.1073/pnas.1503957112}
  {\bibfield  {journal} {\bibinfo  {journal} {PNAS}\ }\textbf {\bibinfo
  {volume} {112}},\ \bibinfo {pages} {9536} (\bibinfo {year}
  {2015})}\BibitemShut {NoStop}%
\bibitem [{\citenamefont {Esposito}(2006)}]{esposito2006}%
  \BibitemOpen
  \bibfield  {author} {\bibinfo {author} {\bibfnamefont {L.}~\bibnamefont
  {Esposito}},\ }\href {https://doi.org/10.1017/CBO9781139236966} {\emph
  {\bibinfo {title} {Planetary Rings}}}\ (\bibinfo  {publisher} {Cambridge
  University Press},\ \bibinfo {address} {Cambridge},\ \bibinfo {year}
  {2006})\BibitemShut {NoStop}%
\bibitem [{\citenamefont {Brilliantov}\ \emph
  {et~al.}(2009{\natexlab{a}})\citenamefont {Brilliantov}, \citenamefont
  {Bodrova},\ and\ \citenamefont {Krapivsky}}]{BBK2009}%
  \BibitemOpen
  \bibfield  {author} {\bibinfo {author} {\bibfnamefont {N.~V.}\ \bibnamefont
  {Brilliantov}}, \bibinfo {author} {\bibfnamefont {A.~S.}\ \bibnamefont
  {Bodrova}},\ and\ \bibinfo {author} {\bibfnamefont {P.~L.}\ \bibnamefont
  {Krapivsky}},\ }\bibfield  {title} {\bibinfo {title} {A model of ballistic
  aggregation and fragmentation},\ }\href {10.1088/1742-5468/2009/06/P06011}
  {\bibfield  {journal} {\bibinfo  {journal} {J. Stat. Mech.}\ }\textbf
  {\bibinfo {volume} {2009}},\ \bibinfo {pages} {P06011} (\bibinfo {year}
  {2009}{\natexlab{a}})}\BibitemShut {NoStop}%
\bibitem [{\citenamefont {Dorogovtsev}\ and\ \citenamefont
  {Mendes}(2003)}]{Dorogov}%
  \BibitemOpen
  \bibfield  {author} {\bibinfo {author} {\bibfnamefont {S.~N.}\ \bibnamefont
  {Dorogovtsev}}\ and\ \bibinfo {author} {\bibfnamefont {J.~F.~F.}\
  \bibnamefont {Mendes}},\ }\href
  {https://doi.org/10.1093/acprof:oso/9780198515906.001.0001} {\emph {\bibinfo
  {title} {Evolution of networks: From biological nets to the Internet and
  {WWW}}}}\ (\bibinfo  {publisher} {Oxford University Press},\ \bibinfo
  {address} {Oxford},\ \bibinfo {year} {2003})\BibitemShut {NoStop}%
\bibitem [{\citenamefont {Grabisch}\ and\ \citenamefont
  {Rusinowska}(2013)}]{Socnet1}%
  \BibitemOpen
  \bibfield  {author} {\bibinfo {author} {\bibfnamefont {M.}~\bibnamefont
  {Grabisch}}\ and\ \bibinfo {author} {\bibfnamefont {A.}~\bibnamefont
  {Rusinowska}},\ }\bibfield  {title} {\bibinfo {title} {A model of influence
  based on aggregation functions},\ }\href
  {https://doi.org/10.1016/j.mathsocsci.2013.07.003} {\bibfield  {journal}
  {\bibinfo  {journal} {Math. Social Sci.}\ }\textbf {\bibinfo {volume} {66}},\
  \bibinfo {pages} {316} (\bibinfo {year} {2013})}\BibitemShut {NoStop}%
\bibitem [{\citenamefont {Skyrms}\ and\ \citenamefont
  {Pemantle}(2000)}]{Socnet2}%
  \BibitemOpen
  \bibfield  {author} {\bibinfo {author} {\bibfnamefont {B.}~\bibnamefont
  {Skyrms}}\ and\ \bibinfo {author} {\bibfnamefont {R.}~\bibnamefont
  {Pemantle}},\ }\bibfield  {title} {\bibinfo {title} {A dynamic model of
  social network formation},\ }\href {https://doi.org/10.1073/pnas.97.16.9340}
  {\bibfield  {journal} {\bibinfo  {journal} {PNAS}\ }\textbf {\bibinfo
  {volume} {97}},\ \bibinfo {pages} {9340} (\bibinfo {year}
  {2000})}\BibitemShut {NoStop}%
\bibitem [{\citenamefont {Ball}\ \emph {et~al.}(1986)\citenamefont {Ball},
  \citenamefont {Carr},\ and\ \citenamefont {Penrose}}]{Ball1986}%
  \BibitemOpen
  \bibfield  {author} {\bibinfo {author} {\bibfnamefont {J.~M.}\ \bibnamefont
  {Ball}}, \bibinfo {author} {\bibfnamefont {J.}~\bibnamefont {Carr}},\ and\
  \bibinfo {author} {\bibfnamefont {O.}~\bibnamefont {Penrose}},\ }\bibfield
  {title} {\bibinfo {title} {The {B}ecker-{D}\"{o}ring cluster equations: Basic
  properties and asymptotic behaviour of solutions},\ }\href
  {https://doi.org/10.1007/BF01211070} {\bibfield  {journal} {\bibinfo
  {journal} {Commun. Math. Phys.}\ }\textbf {\bibinfo {volume} {104}},\
  \bibinfo {pages} {657} (\bibinfo {year} {1986})}\BibitemShut {NoStop}%
\bibitem [{\citenamefont {King}\ and\ \citenamefont
  {Wattis}(2002)}]{KingWattis2002}%
  \BibitemOpen
  \bibfield  {author} {\bibinfo {author} {\bibfnamefont {J.~R.}\ \bibnamefont
  {King}}\ and\ \bibinfo {author} {\bibfnamefont {J.~A.~D.}\ \bibnamefont
  {Wattis}},\ }\bibfield  {title} {\bibinfo {title} {Asymptotic solutions of
  the {B}ecker-{D}\"{o}ring equations with size-dependent rate constants},\
  }\href {https://doi.org/10.1088/0305-4470/35/6/303} {\bibfield  {journal}
  {\bibinfo  {journal} {J. Phys. A}\ }\textbf {\bibinfo {volume} {35}},\
  \bibinfo {pages} {1357} (\bibinfo {year} {2002})}\BibitemShut {NoStop}%
\bibitem [{\citenamefont {Niethammer}(2003)}]{Niethammer2003}%
  \BibitemOpen
  \bibfield  {author} {\bibinfo {author} {\bibfnamefont {B.}~\bibnamefont
  {Niethammer}},\ }\bibfield  {title} {\bibinfo {title} {On the evolution of
  large clusters in the {B}ecker-{D}\"{o}ring model},\ }\href
  {https://doi.org/10.1007/s00332-002-0535-8} {\bibfield  {journal} {\bibinfo
  {journal} {J. Nonlinear Sci.}\ }\textbf {\bibinfo {volume} {13}},\ \bibinfo
  {pages} {115} (\bibinfo {year} {2003})}\BibitemShut {NoStop}%
\bibitem [{\citenamefont {Wattis}(2006)}]{Wattis2006}%
  \BibitemOpen
  \bibfield  {author} {\bibinfo {author} {\bibfnamefont {J.~A.~D.}\
  \bibnamefont {Wattis}},\ }\bibfield  {title} {\bibinfo {title} {An
  introduction to mathematical models of coagulation and fragmentation
  processes: {A} discrete deterministic mean-field approach},\ }\href
  {https://doi.org/10.1016/j.physd.2006.07.024} {\bibfield  {journal} {\bibinfo
   {journal} {Physica D}\ }\textbf {\bibinfo {volume} {222}},\ \bibinfo {pages}
  {1} (\bibinfo {year} {2006})}\BibitemShut {NoStop}%
\bibitem [{\citenamefont {Niethammer}\ and\ \citenamefont
  {Pego}(1999)}]{Niethammer1999}%
  \BibitemOpen
  \bibfield  {author} {\bibinfo {author} {\bibfnamefont {B.}~\bibnamefont
  {Niethammer}}\ and\ \bibinfo {author} {\bibfnamefont {R.~L.}\ \bibnamefont
  {Pego}},\ }\bibfield  {title} {\bibinfo {title} {Non-self-similar behavior in
  the {LSW} theory of {O}stwald ripening},\ }\href
  {https://doi.org/10.1023/A:1004546215920} {\bibfield  {journal} {\bibinfo
  {journal} {J. Stat. Phys.}\ }\textbf {\bibinfo {volume} {95}},\ \bibinfo
  {pages} {867} (\bibinfo {year} {1999})}\BibitemShut {NoStop}%
\bibitem [{\citenamefont {Herrmann}\ \emph {et~al.}(2009)\citenamefont
  {Herrmann}, \citenamefont {Niethammer},\ and\ \citenamefont
  {Velazquez}}]{Herrmann2009}%
  \BibitemOpen
  \bibfield  {author} {\bibinfo {author} {\bibfnamefont {M.}~\bibnamefont
  {Herrmann}}, \bibinfo {author} {\bibfnamefont {B.}~\bibnamefont
  {Niethammer}},\ and\ \bibinfo {author} {\bibfnamefont {J.~J.~L.}\
  \bibnamefont {Velazquez}},\ }\bibfield  {title} {\bibinfo {title}
  {Self-similar solutions for the {LSW} model with encounters},\ }\href
  {https://doi.org/10.1016/j.jde.2009.07.021} {\bibfield  {journal} {\bibinfo
  {journal} {J. Diff. Equations}\ }\textbf {\bibinfo {volume} {247}},\ \bibinfo
  {pages} {2282} (\bibinfo {year} {2009})}\BibitemShut {NoStop}%
\bibitem [{\citenamefont {Lauren\c{c}ot}\ and\ \citenamefont
  {Wrzosek}(2001)}]{Laurencot2001}%
  \BibitemOpen
  \bibfield  {author} {\bibinfo {author} {\bibfnamefont {P.}~\bibnamefont
  {Lauren\c{c}ot}}\ and\ \bibinfo {author} {\bibfnamefont {D.}~\bibnamefont
  {Wrzosek}},\ }\bibfield  {title} {\bibinfo {title} {The discrete coagulation
  equations with collisional breakage},\ }\href
  {https://doi.org/10.1023/A:1010309727754} {\bibfield  {journal} {\bibinfo
  {journal} {J. Stat. Phys.}\ }\textbf {\bibinfo {volume} {104}},\ \bibinfo
  {pages} {193} (\bibinfo {year} {2001})}\BibitemShut {NoStop}%
\bibitem [{\citenamefont {Oort}\ and\ \citenamefont {van~de
  Hulst}(1946)}]{OortHulst}%
  \BibitemOpen
  \bibfield  {author} {\bibinfo {author} {\bibfnamefont {J.~H.}\ \bibnamefont
  {Oort}}\ and\ \bibinfo {author} {\bibfnamefont {H.~C.}\ \bibnamefont {van~de
  Hulst}},\ }\bibfield  {title} {\bibinfo {title} {Gas and smoke in
  interstellar space},\ }\href@noop {} {\bibfield  {journal} {\bibinfo
  {journal} {Bull. Astron. Inst. Netherlands}\ }\textbf {\bibinfo {volume}
  {10}},\ \bibinfo {pages} {187} (\bibinfo {year} {1946})}\BibitemShut
  {NoStop}%
\bibitem [{\citenamefont {Bagland}\ and\ \citenamefont
  {Lauren\c{c}ot}(2007)}]{Laurencot2007}%
  \BibitemOpen
  \bibfield  {author} {\bibinfo {author} {\bibfnamefont {V.}~\bibnamefont
  {Bagland}}\ and\ \bibinfo {author} {\bibfnamefont {P.}~\bibnamefont
  {Lauren\c{c}ot}},\ }\bibfield  {title} {\bibinfo {title} {Self-similar
  solutions to the {O}ort--{H}ulst--{S}afronov coagulation equation},\ }\href
  {https://doi.org/10.1137/060658333} {\bibfield  {journal} {\bibinfo
  {journal} {SIAM J. Math. Anal.}\ }\textbf {\bibinfo {volume} {39}},\ \bibinfo
  {pages} {345} (\bibinfo {year} {2007})}\BibitemShut {NoStop}%
\bibitem [{\citenamefont {Dubovski}(1999)}]{Dubovski1999}%
  \BibitemOpen
  \bibfield  {author} {\bibinfo {author} {\bibfnamefont {P.~B.}\ \bibnamefont
  {Dubovski}},\ }\bibfield  {title} {\bibinfo {title} {A `triangle' of
  interconnected coagulation models},\ }\href
  {https://doi.org/10.1088/0305-4470/32/5/010} {\bibfield  {journal} {\bibinfo
  {journal} {J. Phys. A}\ }\textbf {\bibinfo {volume} {32}},\ \bibinfo {pages}
  {781} (\bibinfo {year} {1999})}\BibitemShut {NoStop}%
\bibitem [{\citenamefont {Schr\"{a}pler}\ and\ \citenamefont
  {Blum}(2011)}]{BlumErosion2011}%
  \BibitemOpen
  \bibfield  {author} {\bibinfo {author} {\bibfnamefont {R.}~\bibnamefont
  {Schr\"{a}pler}}\ and\ \bibinfo {author} {\bibfnamefont {J.}~\bibnamefont
  {Blum}},\ }\bibfield  {title} {\bibinfo {title} {The physics of
  protopanetesimal dust agglomerates. {VI}. erosion of large aggregates as a
  source of micrometer-sized particles},\ }\href {10.1088/0004-637X/734/2/108}
  {\bibfield  {journal} {\bibinfo  {journal} {Astrophys. J.}\ }\textbf
  {\bibinfo {volume} {734}},\ \bibinfo {pages} {108} (\bibinfo {year}
  {2011})}\BibitemShut {NoStop}%
\bibitem [{\citenamefont {Krapivsky}\ and\ \citenamefont
  {E.~Ben-Naim}(2003)}]{KrapivskyBenNaim2003}%
  \BibitemOpen
  \bibfield  {author} {\bibinfo {author} {\bibfnamefont {P.~L.}\ \bibnamefont
  {Krapivsky}}\ and\ \bibinfo {author} {\bibfnamefont {E.}~\bibnamefont
  {E.~Ben-Naim}},\ }\bibfield  {title} {\bibinfo {title} {Shattering
  transitions in collision-induced fragmentation},\ }\href
  {https://doi.org/10.1103/PhysRevE.68.021102} {\bibfield  {journal} {\bibinfo
  {journal} {Phys. Rev. E}\ }\textbf {\bibinfo {volume} {68}},\ \bibinfo
  {pages} {021102} (\bibinfo {year} {2003})}\BibitemShut {NoStop}%
\bibitem [{\citenamefont {Krapivsky}\ \emph {et~al.}(2010)\citenamefont
  {Krapivsky}, \citenamefont {Redner},\ and\ \citenamefont {Ben-Naim}}]{KRB}%
  \BibitemOpen
  \bibfield  {author} {\bibinfo {author} {\bibfnamefont {P.~L.}\ \bibnamefont
  {Krapivsky}}, \bibinfo {author} {\bibfnamefont {S.}~\bibnamefont {Redner}},\
  and\ \bibinfo {author} {\bibfnamefont {E.}~\bibnamefont {Ben-Naim}},\ }\href
  {https://doi.org/10.1017/CBO9780511780516} {\emph {\bibinfo {title} {A
  Kinetic View of Statistical Physics}}}\ (\bibinfo  {publisher} {Cambridge
  University Press},\ \bibinfo {address} {Cambridge},\ \bibinfo {year}
  {2010})\BibitemShut {NoStop}%
\bibitem [{\citenamefont {Brilliantov}\ \emph {et~al.}(2021)\citenamefont
  {Brilliantov}, \citenamefont {Otieno},\ and\ \citenamefont
  {Krapivsky}}]{BOK2021}%
  \BibitemOpen
  \bibfield  {author} {\bibinfo {author} {\bibfnamefont {V.}~\bibnamefont
  {Brilliantov}, \bibfnamefont {N}}, \bibinfo {author} {\bibfnamefont
  {W.}~\bibnamefont {Otieno}},\ and\ \bibinfo {author} {\bibfnamefont {P.~L.}\
  \bibnamefont {Krapivsky}},\ }\bibfield  {title} {\bibinfo {title}
  {Nonextensive supercluster states in aggregation with fragmentation},\ }\href
  {https://doi.org/10.1103/PhysRevLett.127.250602} {\bibfield  {journal}
  {\bibinfo  {journal} {Phys. Rev. Lett.}\ }\textbf {\bibinfo {volume} {127}},\
  \bibinfo {pages} {250602} (\bibinfo {year} {2021})}\BibitemShut {NoStop}%
\bibitem [{\citenamefont {van Dongen}(1989)}]{van89}%
  \BibitemOpen
  \bibfield  {author} {\bibinfo {author} {\bibfnamefont {P.~G.~J.}\
  \bibnamefont {van Dongen}},\ }\bibfield  {title} {\bibinfo {title} {Upper
  critical dimension in irreversible aggregation},\ }\href
  {https://doi.org/10.1103/PhysRevLett.63.1281} {\bibfield  {journal} {\bibinfo
   {journal} {Phys. Rev. Lett.}\ }\textbf {\bibinfo {volume} {63}},\ \bibinfo
  {pages} {1281} (\bibinfo {year} {1989})}\BibitemShut {NoStop}%
\bibitem [{\citenamefont {van Dongen}\ and\ \citenamefont
  {Ernst}(1985)}]{Ernst85a}%
  \BibitemOpen
  \bibfield  {author} {\bibinfo {author} {\bibfnamefont {P.~G.~J.}\
  \bibnamefont {van Dongen}}\ and\ \bibinfo {author} {\bibfnamefont {M.~H.}\
  \bibnamefont {Ernst}},\ }\bibfield  {title} {\bibinfo {title} {Dynamic
  scaling in the kinetics of clustering},\ }\href
  {https://doi.org/10.1103/PhysRevLett.54.1396} {\bibfield  {journal} {\bibinfo
   {journal} {Phys. Rev. Lett.}\ }\textbf {\bibinfo {volume} {54}},\ \bibinfo
  {pages} {1396} (\bibinfo {year} {1985})}\BibitemShut {NoStop}%
\bibitem [{\citenamefont {Leyvraz}(2003)}]{Leyvraz2003}%
  \BibitemOpen
  \bibfield  {author} {\bibinfo {author} {\bibfnamefont {F.}~\bibnamefont
  {Leyvraz}},\ }\bibfield  {title} {\bibinfo {title} {Scaling theory and
  exactly solved models in the kinetics of irreversible aggregation},\ }\href
  {https://doi.org/10.1016/S0370-1573(03)00241-2} {\bibfield  {journal}
  {\bibinfo  {journal} {Phys. Reports}\ }\textbf {\bibinfo {volume} {383}},\
  \bibinfo {pages} {95} (\bibinfo {year} {2003})}\BibitemShut {NoStop}%
\bibitem [{\citenamefont {Brilliantov}\ \emph
  {et~al.}(2009{\natexlab{b}})\citenamefont {Brilliantov}, \citenamefont
  {Bodrova},\ and\ \citenamefont {Krapivsky}}]{BrilBodKrap2009}%
  \BibitemOpen
  \bibfield  {author} {\bibinfo {author} {\bibfnamefont {N.~V.}\ \bibnamefont
  {Brilliantov}}, \bibinfo {author} {\bibfnamefont {A.~S.}\ \bibnamefont
  {Bodrova}},\ and\ \bibinfo {author} {\bibfnamefont {P.~L.}\ \bibnamefont
  {Krapivsky}},\ }\bibfield  {title} {\bibinfo {title} {A model of ballistic
  aggregation and fragmentation},\ }\href
  {https://doi.org/10.1088/1742-5468/2009/06/P06011} {\bibfield  {journal}
  {\bibinfo  {journal} {J. Stat. Mech.}\ }\textbf {\bibinfo {volume} {2009}},\
  \bibinfo {pages} {P06011} (\bibinfo {year} {2009}{\natexlab{b}})}\BibitemShut
  {NoStop}%
\bibitem [{\citenamefont {van Dongen}(1987{\natexlab{a}})}]{van87}%
  \BibitemOpen
  \bibfield  {author} {\bibinfo {author} {\bibfnamefont {P.~G.~J.}\
  \bibnamefont {van Dongen}},\ }\bibfield  {title} {\bibinfo {title} {On the
  possible occurrence of instantaneous gelation in {S}moluchowski's coagulation
  equation},\ }\href {https://doi.org/10.1088/0305-4470/20/7/033} {\bibfield
  {journal} {\bibinfo  {journal} {J. Phys. A}\ }\textbf {\bibinfo {volume}
  {20}},\ \bibinfo {pages} {1889} (\bibinfo {year}
  {1987}{\natexlab{a}})}\BibitemShut {NoStop}%
\bibitem [{\citenamefont {Malyshkin}\ and\ \citenamefont
  {Goodman}(2001)}]{Malyshkin01}%
  \BibitemOpen
  \bibfield  {author} {\bibinfo {author} {\bibfnamefont {L.}~\bibnamefont
  {Malyshkin}}\ and\ \bibinfo {author} {\bibfnamefont {J.}~\bibnamefont
  {Goodman}},\ }\bibfield  {title} {\bibinfo {title} {The timescale of runaway
  stochastic coagulation},\ }\href {https://doi.org/10.1006/icar.2001.6587}
  {\bibfield  {journal} {\bibinfo  {journal} {Icarus}\ }\textbf {\bibinfo
  {volume} {150}},\ \bibinfo {pages} {314} (\bibinfo {year}
  {2001})}\BibitemShut {NoStop}%
\bibitem [{\citenamefont {Ball}\ \emph {et~al.}(2011)\citenamefont {Ball},
  \citenamefont {Connaughton}, \citenamefont {Stein},\ and\ \citenamefont
  {Zaboronski}}]{Colm11}%
  \BibitemOpen
  \bibfield  {author} {\bibinfo {author} {\bibfnamefont {R.~C.}\ \bibnamefont
  {Ball}}, \bibinfo {author} {\bibfnamefont {C.}~\bibnamefont {Connaughton}},
  \bibinfo {author} {\bibfnamefont {T.~H.~M.}\ \bibnamefont {Stein}},\ and\
  \bibinfo {author} {\bibfnamefont {O.}~\bibnamefont {Zaboronski}},\ }\bibfield
   {title} {\bibinfo {title} {Instantaneous gelation in {S}moluchowski's
  coagulation equation revisited},\ }\href
  {https://doi.org/10.1103/PhysRevE.84.011111} {\bibfield  {journal} {\bibinfo
  {journal} {Phys. Rev. E}\ }\textbf {\bibinfo {volume} {84}},\ \bibinfo
  {pages} {011111} (\bibinfo {year} {2011})}\BibitemShut {NoStop}%
\bibitem [{\citenamefont {Leyvraz}(2012)}]{Leyvraz12}%
  \BibitemOpen
  \bibfield  {author} {\bibinfo {author} {\bibfnamefont {F.}~\bibnamefont
  {Leyvraz}},\ }\bibfield  {title} {\bibinfo {title} {Scaling theory for
  systems with instantaneous gelation: partial results},\ }\href
  {https://doi.org/10.1088/1751-8113/45/12/125002} {\bibfield  {journal}
  {\bibinfo  {journal} {J. Phys. A}\ }\textbf {\bibinfo {volume} {45}},\
  \bibinfo {pages} {125002} (\bibinfo {year} {2012})}\BibitemShut {NoStop}%
\bibitem [{\citenamefont {Krapivsky}\ and\ \citenamefont
  {Redner}(2001)}]{KrapRedner2001}%
  \BibitemOpen
  \bibfield  {author} {\bibinfo {author} {\bibfnamefont {P.~L.}\ \bibnamefont
  {Krapivsky}}\ and\ \bibinfo {author} {\bibfnamefont {S.}~\bibnamefont
  {Redner}},\ }\bibfield  {title} {\bibinfo {title} {Organization of growing
  random networks},\ }\href {https://doi.org/10.1103/PhysRevE.63.066123}
  {\bibfield  {journal} {\bibinfo  {journal} {Phys. Rev. E}\ }\textbf {\bibinfo
  {volume} {63}},\ \bibinfo {pages} {066123} (\bibinfo {year}
  {2001})}\BibitemShut {NoStop}%
\bibitem [{\citenamefont {Krapivsky}\ \emph {et~al.}(2017)\citenamefont
  {Krapivsky}, \citenamefont {Otieno},\ and\ \citenamefont
  {Brilliantov}}]{KOB}%
  \BibitemOpen
  \bibfield  {author} {\bibinfo {author} {\bibfnamefont {P.~L.}\ \bibnamefont
  {Krapivsky}}, \bibinfo {author} {\bibfnamefont {W.}~\bibnamefont {Otieno}},\
  and\ \bibinfo {author} {\bibfnamefont {N.~V.}\ \bibnamefont {Brilliantov}},\
  }\bibfield  {title} {\bibinfo {title} {Phase transitions in systems with
  aggregation and shattering},\ }\href
  {https://doi.org/10.1103/PhysRevE.96.042138} {\bibfield  {journal} {\bibinfo
  {journal} {Phys. Rev. E}\ }\textbf {\bibinfo {volume} {96}},\ \bibinfo
  {pages} {042138} (\bibinfo {year} {2017})}\BibitemShut {NoStop}%
\bibitem [{\citenamefont {Graham}\ \emph {et~al.}(1994)\citenamefont {Graham},
  \citenamefont {Knuth},\ and\ \citenamefont {Patashnik}}]{Knuth}%
  \BibitemOpen
  \bibfield  {author} {\bibinfo {author} {\bibfnamefont {R.~L.}\ \bibnamefont
  {Graham}}, \bibinfo {author} {\bibfnamefont {D.~E.}\ \bibnamefont {Knuth}},\
  and\ \bibinfo {author} {\bibfnamefont {O.}~\bibnamefont {Patashnik}},\
  }\href@noop {} {\emph {\bibinfo {title} {Concrete Mathematics: A Foundation
  for Computer Science}}}\ (\bibinfo  {publisher} {Addison-Wesley},\ \bibinfo
  {address} {Reading, Massachusetts},\ \bibinfo {year} {1994})\BibitemShut
  {NoStop}%
\bibitem [{\citenamefont {Gillespie}(1976)}]{Gillespie}%
  \BibitemOpen
  \bibfield  {author} {\bibinfo {author} {\bibfnamefont {D.~T.}\ \bibnamefont
  {Gillespie}},\ }\bibfield  {title} {\bibinfo {title} {A general method for
  numerically simulating the stochastic time evolution of coupled chemical
  reactions},\ }\href {https://doi.org/10.1016/0021-9991(76)90041-3} {\bibfield
   {journal} {\bibinfo  {journal} {J. Comput. Phys.}\ }\textbf {\bibinfo
  {volume} {22}},\ \bibinfo {pages} {403} (\bibinfo {year} {1976})}\BibitemShut
  {NoStop}%
\bibitem [{\citenamefont {Van~Kampen}(2007)}]{vanKampen}%
  \BibitemOpen
  \bibfield  {author} {\bibinfo {author} {\bibfnamefont {N.~G.}\ \bibnamefont
  {Van~Kampen}},\ }\href {https://doi.org/10.1016/B978-0-444-52965-7.X5000-4}
  {\emph {\bibinfo {title} {Stochastic Processes in Physics and Chemistry}}}\
  (\bibinfo  {publisher} {North Holland},\ \bibinfo {address} {Amsterdam},\
  \bibinfo {year} {2007})\BibitemShut {NoStop}%
\bibitem [{\citenamefont {McKane}\ and\ \citenamefont
  {Newman}(2004)}]{McKane04}%
  \BibitemOpen
  \bibfield  {author} {\bibinfo {author} {\bibfnamefont {A.~J.}\ \bibnamefont
  {McKane}}\ and\ \bibinfo {author} {\bibfnamefont {T.~J.}\ \bibnamefont
  {Newman}},\ }\bibfield  {title} {\bibinfo {title} {Stochastic models in
  population biology and their deterministic analogs},\ }\href
  {https://doi.org/10.1103/PhysRevE.70.041902} {\bibfield  {journal} {\bibinfo
  {journal} {Phys. Rev. E}\ }\textbf {\bibinfo {volume} {70}},\ \bibinfo
  {pages} {041902} (\bibinfo {year} {2004})}\BibitemShut {NoStop}%
\bibitem [{\citenamefont {Reichenbach}\ \emph {et~al.}(2006)\citenamefont
  {Reichenbach}, \citenamefont {Mobilia},\ and\ \citenamefont
  {Frey}}]{Mauro07}%
  \BibitemOpen
  \bibfield  {author} {\bibinfo {author} {\bibfnamefont {T.}~\bibnamefont
  {Reichenbach}}, \bibinfo {author} {\bibfnamefont {M.}~\bibnamefont
  {Mobilia}},\ and\ \bibinfo {author} {\bibfnamefont {E.}~\bibnamefont
  {Frey}},\ }\bibfield  {title} {\bibinfo {title} {Coexistence versus
  extinction in the stochastic cyclic lotka-volterra model},\ }\href
  {https://doi.org/10.1103/PhysRevE.74.051907} {\bibfield  {journal} {\bibinfo
  {journal} {Phys. Rev. E}\ }\textbf {\bibinfo {volume} {74}},\ \bibinfo
  {pages} {051907} (\bibinfo {year} {2006})}\BibitemShut {NoStop}%
\bibitem [{\citenamefont {Boland}\ \emph {et~al.}(2008)\citenamefont {Boland},
  \citenamefont {Galla},\ and\ \citenamefont {McKane}}]{McKane08}%
  \BibitemOpen
  \bibfield  {author} {\bibinfo {author} {\bibfnamefont {R.~B.}\ \bibnamefont
  {Boland}}, \bibinfo {author} {\bibfnamefont {T.}~\bibnamefont {Galla}},\ and\
  \bibinfo {author} {\bibfnamefont {A.~J.}\ \bibnamefont {McKane}},\ }\bibfield
   {title} {\bibinfo {title} {How limit cycles and quasi-cycles are related in
  systems with intrinsic noise},\ }\href
  {https://doi.org/10.1088/1742-5468/2008/09/P09001} {\bibfield  {journal}
  {\bibinfo  {journal} {J. Stat. Mech.}\ }\textbf {\bibinfo {volume} {2008}},\
  \bibinfo {pages} {P09001} (\bibinfo {year} {2008})}\BibitemShut {NoStop}%
\bibitem [{\citenamefont {Lushnikov}(1978)}]{Lushnikov78}%
  \BibitemOpen
  \bibfield  {author} {\bibinfo {author} {\bibfnamefont {A.~A.}\ \bibnamefont
  {Lushnikov}},\ }\bibfield  {title} {\bibinfo {title} {Coagulation in finite
  systems},\ }\href {https://doi.org/10.1016/0021-9797(78)90158-3} {\bibfield
  {journal} {\bibinfo  {journal} {J. Coll. Inter. Sci.}\ }\textbf {\bibinfo
  {volume} {65}},\ \bibinfo {pages} {276} (\bibinfo {year} {1978})}\BibitemShut
  {NoStop}%
\bibitem [{\citenamefont {Hendriks}\ \emph {et~al.}(1985)\citenamefont
  {Hendriks}, \citenamefont {Spouge}, \citenamefont {Eibl},\ and\ \citenamefont
  {Schreckenberg}}]{Spouge}%
  \BibitemOpen
  \bibfield  {author} {\bibinfo {author} {\bibfnamefont {E.~M.}\ \bibnamefont
  {Hendriks}}, \bibinfo {author} {\bibfnamefont {J.~L.}\ \bibnamefont
  {Spouge}}, \bibinfo {author} {\bibfnamefont {M.}~\bibnamefont {Eibl}},\ and\
  \bibinfo {author} {\bibfnamefont {M.}~\bibnamefont {Schreckenberg}},\
  }\bibfield  {title} {\bibinfo {title} {Exact solutions for random coagulation
  processes},\ }\href {https://doi.org/10.1007/BF01309254} {\bibfield
  {journal} {\bibinfo  {journal} {Z. Phys. B}\ }\textbf {\bibinfo {volume}
  {58}},\ \bibinfo {pages} {219} (\bibinfo {year} {1985})}\BibitemShut
  {NoStop}%
\bibitem [{\citenamefont {Ben-Avraham}\ and\ \citenamefont
  {Redner}(1986)}]{Dani86}%
  \BibitemOpen
  \bibfield  {author} {\bibinfo {author} {\bibfnamefont {D.}~\bibnamefont
  {Ben-Avraham}}\ and\ \bibinfo {author} {\bibfnamefont {S.}~\bibnamefont
  {Redner}},\ }\bibfield  {title} {\bibinfo {title} {Kinetics of n-species
  annihilation: Mean-field and diffusion-controlled limits},\ }\href
  {https://doi.org/10.1103/PhysRevA.34.501} {\bibfield  {journal} {\bibinfo
  {journal} {Phys. Rev. A}\ }\textbf {\bibinfo {volume} {34}},\ \bibinfo
  {pages} {501} (\bibinfo {year} {1986})}\BibitemShut {NoStop}%
\bibitem [{\citenamefont {van Dongen}\ and\ \citenamefont
  {Ernst}(1987)}]{Ernst87a}%
  \BibitemOpen
  \bibfield  {author} {\bibinfo {author} {\bibfnamefont {P.~G.~J.}\
  \bibnamefont {van Dongen}}\ and\ \bibinfo {author} {\bibfnamefont {M.~H.}\
  \bibnamefont {Ernst}},\ }\bibfield  {title} {\bibinfo {title} {Fluctuations
  in coagulating systems},\ }\href {https://doi.org/10.1007/BF01017553}
  {\bibfield  {journal} {\bibinfo  {journal} {J. Stat. Phys.}\ }\textbf
  {\bibinfo {volume} {49}},\ \bibinfo {pages} {879} (\bibinfo {year}
  {1987})}\BibitemShut {NoStop}%
\bibitem [{\citenamefont {van Dongen}(1987{\natexlab{b}})}]{Ernst87b}%
  \BibitemOpen
  \bibfield  {author} {\bibinfo {author} {\bibfnamefont {P.~G.~J.}\
  \bibnamefont {van Dongen}},\ }\bibfield  {title} {\bibinfo {title}
  {Fluctuations in coagulating systems. {II}},\ }\href
  {https://doi.org/10.1007/BF01017554} {\bibfield  {journal} {\bibinfo
  {journal} {J. Stat. Phys.}\ }\textbf {\bibinfo {volume} {49}},\ \bibinfo
  {pages} {927} (\bibinfo {year} {1987}{\natexlab{b}})}\BibitemShut {NoStop}%
\bibitem [{\citenamefont {Ben-Naim}\ and\ \citenamefont
  {Krapivsky}(2004)}]{BK-van}%
  \BibitemOpen
  \bibfield  {author} {\bibinfo {author} {\bibfnamefont {E.}~\bibnamefont
  {Ben-Naim}}\ and\ \bibinfo {author} {\bibfnamefont {P.~L.}\ \bibnamefont
  {Krapivsky}},\ }\bibfield  {title} {\bibinfo {title} {Finite-size
  fluctuations in interacting particle systems},\ }\href
  {https://doi.org/10.1103/PhysRevE.69.046113} {\bibfield  {journal} {\bibinfo
  {journal} {Phys. Rev. E}\ }\textbf {\bibinfo {volume} {69}},\ \bibinfo
  {pages} {046113} (\bibinfo {year} {2004})}\BibitemShut {NoStop}%
\bibitem [{\citenamefont {Dykman}\ \emph {et~al.}(1994)\citenamefont {Dykman},
  \citenamefont {Mori}, \citenamefont {Ross},\ and\ \citenamefont
  {Hunt}}]{WKB:Dykman}%
  \BibitemOpen
  \bibfield  {author} {\bibinfo {author} {\bibfnamefont {M.~I.}\ \bibnamefont
  {Dykman}}, \bibinfo {author} {\bibfnamefont {E.}~\bibnamefont {Mori}},
  \bibinfo {author} {\bibfnamefont {J.}~\bibnamefont {Ross}},\ and\ \bibinfo
  {author} {\bibfnamefont {P.~M.}\ \bibnamefont {Hunt}},\ }\bibfield  {title}
  {\bibinfo {title} {Large fluctuations and optimal paths in chemical
  kinetics},\ }\href {https://doi.org/10.1063/1.467139} {\bibfield  {journal}
  {\bibinfo  {journal} {J. Chem. Phys.}\ }\textbf {\bibinfo {volume} {100}},\
  \bibinfo {pages} {5735} (\bibinfo {year} {1994})}\BibitemShut {NoStop}%
\bibitem [{\citenamefont {Elgart}\ and\ \citenamefont
  {Kamenev}(2004)}]{WKB:Kamenev}%
  \BibitemOpen
  \bibfield  {author} {\bibinfo {author} {\bibfnamefont {V.}~\bibnamefont
  {Elgart}}\ and\ \bibinfo {author} {\bibfnamefont {A.}~\bibnamefont
  {Kamenev}},\ }\bibfield  {title} {\bibinfo {title} {Rare event statistics in
  reaction-diffusion systems},\ }\href
  {https://doi.org/10.1103/PhysRevE.70.041106} {\bibfield  {journal} {\bibinfo
  {journal} {Phys. Rev. E}\ }\textbf {\bibinfo {volume} {70}},\ \bibinfo
  {pages} {041106} (\bibinfo {year} {2004})}\BibitemShut {NoStop}%
\bibitem [{\citenamefont {Assaf}\ and\ \citenamefont
  {Meerson}(2006)}]{Meerson06}%
  \BibitemOpen
  \bibfield  {author} {\bibinfo {author} {\bibfnamefont {M.}~\bibnamefont
  {Assaf}}\ and\ \bibinfo {author} {\bibfnamefont {B.}~\bibnamefont
  {Meerson}},\ }\bibfield  {title} {\bibinfo {title} {Spectral formulation and
  wkb approximation for rare-event statistics in reaction systems},\ }\href
  {https://doi.org/10.1103/PhysRevE.74.041115} {\bibfield  {journal} {\bibinfo
  {journal} {Phys. Rev. E}\ }\textbf {\bibinfo {volume} {74}},\ \bibinfo
  {pages} {041115} (\bibinfo {year} {2006})}\BibitemShut {NoStop}%
\bibitem [{\citenamefont {Kessler}\ and\ \citenamefont
  {Shnerb}(2007)}]{WKB:KS}%
  \BibitemOpen
  \bibfield  {author} {\bibinfo {author} {\bibfnamefont {D.~A.}\ \bibnamefont
  {Kessler}}\ and\ \bibinfo {author} {\bibfnamefont {N.~M.}\ \bibnamefont
  {Shnerb}},\ }\bibfield  {title} {\bibinfo {title} {Extinction rates for
  fluctuation-induced metastabilities: A real-space {WKB} approach},\ }\href
  {https://doi.org/10.1007/s10955-007-9312-2} {\bibfield  {journal} {\bibinfo
  {journal} {J. Stat. Phys.}\ }\textbf {\bibinfo {volume} {127}},\ \bibinfo
  {pages} {861} (\bibinfo {year} {2007})}\BibitemShut {NoStop}%
\bibitem [{\citenamefont {Assaf}\ and\ \citenamefont
  {Meerson}(2010)}]{Meerson10}%
  \BibitemOpen
  \bibfield  {author} {\bibinfo {author} {\bibfnamefont {M.}~\bibnamefont
  {Assaf}}\ and\ \bibinfo {author} {\bibfnamefont {B.}~\bibnamefont
  {Meerson}},\ }\bibfield  {title} {\bibinfo {title} {Extinction of metastable
  stochastic populations},\ }\href {https://doi.org/10.1103/PhysRevE.81.021116}
  {\bibfield  {journal} {\bibinfo  {journal} {Phys. Rev. E}\ }\textbf {\bibinfo
  {volume} {81}},\ \bibinfo {pages} {021116} (\bibinfo {year}
  {2010})}\BibitemShut {NoStop}%
\bibitem [{\citenamefont {Krapivsky}\ and\ \citenamefont
  {Redner}(2021)}]{WKB:KR}%
  \BibitemOpen
  \bibfield  {author} {\bibinfo {author} {\bibfnamefont {P.~L.}\ \bibnamefont
  {Krapivsky}}\ and\ \bibinfo {author} {\bibfnamefont {S.}~\bibnamefont
  {Redner}},\ }\bibfield  {title} {\bibinfo {title} {Divergence and consensus
  in majority rule},\ }\href {https://doi.org/10.1103/PhysRevE.103.L060301}
  {\bibfield  {journal} {\bibinfo  {journal} {Phys. Rev. E}\ }\textbf {\bibinfo
  {volume} {103}},\ \bibinfo {pages} {L060301} (\bibinfo {year}
  {2021})}\BibitemShut {NoStop}%
\bibitem [{\citenamefont {Assaf}\ and\ \citenamefont
  {Meerson}(2017)}]{WKB:rev}%
  \BibitemOpen
  \bibfield  {author} {\bibinfo {author} {\bibfnamefont {M.}~\bibnamefont
  {Assaf}}\ and\ \bibinfo {author} {\bibfnamefont {B.}~\bibnamefont
  {Meerson}},\ }\bibfield  {title} {\bibinfo {title} {{WKB} theory of large
  deviations in stochastic populations},\ }\href
  {https://doi.org/10.1088/1751-8121/aa669a} {\bibfield  {journal} {\bibinfo
  {journal} {J. Phys. A}\ }\textbf {\bibinfo {volume} {50}},\ \bibinfo {pages}
  {263001} (\bibinfo {year} {2017})}\BibitemShut {NoStop}%
\bibitem [{\citenamefont {Dyachenko}\ \emph {et~al.}(2023)\citenamefont
  {Dyachenko}, \citenamefont {Matveev},\ and\ \citenamefont
  {Krapivsky}}]{Sergei23}%
  \BibitemOpen
  \bibfield  {author} {\bibinfo {author} {\bibfnamefont {R.~R.}\ \bibnamefont
  {Dyachenko}}, \bibinfo {author} {\bibfnamefont {S.~A.}\ \bibnamefont
  {Matveev}},\ and\ \bibinfo {author} {\bibfnamefont {P.~L.}\ \bibnamefont
  {Krapivsky}},\ }\bibfield  {title} {\bibinfo {title} {Finite size effects in
  addition and chipping processes},\ }\href {https://arxiv.org/abs/2307.13111}
  {\bibfield  {journal} {\bibinfo  {journal} {arXiv:2304.14661}\ } (\bibinfo
  {year} {2023})}\BibitemShut {NoStop}%
\end{thebibliography}%

\end{document}